\documentclass[notitlepage,10pt,onecolumn,superscriptaddress,aps,prx,balancelastpage,nofootinbib]{revtex4-2}
\usepackage[utf8]{inputenc}
\usepackage{amsmath,amsfonts,amssymb,fullpage,graphicx,physics,braket,mathtools}
\usepackage[dvipsnames]{xcolor,colortbl}
\usepackage[colorlinks]{hyperref}
\hypersetup{
  breaklinks=true,
  colorlinks=true,
  linkcolor=NavyBlue,
  citecolor=NavyBlue,
  urlcolor=NavyBlue,
  plainpages=false
}
\usepackage{CJKutf8}

\usepackage{algpseudocode}
\usepackage{algorithm}

\newcommand{\budgetlb}{{\text{b}}}

\let\originalleft\left
\let\originalright\right
\renewcommand{\left}{\mathopen{}\mathclose\bgroup\originalleft}
\renewcommand{\right}{\aftergroup\egroup\originalright}

\begin{document}
\title{Bosonic unstructured search: quantum enhancement without entanglement}

\author{\begin{CJK}{UTF8}{gbsn}Yu-Xin~Wang (王语馨)\end{CJK}}
\affiliation{Joint Center for Quantum Information and Computer Science, NIST/University of Maryland, College Park, Maryland 20742, USA}

\author{Erfan~Abbasgholinejad}
\affiliation{Joint Center for Quantum Information and Computer Science, NIST/University of Maryland, College Park, Maryland 20742, USA}
\affiliation{Joint Quantum Institute, NIST/University of Maryland, College Park, Maryland 20742, USA}

\author{Supratik~Sarkar}
\affiliation{Joint Quantum Institute, NIST/University of Maryland, College Park, Maryland 20742, USA}

\author{Sharoon~Austin}
\affiliation{Joint Center for Quantum Information and Computer Science, NIST/University of Maryland, College Park, Maryland 20742, USA}
\affiliation{Joint Quantum Institute, NIST/University of Maryland, College Park, Maryland 20742, USA}

\author{Andrew~M.~Childs}
\affiliation{Joint Center for Quantum Information and Computer Science, NIST/University of Maryland, College Park, Maryland 20742, USA}
\affiliation{Department of Computer Science and Institute for Advanced Computer Studies, University of Maryland, College Park, Maryland 20742, USA}

\author{Alexey~V.~Gorshkov}
\affiliation{Joint Center for Quantum Information and Computer Science, NIST/University of Maryland, College Park, Maryland 20742, USA}
\affiliation{Joint Quantum Institute, NIST/University of Maryland, College Park, Maryland 20742, USA}

\author{Mohammad~Hafezi}
\affiliation{Joint Quantum Institute, NIST/University of Maryland, College Park, Maryland 20742, USA}
\affiliation{Joint Center for Quantum Information and Computer Science, NIST/University of Maryland, College Park, Maryland 20742, USA}

\date{\today}
\begin{abstract}
We consider unstructured search using multimode bosonic quantum systems under energy constraints. The task is to locate an unknown marked mode among $N$ possibilities using arbitrary control, subject to a time-averaged mean-boson-number budget $n_{{\text{b}}}\gg 1$. We analyze a continuous-time phase-shift oracle of strength $B_{0}$, with cost given by the total oracle time, and a discrete oracle with fixed phase $\phi_{0}$, with cost given by the number of queries. We give algorithms that identify the target with query complexity $O(\sqrt{N}/{n_{{\text{b}}}}B_{0})$ for continuous-time oracles or $O(\sqrt{N}/n_{{\text{b}}} |{\sin(\phi_{0}/{2})}|)$ for discrete oracles with $n_{{\text{b}}} |{\sin(\phi_{0}/{2})}|=O(1)$. In contrast, for discrete oracles with $n_{{\text{b}}} |{\sin(\phi_{0}/{2})}|=\Omega(1)$, we provide an algorithm using $O(\max\{1,\sqrt{N/n_{{\text{b}}}|{\sin(\phi_{0}/{2})}|}\})$ queries. We also prove matching lower bounds on the query complexity, showing that these algorithms are asymptotically optimal. Furthermore, we show that, in the absence of loss, non-Gaussianity is necessary to achieve the optimal joint query scaling in $N$ and $n_{{\text{b}}}$, but entanglement is not. Under uniform boson loss at rate $\gamma$, we prove a query lower bound of $\Omega ( \gamma N / n _{{\text{b}}} B_{0}^{2} )$, attained for $\gamma = \Omega(B_{0} \sqrt{n _{{\text{b}}}/N})$ using coherent-state probes. Finally, we propose experimental implementations using multipass optical microscopy or superconducting circuits as potential routes to speeding up imaging or device calibration with a low boson-number budget.

\end{abstract}

\maketitle

\tableofcontents

\section{Introduction}

Unstructured search is one of the most basic primitives in quantum computing. In this task, the goal is to identify an unknown marked element $x_{0}$ via oracle access to the unitary operator $\hat U _{x_{0}} = \hat {\mathbb{I} } - 2\ketbra{x_{0}}$ acting on an $N$-dimensional quantum system. The Grover algorithm identifies the value of $x_{0}$ using only $\Theta(\sqrt{N})$ queries to $\hat U _{x_{0}} $~\cite{Grover1996}, offering a $\Theta(\sqrt{N})$ speedup over classical strategies, which is optimal~\cite{BBBV1994}. The mechanism of the Grover algorithm also enables many quantum algorithms and subroutines, including amplitude amplification and quantum-walk search.

While the unstructured search problem is typically considered in a finite-dimensional setting, bosons with an infinite-dimensional Hilbert space present a useful and qualitatively different alternative~\cite{RMP.CVQI2005}. Indeed, bosonic systems including optical~\cite{Furusawa2012,Bowen2013,Bowen2021}, photonic~\cite{SafaviNaeini2023}, microwave~\cite{sensing_axion2021,Chou2021,SunLuyan2022}, and acoustic~\cite{Schoelkopf2018,Cleland2018,SafaviNaeini2019} systems provide sensors for a wide variety of physical signals including temperature, electric and magnetic fields, mechanical force and motion, and charge~\cite{Mitchell2010,StamperKurn2014,SafaviNaeini2019,Schmidt2019,Leibfried2019,Rimberg2021}. Although practical sensors rarely map exactly to the problem of unstructured search, the ubiquity of bosonic quantum systems motivates bosonic unstructured search. Specifically, in many settings, locating an unknown marked site directly involves bosonic sensors. In these platforms, the natural search target is often a target mode instead of a specific quantum level. Correspondingly, a Gaussian phase shift $e^{-i\phi\, \hat{a}^{\dagger}_{x_0}\hat{a}_{x_0}}$
provides a physically relevant oracle unitary for bosons. Here, $\hat{a}_{x}$ denotes the bosonic lowering operator of probe mode ${x} \in \{ 1,2,\ldots, N \}$ satisfying the canonical commutation relations $[\hat{a}_{x }, \hat{a} ^\dag_{x'}] = \delta_{xx'} $ and $[\hat{a}_{x }, \hat{a} _{x'}] = 0 $, and $x_{0}$ labels the marked mode.

The bosonic unstructured search problem directly relates to a variety of applications including scanning optical imaging and microscopy~\cite{Bowen2013,Bowen2021,Kasevich2016}, distributed phase sensing~\cite{Furusawa2012,ZhangZheshen2021,SafaviNaeini2023}, and the calibration of multimode quantum devices~\cite{SafaviNaeini2019}, where one searches an array of pixels, optical cavities, or resonators for the one that carries the signal. In contrast to the qubit case, because the Hilbert space now is infinite-dimensional, any meaningful query complexity statement for the bosonic unstructured search problem must be made under a boson-number constraint. Indeed, in many of the applications of the bosonic search problem, the number of bosons (e.g., photons) that may be sent into the sample constitutes a constrained resource, which can be limited by photodamaging or photobleaching effects in biological imaging~\cite{Bowen2013,Bowen2021,Kasevich2016,Kasevich2025} or unwanted higher-order nonlinearities and heating in nonlinear optical devices and superconducting circuits~\cite{Martinis2016}. In this context, the energy-constrained unstructured search problem is also practically relevant.

Prior works have considered two different extensions of a Grover-style problem to bosonic systems: first, a search over multiple bosonic modes with an oracle unitary acting on an unknown bosonic mode~\cite{Spreeuw2002}; second, using a single bosonic mode and a dense encoding to simulate the Grover algorithm~\cite{Lloyd2000,Braedler2018,Sargsyan2021}. While the second option is algorithmically more efficient and directly maps to the standard unstructured search problem, the oracle and control unitaries are typically not compatible with natural image or sensing processes in continuous-variable (CV) systems. In contrast, the first option (i.e., a search over multiple bosonic modes) is directly compatible with existing continuous-variable systems and provides a candidate for speeding up imaging applications in the low boson-budget regime. However, existing works focus on specific protocols or experimental implementations, and a systematic and rigorous study of the query complexity under a constraint on energy (the
resource that is actually constrained in practice) and optimal protocols for the physically motivated bosonic unstructured search problem are still lacking. In particular, it is known that even a purely coherent-state strategy can realize a Grover speedup in such a search problem~\cite{Spreeuw2002}. However, the existence and nature of quantum speedup using resources such as entanglement and non-Gaussianity for the unstructured search problem in the bosonic context remain an open problem.

In this work, we present a complete query-complexity analysis for the $N$-mode bosonic search problem with oracle strength $B_{0}$ under an upper bound $n_{\budgetlb}$ on the time-averaged mean boson number. Our contributions are as follows:
\begin{enumerate}
    \item {In the noiseless regime, we prove a lower bound $T_{\mathrm{tot}} = \Omega(\sqrt{N}/n_{\budgetlb}B_0)$ on the total oracle time, valid for arbitrary ancillas, adaptive control, and measurements (Sec.~\ref{nsec:Ttot.gen.bound}). The proof requires a technique that tolerates an unbounded oracle generator with only a constraint on its mean, and hence does not follow from existing finite-dimensional arguments.}
    \item {We prove a strictly stronger bound $T_{\mathrm{Gaussian}} = \Omega(\min\{\sqrt{N/n_{\budgetlb}},N/n_{\budgetlb}\}B_0^{-1})$ for $n_{\budgetlb}\gg 1$, for every strategy whose reduced, unconditional multimode probe state remains Gaussian throughout the search, including adaptive strategies satisfying this condition
    (Sec.~\ref{nsec:Gaus.bound}).
    Here ``unconditional'' means averaged over any intermediate
measurement outcomes.
    For $1\ll n_{\budgetlb}=O(N)$, active squeezing therefore cannot improve the asymptotic query-time scaling once the energy it injects is accounted for.}
    \item {Under uniform (linear) boson loss at rate $\gamma$, we prove the additional query lower bound $T_{\mathrm{tot}} = \Omega(N\gamma/n_{\budgetlb}B_0^{2})$ (Sec.~\ref{nsec:loss.bound}). For $\gamma=\omega(B_0/\sqrt{N})$, the loss-dependent lower bound dominates, while for $\gamma=O(B_0/\sqrt{N})$ the Grover-style speedup survives. For $\gamma = \Omega(B_{0} \sqrt{n _{{\text{b}}}/N})$, the bound is attained by probing 
    clusters of modes in parallel using only coherent states and optical linear operations (Sec.~\ref{nsec:opt.protocol.ent}), so that neither non-Gaussianity nor entanglement provides any query-complexity scaling advantage in this regime.}
    \item {We also extend the analysis to fixed-phase oracles and prove a tighter query lower bound in this case, with the total number of queries $K _{\text{tot}} = \Omega ( \sqrt{N} / \min \{ n _{\budgetlb} |{\sin (\phi_{0}/{2})}| , \sqrt{n  _{\budgetlb} | \sin (\phi_{0}/{2})|} \} )$ (Sec.~\ref{nsec:oracle.phi0}). In this case, the optimal boson-number scaling only holds for sufficiently small oracle phase satisfying $n _{\budgetlb} |{\sin (\phi_{0}/{2})}| = O(1)$. }
    \item {We propose experimental implementations attaining the general bound, including an entangled protocol using a Fock-analog of W states (Sec.~\ref{nsec:nonG.ent.expt}). Surprisingly, a fully unentangled product-state protocol built from single-mode superpositions also achieves the same scaling (Sec.~\ref{nsec:nonG.prod.expt}). Our results identify non-Gaussianity rather than entanglement as the resource responsible for the optimal joint mode-number and boson-number scaling.}
\end{enumerate}

The remainder of the paper is organized as follows. Section~\ref{nsec:setup} defines the oracle models and the energy constraint, Sec.~\ref{nsec:summary} summarizes our results in more detail, and Sec.~\ref{nsec:literature} reviews related work. Section~\ref{nsec:query.lower.bounds} contains the proofs, and Secs.~\ref{nsec:nonG.prod.expt} and~\ref{nsec:nonG.ent.expt} present the experimental proposals for the unentangled and entangled strategies, respectively.

\section{Problem setup}
\label{nsec:setup}

\begin{figure}[hbt]
\centering
\includegraphics[width=.4\textwidth]{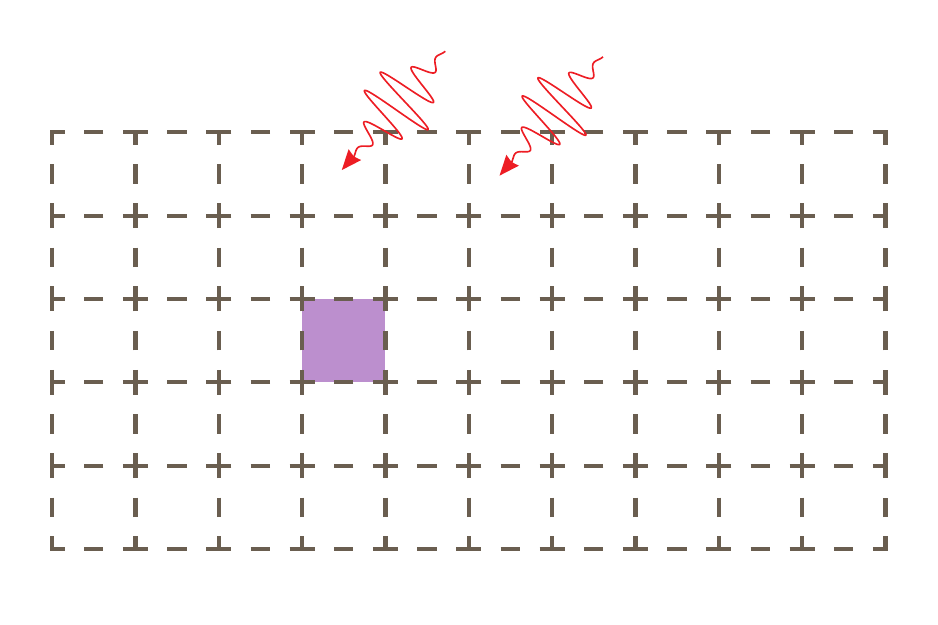}
\caption{Schematic showing the sample space with a marked element that applies the unitary $\hat U _{x_{0} } (\phi_{0}) $ on any probe light reflected off site $x_{0} $. Because our analysis allows nonlocal control operations on the probes, our results are independent of the dimension and connectivity of the sample space. 
}
\label{fig:schmatic}
\end{figure}

Consider a sample space containing $N$ pixels. There exists a marked pixel, whose location (labeled by mode index $x_{0}\in \{ 1,2,\ldots,N\}$) is unknown at the beginning of the algorithm (see Fig.~\ref{fig:schmatic}). The goal is to identify the location of this marked pixel with a high probability.
The sample space can be probed using light, or more generally, a bosonic multimode quantum system.\footnote{We use photon and boson interchangeably in the rest of this paper when it does not cause confusion.}
The target induces a phase shift $\Delta \phi = B _{0} \Delta t$ per boson
to any light reflected from or transmitted through it, where $B _{0} $ denotes a known single-boson coupling constant between the target and the probe and $\Delta t$ the continuous interaction time. Alternatively, the target-probe interaction time can be fixed to be a given constant, corresponding to an alternative fixed-phase oracle model with a constant phase rotation $\phi_{0}$.

Formally, we consider query access to a unitary oracle. The unitary can be either a continuous-time rotation, or alternatively, a fixed-phase rotation generated by the number operator at $x_{0}$, $\hat n _{x_{0}} \coloneqq \hat a _{x_{0}} ^\dag \hat a _{x_{0}} $: 
\begin{align}
\text{Continuous-time unitary query: }&
\hat U _{x_{0}} (\Delta \phi = B _{0} \Delta t ) = e^{-i (B _{0} \Delta t) \hat n _{x_{0}}} \tag{i}
\label{neq:time.oracle.def}
; \\
\text{Fixed-phase unitary query: }&
\hat U _{x_{0} } (\phi_{0}) = e^{-i \phi_{0} \hat n _{x_{0}} } \tag{ii}
\label{neq:phi.oracle.def}
.
\end{align}
For a given search protocol, we call the time-dependent state of the probe and ancillas generated when the oracle acts on mode $x_{0}$ the \textit{oracle trajectory} associated with $x_{0}$; identifying the marked mode amounts to distinguishing the $N$ oracle trajectories at the end of the protocol.
In model~\eqref{neq:time.oracle.def}, we assume $B _{0} >0$, since complex-conjugating all controls and measurements in the Fock basis maps any protocol for $B _{0}<0$ to one for $B _{0}>0$ with the same success probability and boson-number usage 
(see the discussion following Eq.~\eqref{neq:pth.1norm.coh} for detail). 
Without loss of generality, we also assume $\phi_{0} \in (0,  \pi]$ in model~\eqref{neq:phi.oracle.def}, since the fixed-phase oracle is periodic in $\phi_{0} $ with a $2\pi$ period, and $\phi_{0} $ and $2\pi -\phi_{0} $ are related by the same complex-conjugation argument as $\pm B_{0}$. Here, we exclude the edge case where the oracle does not induce any observable impact on the probes.
The search protocol cost (i.e., the \emph{query complexity}) is measured by the total time $T_{\text{tot}}$ querying the oracle for model~\eqref{neq:time.oracle.def}, or the total number of fixed-phase queries $K _{\text{tot}}$ for model~\eqref{neq:phi.oracle.def}. 

By definition, the fixed-phase query models~\eqref{neq:phi.oracle.def} form a subset of the continuous-query model~\eqref{neq:time.oracle.def}.
Thus, for any fixed pair of parameters $B _{0} $ and $\phi_{0}$, the corresponding query complexity satisfies 
\begin{align}
\label{neq:Ktot.phi0.bound.inT}
\left (K_{\text{tot}}\right ) _{\min}  \ge \frac{B _{0} }{\phi_{0} }
\left ( T _{\text{tot}}\right ) _{\min}
\end{align}
(where $\min$ indicates minimization over protocols), so that the discrete oracle model~\eqref{neq:phi.oracle.def} has a query complexity scaling that is at least as large as $(B _{0} /\phi_{0} ) \left ( T _{\text{tot}}\right ) _{\min} $. 
In Sec.~\ref{nsec:oracle.phi0}, we show that the oracle model~\eqref{neq:phi.oracle.def} leads to a similar query complexity scaling in $n_{\budgetlb} $ when the phase parameter satisfies 
\begin{align}
n _{\budgetlb} |{\sin(\phi_{0}/{2})}| = O(1).
\end{align}
In contrast, when $|{\sin(\phi_{0}/{2})}|=\Theta(1)$, the query complexity scaling with $n _{\budgetlb}$ for model~\eqref{neq:phi.oracle.def} is strictly higher than~\eqref{neq:time.oracle.def}. Unless stated otherwise, we focus on the continuous-time oracle model~\eqref{neq:time.oracle.def} throughout; the fixed-phase model~\eqref{neq:phi.oracle.def} is analyzed separately in Sec.~\ref{nsec:oracle.phi0}. 
To model imperfections in realistic implementations, we also assume that all probe bosons are subject to linear boson loss at rate $\gamma$ for continuous-time queries (see Eq.~\eqref{neq:time.oracle.gamma.def} and the associated discussion in Sec.~\ref{nsec:loss.bound} for detail).

Our goal is to identify $x_{0}$ using the shortest total time querying the oracle (or alternatively, the smallest number of queries for model \eqref{neq:phi.oracle.def}). We say a protocol solves the search problem if, for every $x_{0}$, it outputs $x_{0}$ with probability at least\footnote{Note that this threshold probability can be taken as arbitrary constant $>1/2$; here, we assume its value is $2/3$ for concreteness.} $2/3$. By majority vote over $O(\log (1/\epsilon))$ independent repetitions, the error can be reduced to $\epsilon$. 
We first note that without a constraint on the total probe-boson number, the unstructured search problem can be solved in arbitrarily short time, or with $O(1)$ discrete queries. For example, preparing each pixel's probe in a superposition of Fock states $(\ket{0} + \ket{M})/\sqrt{2}$ enables detection of the marked element in $O(1/M B_{0})$ query time. This search time does not depend on the total number of modes $N$. However, this strategy involves sending $\Theta(NM)$ total bosons to the sample. In optical imaging, this can cause significant photo damaging or bleaching on samples that are sensitive to the peak or time-integrated photon number. Moreover, large boson numbers in the probe modes can themselves be a costly resource: preparing nonclassical probe states, such as high-number Fock states or strongly squeezed states, becomes increasingly demanding in time and fidelity as the boson number grows, while strong driving in nonlinear platforms such as superconducting circuits can induce unwanted higher-order nonlinearities and heating~\cite{Martinis2016}. 
Motivated by the desire to use a low total probe-boson-number budget in many applications, and to derive a meaningful query lower bound, in the rest of this work we focus on bosonic unstructured search subject to a constraint on the time-averaged expectation value of the total boson number: 
\begin{align}
\label{neq:def.nS.avg.var}
\frac{1}{T_{\text{tot}}}\int ^{T_{\text{tot}}} _{0}
\braket{ \hat n _{\text{S}} (t) } dt \le n _{\budgetlb} 
,
\end{align}
where $\hat n _{\text{S}} = \sum _{x=1}^{N} \hat a _{x} ^\dag \hat a _{x}$ denotes the total probe boson number operator and $T_{\text{tot}}$ denotes the total query time. As Eq.~\eqref{neq:def.nS.avg.var} limits the time-averaged total energy in the probe modes, we also refer to it as an energy constraint, and we use the terms \textit{boson-number constraint} and \textit{energy constraint} interchangeably in this work. This time-averaged boson-number constraint is weaker than a constraint on the peak power (i.e., the maximal instantaneous mean boson number), so all our lower bounds 
automatically apply to the case where the peak power is limited. Further, the optimal protocols in Sec.~\ref{nsec:summary} (see Sec.~\ref{nsec:opt.protocol.ent}) satisfy a peak-power constraint: in all protocols, the instantaneous mean boson number stays at or below $n _{\budgetlb} $ 
at every moment by construction (see Sec.~\ref{nsec:query.alt.limit} for detail), so they remain optimal in that setting as well. 
In Sec.~\ref{nsec:loss.bound}, we also assume the regularity condition
\begin{align}
\label{neq:nS.regular.cond}
\forall t, \quad 
\braket{ \hat n _{\text{S}} ^{2} (t) } < \infty
,
\end{align}
i.e., the second-order moment of the total probe-boson number stays finite throughout the protocol.
This is a mild regularity assumption rather than a resource constraint: it does not bound $\langle \hat n_{S}^{2} \rangle$ by any particular value and imposes no cutoff on the boson number. It holds automatically for any state supported on finitely many Fock states, for all Gaussian states, and for every protocol considered in this work; it excludes only states with heavy-tailed boson-number distributions, e.g., $p(n) \propto n^{-3}$, which has a finite mean but an infinite second moment. We use Eq.~\eqref{neq:nS.regular.cond} only in the proof of the lossy lower bound in Sec.~\ref{nsec:loss.bound}, where it ensures that the stochastic dilation of the loss channel is well defined.

\section{Summary of results}
\label{nsec:summary}

In this section, we summarize the main results on query complexity and optimal protocols for bosonic unstructured search under an energy constraint.
In a search protocol, we allow an arbitrary number of fresh (qubit and bosonic) ancillas
and general quantum and classical control operations. However, we assume that the time-averaged total boson number in the probe modes throughout the protocol is upper bounded by a constant
as in Eq.~\eqref{neq:def.nS.avg.var}.
We also assume that $n _{\budgetlb} \gtrsim 1 $ unless explicitly noted otherwise.\footnote{\label{footnote:small_nb}In the opposite limit $n _{\budgetlb} \ll 1 $, the general bound~\eqref{neq:Ttot.lower.bound.0} is attained by coherent states, so non-Gaussian resources offer no advantage. In this regime, the bosonic modes behave similarly as qubits subject to a corresponding energy constraint.}
We obtain the following lower bound on the complexity of bosonic unstructured search with a boson-loss rate $\gamma$, under the regularity condition~\eqref{neq:nS.regular.cond} (see Sec.~\ref{nsec:loss.bound} for a detailed proof):
\begin{align}
\label{neq:Ttot.lower.bound.0}
T_{\text{tot}}  = \Omega \left (\max \left \{ \frac{\sqrt{N}}{ n _{\budgetlb} B _{0} }
, \frac{N \gamma }{ n _{\budgetlb} B _{0} ^{2}  } 
\right \} 
\right )
.
\end{align}
Thus, when 
$\gamma = \omega(B_0/\sqrt{N})$ (i.e., $\gamma \gg B_0/\sqrt{N}$),
the quadratic speedup in $N$ is lost, and the loss-dependent lower bound dominates the noiseless bound. In particular, for fixed $\gamma/B_{0}$ and fixed boson-number budget, the query lower bound is linear in $N$.
In contrast, if 
$\gamma = O(B_0/\sqrt{N})$
then the first term in Eq.~\eqref{neq:Ttot.lower.bound.0} dominates, and the bound is the same as in the lossless case.

To better understand what protocols saturate the optimal bounds in Eq.~\eqref{neq:Ttot.lower.bound.0}, it is natural to ask whether different sources of non-classicality, including non-Gaussianity and entanglement, are required to achieve the optimal scaling  $ O (\sqrt{N}/n _{\budgetlb}B_{0})$, i.e., Grover scaling in $N$ and Heisenberg scaling in $n _{\budgetlb}$.
For this purpose, we consider the performance of various Grover-style algorithms using Gaussian and non-Gaussian states. In particular, Ref.~\cite{Spreeuw2002} showed that a coherent-state strategy can achieve the Grover scaling $O(\sqrt{N})$, demonstrating that neither non-Gaussianity nor entanglement is required for the Grover speedup in the low-loss regime. However, Ref.~\cite{Spreeuw2002} did not consider any speedup from using more photons, and their algorithm's claimed search time scaling is  suboptimal in $n _{\budgetlb}$ for $n _{\budgetlb}\gg 1$. In contrast, in Sec.~\ref{nsec:Gaus.bound}, we prove that fully Gaussian probes (i.e., Gaussian inputs and fully Gaussian controls) can achieve at best $\Theta(\min\{\sqrt{N/n_{\budgetlb}},N/n_{\budgetlb}\}/B_{0})$, which still allows quadratic speedup in $N$ when $n_{\budgetlb}=O(N)$ but exhibits suboptimal joint scaling in $N$ and $n_{\budgetlb}$. 
In Sec.~\ref{nsec:loss.bound}, we further prove the $\gamma$-dependent query complexity lower bound [see Eq.~\eqref{neq:Ttot.lower.bound.0}] for bosonic unstructured search under boson loss.
For $\gamma = \Omega(B_{0} \sqrt{n _{{\text{b}}}/N})$, the best possible scaling for bosonic unstructured search is $\Theta (\gamma N/ n _{\budgetlb}B_{0}^{2})$, and this $n _{\budgetlb}^{-1}$ scaling in $n_{\budgetlb}$ is achieved by a classical strategy using only coherent-state probes, as shown in Sec.~\ref{nsec:opt.protocol.ent}.

We now discuss protocols that achieve the optimal bounds in Eq.~\eqref{neq:Ttot.lower.bound.0}.
First, it is straightforward to give an entangled non-Gaussian algorithm for bosonic unstructured search: we can take the usual Grover search algorithm and promote the single-particle states on each individual site to the $M$-photon Fock state in the $x$th mode, $\ket{ M } _{x} $, where we let $M \coloneqq \lfloor n _{\budgetlb} \rfloor$. The initial state is
\begin{align}
\label{neq:nph.Wstate}
\ket{\psi _{\text{in}}} &= \frac{1}{\sqrt{N}} \sum _{x=1}^{N}    \ket{ M }  _{x}
= \frac{1}{\sqrt{N}} \sum _{x=1}^{N} \frac{1}{\sqrt{ M !}}\hat a _{x}  ^\dag {}^{M} \ket{\text{vac}}
. 
\end{align}
This is a strongly non-Gaussian state, which we call a $W$-Fock state since it resembles the standard $W$ state. The Grover diffusion operator in this case reads $\hat U _{\text{G}} = \hat {\mathbb{I} } - 2\ketbra{\psi _{\text{in}}} $. A Fock-state-analog of the standard Grover search algorithm now reads
\begin{align}
\label{neq:Wstate.Grover}
\left ( \hat U _{\text{G}} e ^{-i B _{0} \Delta t \hat a _{x_{0}} ^\dag \hat a _{x_{0}} } \right ) ^{K} \ket{\psi _{\text{in}}} 
.
\end{align}
Choosing $\Delta t = \frac{\pi}{M B _{0}} $ and $K = O(\sqrt{N})$, we recover the standard Grover search algorithm with a discrete oracle. Measuring all probe modes using photon counting identifies $x _{0}$
with high probability in $ K  \Delta t = O (\sqrt{N} /  n _{\budgetlb} B_{0})$ time.

Surprisingly, the loss-free limit of the lower bound in Eq.~\eqref{neq:Ttot.lower.bound.0} can be achieved without using any entanglement. Specifically, we can use an ancilla-free protocol that only involves fully product states of the probe modes to realize the same performance as the entangled protocol discussed above. The initial probe state is
\begin{align}
\label{neq:nph.FockSum.prod}
\ket{\psi _{\text{prod}}} & = \bigotimes_{x=1}^{N}  \ket{\psi _{M';N}}_{x}  
\end{align}
with
\begin{align}
\ket{\psi _{M';N}}_{x} & \coloneqq \sqrt{1-\frac{1}{N}} \ket{0}_{x}
+ \frac{1}{\sqrt{N}} \frac{1}{\sqrt{ M' !}}\hat a _{x}  ^\dag {}^{ M' } \ket{0}_{x}
\nonumber \\
\label{neq:nph.FockSum.prod.sing}
& = \sqrt{1-\frac{1}{N}} \ket{0}_{x}
+ \frac{1}{\sqrt{N}} \ket{M'}_{x}
,
\end{align}
where we let $M' = \lfloor n _{\budgetlb}/2 \rfloor$.
The search protocol corresponds to a Grover-type amplitude-amplification procedure on each individual mode
with the Grover iteration now generated by a reflection with respect to the states $\ket{\psi _{M';N}}_{x} $:
\begin{align}
\label{neq:prod.Grover}
\left ( e ^{- i \pi \sum _{x=1}^{N} \ketbra{\psi _{M';N} }_{x} }
e ^{-i B _{0} \Delta t \hat a _{x_{0}} ^\dag \hat a _{x_{0}} }
\right ) ^{K} \ket{\psi _{\text{prod}}}
.
\end{align}
Choosing $\Delta t = \frac{\pi}{M' B _{0}} $ and $K = O(\sqrt{N})$, we again achieve the lower bound in Eq.~\eqref{neq:Ttot.lower.bound.0} using fully product probe states.

Search protocols that only involve Gaussian inputs and linear optical operations are typically much simpler to implement in practice compared with ones that use non-Gaussian controls. 
We can achieve a similar scaling with $N$ using only Gaussian inputs and linear optical operations. In fact, as shown in~\cite{Spreeuw2002}, fully coherent-state probes can achieve an $O(\sqrt{N})$ time-complexity in unstructured search.\footnote{However, Ref.~\cite{Spreeuw2002} does not discuss optimal search cost scaling in $n _{\budgetlb}$.} To see this, one can prepare a coherent initial state of the form 
\begin{align}
\ket{\psi _{\text{coh}}} = \ket{\alpha} ^{\otimes N}
,
\end{align}
where $\alpha = \sqrt{n _{\budgetlb} /N}$ is the coherent-state amplitude chosen to satisfy the boson-number constraint.
A general bosonic Grover-style algorithm using only Gaussian controls rotates the initial state to 
\begin{align}
\label{seq:boson.Grover}
\ket{\psi _{\text{out},K}}
&= \left ( e ^{-i  \delta \Delta t \hat n _{ k = {0}}  } e ^{-i B _{0}  \Delta t  \hat n _{x_{0}}   }
\right ) ^{K}
\ket{\psi _{\text{coh}}}
\end{align}
in query time $T _{\text{tot} } = K \Delta t$,
where $\delta$ is a constant that determines the speed of the Grover diffusion, $\Delta t$ denotes the single-step oracle query time, and $\hat n _{ k = {0}} \coloneqq \hat a ^{\dag} _{ k = {0}} \hat a _{ k = {0}} $ denotes the number operator for the center-of-mass mode $\hat a _{ k = {0}}  \coloneqq \frac{1}{\sqrt{N}} \sum _{x } \hat a _{x}$. 
To illustrate the idea, we follow~\cite{Spreeuw2002} and choose $\Delta t = \frac{\pi}{B _{0} }$ and $\delta=B_{0}$. The unitary evolution effectively concentrates the boson population into the target $x_{0}$
mode at a rate $O (N ^{-\frac{1}{2}})$. More specifically, we can analytically derive the boson population in the target and non-target modes, respectively, as
\begin{subequations}
\label{neq:coh.K.pop.sum}
\begin{align}
\braket{\hat n _{x_{0}} ( K \Delta t ) } &
= n _{\budgetlb} \sin ^{2} [ (2 {K} + 1) \varphi]
, \\
\braket{\hat n _{x\ne x_{0}} ( K \Delta t ) } 
& = \frac{n _{\budgetlb} }{N-1} \cos ^{2} [ (2 {K} + 1) \varphi]
,
\end{align}
\end{subequations}
where $\varphi = \arcsin {(} N ^{-\frac{1}{2}} )$ denotes the phase of a beamsplitter operation that transforms between $\hat a_{x_{0}}$ and the collective momentum eigenmode operator $\hat a _{ k = {0}} $ (see Appendix~\ref{app.sec:coh.npop.K} for detail).

In the $N\to \infty$, $n _{\budgetlb} \ll N $ limit, the non-target mode boson number satisfies  
\begin{align}
\label{neq:coh.nbg.K}
\braket{\hat n _{x\ne x_{0}} ( K \Delta t ) } = O \left (\frac{n _{\budgetlb} }{N} \right ) \ll 1
\end{align}
throughout the protocol. In contrast, the mean occupation of the target $x_{0}$ mode is 
\begin{align}
\label{neq:coh.nx0.K}
\braket{\hat n _{x_{0}} ( K \Delta t ) } & = n _{\budgetlb} \sin ^{2} [ (2 {K} + 1) \varphi]
.
\end{align}
We can approximate the target-mode boson number in the $1 \ll K \ll \sqrt{N}$ regime as 
\begin{align}
\label{neq:coh.nx0.fin}
\braket{\hat n _{x_{0}} ( K \Delta t ) } \simeq 4K^{2} \frac{n _{\budgetlb} }{N}
,
\end{align}
which reaches $\Omega(1)$ for $K = \Theta(\sqrt{N/n_{\budgetlb}})$, so the target mode is distinguishable from the background~\eqref{neq:coh.nbg.K} after $K = \Theta(\sqrt{N/n _{\budgetlb} })$ steps.
One can apply a global displacement that uniformly shifts the amplitude of all modes by the same constant amplitude $O(\sqrt{n _{\budgetlb}/{N}})$ to transform all the background modes into vacuum, while the target mode retains a $\Theta(1)$ amplitude. Consequently, one can distinguish the target $x_{0}$ mode from the rest of the probe modes with constant probability, which can be boosted to $1-\epsilon$ for arbitrarily small $\epsilon$ using $O(\log(1/\epsilon))$ repetitions.

In summary, the Gaussian unstructured search protocol presented in Eq.~\eqref{seq:boson.Grover}, using a displaced photon counting detection procedure, succeeds with high probability after time $O(K \Delta t ) = O(B _{0}  ^{-1} \sqrt{N/n _{\budgetlb}})$. 
We also show that this is optimal: for $n_{\budgetlb}=O(N)$, any protocol 
whose reduced, unconditional multimode probe state remains Gaussian throughout the search
must use query time
\begin{align}
\label{neq:Ttot.lb.gen.coh.0}
n_{\budgetlb}=O(N): \quad 
T _{\text{Gaussian}} = \Omega \left ( \frac{\sqrt{N}}{ B _{0} \sqrt{n _{\budgetlb}} } \right ) 
,
\end{align}
indicating that, in this regime, active squeezing during the search cannot improve the asymptotic query-time scaling after we take into account the additional energy injected during squeezing (see Sec.~\ref{nsec:Gaus.bound} for detail). We note that this lower bound holds for arbitrary final readout methods, provided that the reduced, unconditional multimode probe state remains Gaussian throughout the protocol. This result proves that non-Gaussian probes are required to achieve the optimal joint scaling in $N$ and $n _{\budgetlb}$ for unstructured search.

\section{Literature review}
\label{nsec:literature}

Bosonic unstructured search algorithms have been previously considered in two contexts: first, building a bosonic analog of the Grover algorithm; second, using bosonic systems to simulate a Grover algorithm. In the latter case, the oracle and control unitaries are typically not compatible with natural imaging or sensing processes in continuous-variable (CV) systems. In contrast, our work proposes a bosonic implementation of the search algorithm that is directly compatible with existing continuous-variable systems and can potentially speed up imaging applications in the low boson-number regime. Further, we provide a rigorous analysis of the lower bounds on the search time as a function of the boson-number budget [Eq.~\eqref{neq:def.nS.avg.var}], which has not been done in previous works.

Here, we review references on various bosonic generalizations of Grover search and, more broadly, continuous-variable sensing, channel discrimination and quantum algorithm applications.

Several papers study a continuous-variable version of the Grover algorithm~\cite{Lloyd2000,Braedler2018,Sargsyan2021}. Those works consider a search problem with respect to the eigenbasis of the quadrature operators, with the query operator $\hat U_{x_{0}} = \exponential ( -i\pi\int ^{x_{0}+\Delta } _{x_{0}} \ketbra{x} \, dx )$ defined with respect to idealized quadrature eigenstates, which can only be approximated with finite resources, and the corresponding Grover iteration involves non-Gaussian operations. These oracle and encoding models differ from the native phase oracle considered here, so the boson-number-constrained bounds we prove do not directly apply to them.
Further, those works do not rigorously analyze the optimality of the proposed algorithms; a quantum speedup is implied but not proven.

In contrast to Refs.~\cite{Lloyd2000,Braedler2018,Sargsyan2021}, we consider a distinct oracle that is directly compatible with imaging and sensing applications using optical, photonic, and other bosonic systems (such as superconducting circuits and motional, phononic, and acoustic modes of quantum systems). Our oracle involves a simple phase shift unitary, which is ubiquitous in all types of bosonic systems, and our probe state contains a finite number of bosons. A closely related search problem is considered in Ref.~\cite{Spreeuw2002}, which proposes and implements a bosonic unstructured search algorithm that achieves $O(\sqrt{N})$ speedup using only coherent states. In contrast, our coherent-state protocol achieves $O(\sqrt{N/n _{\budgetlb}})$ scaling, which we show is optimal among Gaussian probes for $n_{\budgetlb}=O(N)$. We further prove a complexity separation between non-Gaussian and Gaussian probes and propose two non-Gaussian algorithms that achieve the optimal joint mode-number and boson-number scaling.

Grover and Sengupta proposed using \textit{classical} coupled pendulums to achieve $O(\sqrt{N})$ scaling in a continuous-variable version of Grover search~\cite{Sengupta2002}. Our analysis further shows that the performance of a continuous-variable algorithm crucially relies on the total boson number used. The non-Gaussian $W$-Fock state \eqref{neq:nph.Wstate} used in the non-Gaussian Grover algorithm has also been studied in metrological settings~\cite{Chan2018}. However, its use and optimality in bosonic unstructured search, and the strategies for deterministically implementing the non-Gaussian Grover diffusion operator using experimentally accessible ingredients, are novel.

Another set of references considers imaging or bosonic channel discrimination problems. Specifically, Ref.~\cite{Yen2011} considers a single-pass imaging setup, where probe light is sent to a detector immediately after being reflected off a target. Further, Ref.~\cite{Pirandola2020} discusses an entanglement enhancement in hypothesis testing of multiple bosonic channels, Ref.~\cite{NairGu2023} considers parameter estimation for phase-covariant optical channels, Refs.~\cite{Yuasa2019,Guha2026} study quantum Fisher information of Gaussian metrology, and Ref.~\cite{Wilde2026} derives upper bounds on error exponents for bosonic channel discrimination subject to energy constraints. In all of those cases, the target channel is motivated by physically relevant sensing problems, but a Grover speedup is not considered (and sometimes not possible) because it is beyond the scope of the specific sensing or detection scheme. In a different context, Refs.~\cite{Maccone2025,Yang2026} derive energy-constrained bounds for bosonic channel estimation. However, the metrological bounds in those works are distinct and not directly applicable to the unstructured search problem considered here.

There are also several papers that investigate more general quantum algorithms using continuous-variable systems or CV-DV hybrid systems.  This includes an all-optical Grover search over a four-element database with qubits encoded in polarization and spatial modes~\cite{White2000},
continuous-variable quantum computing with encrypted data~\cite{Andersen2016}, quantum computing with photonic cluster states~\cite{Braedler2018} and GKP qubits~\cite{Zinner2025}, a continuous-variable version of quantum approximate optimization algorithms~\cite{Killoran2019}, and quantum signal processing with continuous-variable systems~\cite{Chuang2023} and hybrid DV-CV systems~\cite{Wiebe2026}. Those works provide a more algorithmic perspective to the problem and some include rigorous performance analysis for the corresponding task. However, these works either address different computational tasks or implement Grover search using different encodings and oracle models. Specifically, the Grover-search constructions using photonic cluster states~\cite{Braedler2018}, GKP qubits~\cite{Zinner2025}, and continuous-variable QAOA~\cite{Killoran2019} employ non-Gaussian resource states or operations. Their search oracles act on encoded logical states or distinguish regions of quadrature space.\footnote{Universal continuous-variable quantum computation requires resources beyond Gaussian states, Gaussian operations, and Gaussian measurements~\cite{Braunstein1999,Nemoto2002}. In this context, the fact that purely coherent states can already achieve the Grover speedup may seem surprising. However, the implementation of an unstructured search algorithm does not require a universal quantum gate set; indeed, efficient classical simulability of Gaussian circuits~\cite{Nemoto2002} does not preclude a quadratic reduction in oracle queries, which is achievable even with classical waves~\cite{Spreeuw2002}.} This differs from the unstructured search problem considered here, where the marked element produces a phase shift on an unknown bosonic mode, and the lossless Grover speedup can be achieved using coherent probes and Gaussian oracle and control operations.

On a related note, Ref.~\cite{Rouze2021} proves a general upper bound on bosonic state distinguishability that only involves averaged Hamiltonian energy $E$ generated by, e.g., the bosonic number operator, but does not require constraining the variance or any higher-order moments of the bosonic state. However, that bound does not directly apply to the bosonic unstructured search problem considered in our work: it provides a state distinguishability bound that scales as $\sqrt{E t}$, and the technique does not generate a meaningful bound when directly applied to the continuous-time limit.

Lastly, we note that discrete- and continuous-time unstructured search are both well-studied, and prior works have provided both query lower bounds and efficient algorithms available for finite-dimensional systems~\cite{BBBV1994,Gutmann1998}. Further, finite-dimensional continuous-time oracles can be simulated using discrete queries~\cite{Cleve2009} with only constant overhead~\cite{Szegedy2011}, so the two oracle models give the same query complexity scaling for finite-dimensional systems. 
In contrast, as we prove in Sec.~\ref{nsec:oracle.phi0}, the two oracle models can exhibit qualitatively different query complexity scaling for bosonic systems.

As for unstructured search in the presence of noise and dissipation, Regev and Schiff proved that no quantum speedup exists for a special kind of faulty oracle~\cite{Regev2008}, and later works~\cite{Temme2014,Rosmanis2023} established no-go results for other error channels. The proof technique in~\cite{Regev2008,Rosmanis2023} uses channel purification, which is also used in our proof (see Sec.~\ref{nsec:loss.bound}). More specifically, we adapt the proof technique in Ref.~\cite{Regev2008} to oracle models where only the mean boson number is constrained but no hard cutoff on the probe state's Hilbert space dimension is imposed.

\section{Proofs of query lower bounds for the bosonic unstructured search problem}
\label{nsec:query.lower.bounds}

\subsection{General query lower bound for bosonic unstructured search without loss}

\label{nsec:Ttot.gen.bound}

Here, we prove a lower bound on the query complexity of bosonic unstructured search in the continuous-time model~\eqref{neq:time.oracle.def} subject to a time-averaged probe energy constraint. This result immediately leads to a lower bound for the discrete oracle model, as the time-continuous oracle includes discrete oracles as special instances.\footnote{For bosonic systems subject to a fixed boson-number budget, the optimal query complexities of the two oracle models have the same scaling in the small-signal regime $n _{\budgetlb}|{\sin(\phi _{0}/2)}| = O(1)$ after converting query number to interrogation time by $T_{\text{tot}}=(\phi_{0}/B _{0}) K_{\text{tot}}$. In contrast, the two oracle models' query complexities exhibit qualitatively different scaling in the boson-number budget $n _{\budgetlb}$ when $n _{\budgetlb} |{\sin(\phi _{0}/2)}| \gg 1$, see Sec.~\ref{nsec:oracle.phi0} for detail.} We apply a lower-bound technique reminiscent of those in~\cite{BBBV1994,Gutmann1998} and derive an upper bound on the rate of change of the state distinguishability (measured by the trace distance) between two probe states evolving under any two distinct oracles. However, our proof differs from the approaches of Refs.~\cite{BBBV1994,Gutmann1998} (which compare evolution under the oracle with a reference evolution without a marked element) in two crucial aspects. First, we compare probe evolution corresponding to distinct oracles directly, which enables incorporating the time-averaged boson-number constraint without having to limit the boson-number dynamics of an extra reference state. Second, we use a different method to bound the time-derivative of the probe evolution under an oracle, which does not require any hard cutoff on the probe's Hilbert space and applies to an oracle unitary generated by an unbounded number operator.

Bosonic unstructured search can be viewed as a hypothesis testing task over $N$ different hypotheses associated with $N$ single-mode oracle Hamiltonians (i.e., generators of the oracle unitaries), subject to a time-averaged total energy constraint. At the beginning of the protocol, no prior information about the target is given, so we can assume a flat prior distribution.
To derive a query lower bound, we consider the success probability of distinguishing between an arbitrary pair of oracle Hamiltonians, averaged over a flat prior distribution of all possible pairs. A related decision problem is considered in~\cite{Choi2025}, though without the energy constraint necessary for the bosonic case.

We consider a target field on site $x _{0} \in \{ 1,2,\ldots , N\} $. The corresponding Hamiltonian reads $\hat H _{x_{0}} = B _{0} \hat n  _{x _{0}}  = B _{0} \hat a _{x_{0}} ^\dag \hat a _{x_{0}}  $. We do not know $x_{0} $ at the beginning of the sensing protocol, and our goal is to learn its value using as little time as possible, subject to the aforementioned energy constraint. Specifically, in any given search protocol, we allow arbitrary ancillas and probe-ancilla quantum control operations and measurements
satisfying Eq.~\eqref{neq:def.nS.avg.var}.
Without loss of generality, we incorporate measurement as deferred detection on some appropriately defined ancilla mode.

Generally, we can write any given search strategy as a sequence of probe-ancilla control unitaries $\hat V_{j}$ ($j=1,2,\ldots, K$) applied at times $j \Delta t$, so that the final sensor state reads
\begin{align}
\label{neq:Psi.xo.def}
\ket{\Psi _{x _{0}}} & = \hat V _{K+1} \mathcal{T} \prod _{j=1} ^{K} \left( e ^{-i \hat H _{x _{0}} \Delta t  } \hat V _{j} \right) (\ket{\psi _{0}} \otimes \ket{\text{vac}}_{\text{A}} )
,
\end{align}
where the total protocol time is $T_{\text{tot}} = K \Delta t $. For convenience, we also define the instantaneous state after the application of the $(\ell+1)$st control unitary as 
\begin{align}
& \ket{\Psi _{x _{0}} (\ell \Delta t )} 
\label{neq:Psi.xo.time.def}
= \hat V _{\ell+1} \mathcal{T} \prod _{j=1} ^{\ell} \left(e ^{-i \hat H _{x _{0}} \Delta t  } \hat V _{j} \right) (\ket{\psi _{0}} \otimes \ket{\text{vac}}_{\text{A}} )
.\end{align}
Without loss of generality, we consider the limit $K\to \infty$, which also covers the scenario with longer durations of query evolution.
For a protocol to succeed in total time $T_{\text{tot}}$ with high probability, a necessary condition is for the probability of distinguishing states $\ket{\Psi_{x _{1}}} $ and $\ket{\Psi_{x _{2}}} $ averaged over all pairs of $x _{1}$ and $x _{2}$ to be larger than a threshold value:
\begin{align}
\label{neq:gen.pth.def.0}
p_{\text{suc}} \coloneqq \frac{1}{N (N-1)} \sum _{x _{1} \ne x _{2}} 
\frac{1+\frac{1}{2} \| \ket{\Psi _{x _{1}}}\!\bra{\Psi _{x _{1}}}  -  \ket{\Psi _{x _{2}}}\!\bra{\Psi _{x_{2}}}   \| _{1}}{2}  
> \frac{1+ p ^{(0)}_{\text{th}} }{2}
,
\end{align}
where $p ^{(0)}_{\text{th}} \in (0,1)$ denotes a threshold constant. For any protocol meeting the success criterion of Sec.~\ref{nsec:setup}, the test ``output $x_1$'' gives $\frac12\|\hat\rho_{x_1}-\hat\rho_{x_2}\|_1\ge1/3$ for every pair, so Eq.~\eqref{neq:gen.pth.def.0} holds with, e.g., $p^{(0)}_{\text{th}}=1/4$; the same choice applies to $p^{(1)}_{\text{th}}$ in Sec.~\ref{nsec:loss.bound} and to the fixed-phase bounds in Sec.~\ref{nsec:oracle.phi0}.
We can rewrite the above condition as
\begin{align}
p ^{(0)}_{\text{th}} &< \frac{1}{N (N-1)} \sum _{x _{1} \ne x _{2}} \frac{1}{2} \| \ket{\Psi _{x _{1}}}\!\bra{\Psi _{x _{1}}}  -  \ket{\Psi _{x _{2}}}\!\bra{\Psi _{x _{2}}}   \| _{1} 
\nonumber \\
\label{neq:pth.1norm.coh}
&= \frac{1}{N (N-1)} \sum _{x _{1} \ne x _{2}} \sqrt{1- |\! \bra{\Psi  _{x _{1}}} \Psi  _{x _{2}} \rangle | ^{2}}
.
\end{align}
Formally, the above-threshold probability condition can also be justified by considering the success probability of the search problem over a uniform prior distribution~\cite{Lax1975}. We also note that, for any given search protocol that satisfies Eq.~\eqref{neq:pth.1norm.coh} for a continuous-time oracle~\eqref{neq:time.oracle.def} with $B_{0}<0$, there exists a complex-conjugated control strategy (i.e., conjugating all controls and measurement operators in the Fock basis) that can distinguish similar oracle models with $B_{0}>0$. As such, we assume $B_{0}>0$ throughout this paper.

To bound the right-hand side of Eq.~\eqref{neq:pth.1norm.coh}, we first define the interaction-picture probe-ancilla states
\begin{align}
\ket{\Psi ^{(\text{i})}_{x _{1}} } & \coloneqq \left ( \mathcal{T} \prod _{j=1} ^{K+1} \hat V _{j} \right )^{\dag} \ket{\Psi _{x _{1}}}  \\
& =  \mathcal{T} \prod _{\ell=1} ^{K} e ^{-i \Delta t \hat H ^{\text{(i)}} _{x _{1}} (\ell \Delta t)  }  (\ket{\psi _{0}} \otimes \ket{\text{vac}}_{\text{A}} )
,
\end{align}
where we define the interaction-picture query Hamiltonian as 
\begin{align}
\hat H ^{\text{(i)}} _{x } (\ell \Delta t) \coloneqq  \left ( \tilde{\mathcal{T}} \prod _{j=1} ^{\ell} \hat V _{j} ^{\dag}  \right  )  \hat H _{x }  \left ( \mathcal{T} \prod _{j=1} ^{\ell} \hat V _{j} \right )
.
\end{align}
Here, $\tilde{\mathcal{T}}$ signifies an anti-time-ordered product of the control unitaries. 
In the interaction picture, we rewrite the boson-number constraint in Eq.~\eqref{neq:def.nS.avg.var} in terms of a discrete sum over contributions from each time steps as 
\begin{align}
\forall \, x_{1}: \quad 
& \frac{1}{K} \sum _{\ell=1}^K
\sum _{x_{2}} \braket{\Psi ^{(\text{i})}_{x _{1}} (\ell \Delta t) | \hat H ^{\text{(i)}} _{x _{2}} (\ell \Delta t)  |\Psi ^{(\text{i})}_{x _{1}} (\ell \Delta t) } 
\le  B _{0}  n _{\budgetlb}  , 
\end{align}
where we define the interaction-picture instantaneous state immediately after the $\ell$th query as 
\begin{align}
\ket{\Psi ^{(\text{i})}_{x _{1}} (\ell \Delta t) }
& \coloneqq \left ( \mathcal{T} \prod _{j=1} ^{\ell+1} \hat V _{j} \right ) ^{\dag} \ket{\Psi _{x _{1}}(\ell \Delta t)  }  \\
& =  \mathcal{T} \prod _{j=1} ^{\ell} e ^{-i \Delta t \hat H ^{\text{(i)}} _{x _{1}} (j \Delta t)  }  (\ket{\psi _{0}} \otimes \ket{\text{vac}}_{\text{A}} )
,
\end{align}
where the Schr{\"{o}}dinger-picture state $\ket{\Psi _{x _{1}}(\ell \Delta t)  } $ is defined in Eq.~\eqref{neq:Psi.xo.time.def}.
Further, using the nonnegativity of the probe Hamiltonian, we obtain the following condition on the expectation value of the time-averaged query Hamiltonian:
\begin{align}
\forall \, x_{1}: \quad 
& \frac{1}{K} \sum _{\ell=1}^K
\braket{\Psi ^{(\text{i})}_{x_{1}} (\ell \Delta t) | \hat H ^{\text{(i)}} _{x _{1}} (\ell \Delta t)  |\Psi ^{(\text{i})}_{x_{1}} (\ell \Delta t) } 
\le B _{0}  n _{\budgetlb} 
.
\end{align}
Taking the $K \to \infty$ continuous-time limit, we obtain the following bound on the time-averaged expectation value of total boson number:
\begin{align}
\label{neq:tavg.nph}
\forall \, x_{1}: \quad 
& \frac{1}{T_{\text{tot}}} \sum _{x_{2}} \int ^{T_{\text{tot}}} _{0} 
\braket{\Psi ^{(\text{i})}_{x _{1}} | \hat H ^{\text{(i)}} _{x _{2}} (\tau ) | \Psi ^{(\text{i})}_{x _{1}} }  d \tau 
\le B _{0}  n _{\budgetlb} 
,\\
& \frac{1}{T_{\text{tot}}}  \int ^{T_{\text{tot}}} _{0} 
\braket{\Psi ^{(\text{i})}_{x_{1}} | \hat H ^{\text{(i)}} _{x _{1}} (\tau ) | \Psi ^{(\text{i})}_{x_{1}}  } d \tau 
\le B _{0}  n _{\budgetlb} 
,
\end{align}
where we define the instantaneous state 
\begin{align}
& \tau \in [0,T_{\text{tot}} =K\Delta t ]: \quad
\ket{\Psi ^{(\text{i})}_{x _{1}}  (\tau )} \coloneqq \mathcal{T} e ^{-i\int ^{\tau} _{0} \hat H ^{\text{(i)}} _{x _{1}} (t' ) \, dt' }  (\ket{\psi _{0}} \otimes \ket{\text{vac}}_{\text{A}} )
.
\end{align}

We rewrite the state overlaps in the interaction picture as
\begin{align}
\bra{\Psi  _{x _{1}}} \Psi  _{x _{2}} \rangle 
= \braket{\Psi ^{(\text{i})}_{x _{1}} | \Psi ^{(\text{i})}_{x_{2}}  } 
,
\end{align}
so that in the $K\to \infty$ (i.e., $\Delta t \to 0$) limit, we can derive its time derivative via the change of the state overlap per time step as
\begin{align}
\label{neq:int.overlap.time.deriv}
& \partial _{\tau } \braket{\Psi ^{(\text{i})}_{x_{1}} (\tau ) | \Psi ^{(\text{i})}_{x_{2}}  (\tau ) } 
= i \braket{\Psi ^{(\text{i})}_{x_{1}} (\tau )| [\hat H ^{\text{(i)}} _{x _{1}} (\tau ) - \hat H ^{\text{(i)}} _{x _{2}} (\tau ) ] | \Psi ^{(\text{i})}_{x_{2}} (\tau) }
.
\end{align}
Note that we only require the above equation to hold for almost every $\tau$, in the integrated sense
\begin{align}
\braket{\Psi ^{(\text{i})}_{x_{1}} (\tau_{2} ) | \Psi ^{(\text{i})}_{x_{2}}  (\tau_{2} ) } 
- \braket{\Psi ^{(\text{i})}_{x_{1}} (\tau_{1} ) | \Psi ^{(\text{i})}_{x_{2}}  (\tau_{1} ) } 
= i \int^{\tau _{2}} _{\tau_{1}} \braket{\Psi ^{(\text{i})}_{x_{1}} (\tau )| [\hat H ^{\text{(i)}} _{x _{1}} (\tau ) - \hat H ^{\text{(i)}} _{x _{2}} (\tau ) ] | \Psi ^{(\text{i})}_{x_{2}} (\tau) } d \tau
,
\end{align}
which automatically holds when [since $|e^{-i \theta \hat n }-\hat {\mathbb{I}}| \le \theta \hat n $, cf.~Eq.~\eqref{neq:phi0.evol.bound.n}]
\begin{align}
\forall \tau \in [0,T_{\text{tot}}]
,\
x, y\in \{x_{1},x_{2}\}: \quad \int ^{\tau} _{0} d \tau_{0} \braket{\Psi ^{(\text{i})}_{x } (\tau_{0}  ) | \hat n ^{\text{(i)}} _{y} (\tau_{0}  )| \Psi ^{(\text{i})}_{x } (\tau_{0}  )} < \infty
,
\end{align}
which is guaranteed by the boson-number constraint~\eqref{neq:def.nS.avg.var}, and does not assume any additional regularity condition. Thus, we can use the triangle inequality and the non-negativity of the query Hamiltonians $\hat H ^{\text{(i)}} _{x _{1}} (\tau )$ and $\hat H ^{\text{(i)}} _{x _{2}} (\tau )$ to bound the norm of the time derivative as
\begin{align}
| \partial _{\tau } \braket{\Psi ^{(\text{i})}_{x_{1}} (\tau ) | \Psi ^{(\text{i})}_{x_{2}}  (\tau ) }  |
\le {}& |\braket{\Psi ^{(\text{i})}_{x_{1}} | \hat H ^{\text{(i)}} _{x _{1}} (\tau )    | \Psi ^{(\text{i})}_{x_{2}}  }| + |\braket{\Psi ^{(\text{i})}_{x_{1}} | \hat H ^{\text{(i)}} _{x _{2}} (\tau )  | \Psi ^{(\text{i})}_{x_{2}}  }|
\nonumber \\
\le {}& \sqrt{ \braket{\Psi ^{(\text{i})}_{x _{2}} | \hat H ^{\text{(i)}} _{x _{1}} (\tau ) | \Psi ^{(\text{i})}_{x _{2}}   } 
\braket{\Psi ^{(\text{i})}_{x_{1}} | \hat H ^{\text{(i)}} _{x _{1}} (\tau ) | \Psi ^{(\text{i})}_{x_{1}}  }}
\nonumber \\& 
\label{neq:gen.qsl.bound}
+ \sqrt{ \braket{\Psi ^{(\text{i})}_{x _{1}} | \hat H ^{\text{(i)}} _{x _{2}} (\tau ) | \Psi ^{(\text{i})}_{x _{1}}   } 
\braket{\Psi ^{(\text{i})}_{x_{2}}  | \hat H ^{\text{(i)}} _{x _{2}} (\tau ) | \Psi ^{(\text{i})}_{x_{2}}   }}
,
\end{align}
where we omit the time argument for the state vectors when it does not cause confusion.

The right-hand side of Eq.~\eqref{neq:pth.1norm.coh} can be rewritten in terms of interaction-picture states as
\begin{align}
p ^{(0)}_{\text{th}} &<  \frac{1}{N (N-1)} \sum _{x _{1} \ne x _{2}} \sqrt{1- |\! \bra{\Psi  _{x _{1}}} \Psi  _{x _{2}} \rangle | ^{2}}
\nonumber \\
& = \frac{1}{N (N-1) } \sum _{x _{1} \ne x _{2}}  \sqrt{1- |\! \braket{  \Psi ^{(\text{i})}_{x_{1}} (T_{\text{tot}} ) | \Psi ^{(\text{i})}_{x_{2}} (T_{\text{tot}} ) }\! |^{2} }
\nonumber \\
\label{neq:avg.bures.int.0}
&\le \sum _{x _{1} \ne x _{2}} \frac{\sqrt{ 2\left ( 1- |\! \braket{  \Psi ^{(\text{i})}_{x_{1}} (T_{\text{tot}} ) | \Psi ^{(\text{i})}_{x_{2}} (T_{\text{tot}} ) }\! | \right )  }}{N (N-1) }  
.
\end{align}
To bound the terms contributing to the sum on the right-hand side of the above equation, we first rewrite the overlaps between the final states under two different oracles as 
\begin{align}
\braket{  \Psi ^{(\text{i})}_{x_{1}} (T_{\text{tot}} ) | \Psi ^{(\text{i})}_{x_{2}} (T_{\text{tot}} ) }
= {}& \braket{\Psi ^{(\text{i})}_{x_{1}} ( 0 ) | \Psi ^{(\text{i})}_{x_{2}}  ( 0 ) }
+ \int ^{T_{\text{tot}}} _{0}
\partial _{\tau } \braket{\Psi ^{(\text{i})}_{x_{1}} (\tau ) | \Psi ^{(\text{i})}_{x_{2}}  (\tau ) }   d \tau 
\nonumber \\
= {}& 1 + \int ^{T_{\text{tot}}} _{0}
\partial _{\tau } \braket{\Psi ^{(\text{i})}_{x_{1}} (\tau ) | \Psi ^{(\text{i})}_{x_{2}}  (\tau ) }   d \tau ,
\end{align}
so that using the triangle inequality, we derive the following upper bound on the change in the overlap function between the two probe-ancilla states:
\begin{align}
1- |\! \braket{  \Psi ^{(\text{i})}_{x_{1}} (T_{\text{tot}} ) | \Psi ^{(\text{i})}_{x_{2}} (T_{\text{tot}} ) } \! | 
\le {}& \int ^{T_{\text{tot}}} _{0}
|{\partial _{\tau } \braket{\Psi ^{(\text{i})}_{x_{1}} (\tau ) | \Psi ^{(\text{i})}_{x_{2}}  (\tau ) }}  | \, d \tau 
.
\end{align}
Plugging the above equation into Eq.~\eqref{neq:avg.bures.int.0}, we have
\begin{align}
p ^{(0)}_{\text{th}} &<  \sum _{x _{1} \ne x _{2}}   \frac{\sqrt{ 2\left ( 1- |\! \braket{  \Psi ^{(\text{i})}_{x_{1}} (T_{\text{tot}} ) | \Psi ^{(\text{i})}_{x_{2}} (T_{\text{tot}} ) }\! | \right )  }}{N (N-1) } 
\nonumber \\
& \le \sum _{x _{1} \ne x _{2}} \frac{\sqrt{2 \int ^{T_{\text{tot}}} _{0}
|{\partial _{\tau } \braket{\Psi ^{(\text{i})}_{x_{1}} (\tau ) | \Psi ^{(\text{i})}_{x_{2}}  (\tau ) }  }| \, d \tau  }}{N (N-1) } 
\label{neq:gen.bound.coh.deriv}
.
\end{align}
Using the Cauchy-Schwarz inequality and the upper bound in Eq.~\eqref{neq:gen.qsl.bound}, we can further bound the right-hand side of the above equation as
\begin{align}
p ^{(0)}_{\text{th}} & <  \sqrt{\frac{2  }{N (N-1) } \sum _{x _{1} \ne x _{2}}    \int ^{T_{\text{tot}}} _{0}
|{\partial _{\tau } \braket{\Psi ^{(\text{i})}_{x_{1}} (\tau ) | \Psi ^{(\text{i})}_{x_{2}}  (\tau ) }}  | \, d \tau   }
\\
& \le  \sqrt{\frac{4  }{N (N-1) } \sum _{x _{1} \ne x _{2}}    \int ^{T_{\text{tot}}} _{0}
\sqrt{ \braket{\Psi ^{(\text{i})}_{x _{2}} (\tau ) | \hat H ^{\text{(i)}} _{x _{1}} (\tau ) | \Psi ^{(\text{i})}_{x _{2}} (\tau )   } 
\braket{\Psi ^{(\text{i})}_{x_{1}} (\tau ) | \hat H ^{\text{(i)}} _{x _{1}} (\tau ) | \Psi ^{(\text{i})}_{x_{1}}(\tau )   }}  \, d \tau  }
\\
& \le 2 \sqrt{\frac{1 }{N (N-1)  } \sum _{x _{1} \ne x _{2}}    
\sqrt{ \left [ \int ^{T_{\text{tot}}} _{0}  \braket{\Psi ^{(\text{i})}_{x _{2}} (\tau ) | \hat H ^{\text{(i)}} _{x _{1}} (\tau ) | \Psi ^{(\text{i})}_{x _{2}}  (\tau )  } 
d \tau \right ] \left [ \int ^{T_{\text{tot}}} _{0} \braket{\Psi ^{(\text{i})}_{x_{1}} (\tau ) | \hat H ^{\text{(i)}} _{x _{1}} (\tau ) | \Psi ^{(\text{i})}_{x_{1}} (\tau )  }  d \tau \right ] }  }
\\
& \le 2 \sqrt{\frac{1 }{N (N-1)  } \sum _{x _{1} \ne x _{2}}    
\sqrt{ n _{\budgetlb} B _{0} T_{\text{tot}} \int ^{T_{\text{tot}}} _{0}  \braket{\Psi ^{(\text{i})}_{x _{2}} (\tau ) | \hat H ^{\text{(i)}} _{x _{1}} (\tau ) | \Psi ^{(\text{i})}_{x _{2}} (\tau )   } 
d \tau  }  }
\\
& \le 2 \sqrt{\frac{1 }{N-1 } \sum _{x _{2}  }  
\sqrt{ \frac{n _{\budgetlb} B _{0}  T_{\text{tot}} }{N}   \int ^{T_{\text{tot}}} _{0}  \sum _{x _{1} \ne x _{2}} \braket{\Psi ^{(\text{i})}_{x _{2}} (\tau ) | \hat H ^{\text{(i)}} _{x _{1}} (\tau ) | \Psi ^{(\text{i})}_{x _{2}} (\tau ) } d \tau 
} }
\\
& \le 2 \sqrt{ \frac{N}{N-1} \frac{B _{0}  n _{\budgetlb} T_{\text{tot}} }{\sqrt{N}}}
,
\end{align} 
so that we have proved the following lower bound on the total search time:
\begin{align}
T_{\text{tot}} > \frac{(p ^{(0)}_{\text{th}})^{2}}{4} 
\frac{\left (1-\frac{1}{N} \right )\sqrt{N}}{B _{0}  n _{\budgetlb} }
.
\end{align}
Up to a constant factor and an $O(N^{-1})$ correction, the lower bound on the total query time reads
\begin{align}
\label{neq:Ttot.lb.gen.coh}
T_{\text{tot}}  = \Omega \left (\frac{\sqrt{N}}{ n _{\budgetlb} B _{0} }\right )
.
\end{align}

This lower bound is reminiscent of the continuous-time query lower bound for unstructured search with finite-dimensional quantum systems~\cite{Gutmann1998}. However, the lower bound in \cite{Gutmann1998} has the form $\Omega(\sqrt{N}/E)$, where $E$ denotes the oracle Hamiltonian's operator norm. In contrast, Eq.~\eqref{neq:Ttot.lb.gen.coh} applies to unstructured search over oracles generated by an unbounded bosonic number operator, subject to a time-averaged total energy constraint. As such, the lower bound in~\cite{Gutmann1998} is not directly applicable. Also, as mentioned at the beginning of this section, our proof  differs technically from the method used to prove the original query lower-bound result for unstructured search in~\cite{BBBV1994}.

\subsection{Proof for lower bound on query complexity for bosonic search using purely Gaussian probes}

\label{nsec:Gaus.bound}

To achieve the bound in Eq.~\eqref{neq:Ttot.lb.gen.coh}, non-Gaussian probe states are necessary. To show this, we use a different way to bound the right-hand side of Eq.~\eqref{neq:gen.bound.coh.deriv} for Gaussian-only probes, which leads to a tighter bound. Our proof only requires Gaussianity of the reduced multimode probe states at each time (with measurements deferred), so it also covers adaptive strategies whose unconditional probe states remain Gaussian, but not general outcome-dependent feedforward.

First recall Eq.~\eqref{neq:gen.bound.coh.deriv}, which holds for general sensing strategies using arbitrary probe states:
\begin{align}
p ^{(0)}_{\text{th}} <  \frac{1}{N (N-1) } 
\label{neq:gen.bound.coh.deriv.1}
\sum _{x _{1} \ne x _{2}} {}&\sqrt{2 \int ^{T_{\text{tot}}} _{0}
|{\partial _{\tau } \braket{\Psi ^{(\text{i})}_{x_{1}} (\tau ) | \Psi ^{(\text{i})}_{x_{2}}  (\tau ) }}  | d \tau  }
.
\end{align}
Substituting  Eq.~\eqref{neq:int.overlap.time.deriv} into the integrand in the above equation and using the Cauchy-Schwarz and triangle inequalities, we can bound the time derivative of the state overlap as
\begin{align}
& \quad |{\partial _{\tau } \braket{\Psi ^{(\text{i})}_{x_{1}} (\tau ) | \Psi ^{(\text{i})}_{x_{2}}  (\tau ) } } |
\nonumber \\
& = | \braket{\Psi ^{(\text{i})}_{x_{1}} | [\hat H ^{\text{(i)}} _{x _{1}} (\tau ) - \hat H ^{\text{(i)}} _{x _{2}} (\tau ) ] | \Psi ^{(\text{i})}_{x_{2}} (\tau) } |
\nonumber \\
\label{neq:tderiv.overlap.lin}
& \le | \braket{\Psi ^{(\text{i})}_{x_{1}} | \hat H ^{\text{(i)}} _{x _{1}} (\tau ) | \Psi ^{(\text{i})}_{x_{2}} (\tau) } |
+ | \braket{\Psi ^{(\text{i})}_{x_{1}} | \hat H ^{\text{(i)}} _{x _{2}} (\tau )  | \Psi ^{(\text{i})}_{x_{2}} (\tau) } |
\nonumber\\
& \le \sqrt{
\braket{\Psi ^{(\text{i})} _{x _{2}} (\tau) | [\hat H ^{\text{(i)}} _{x _{1}}  (\tau ) ]^{2} | \Psi ^{(\text{i})}_{x _{2}} (\tau) }
}
+ \sqrt{
\braket{\Psi ^{(\text{i})} _{x _{1}} (\tau) | [\hat H ^{\text{(i)}} _{x _{2}}  (\tau ) ]^{2} | \Psi ^{(\text{i})}_{x _{1}} (\tau) }
}
\\
\label{neq:tderiv.overlap.4m}
& = B _{0}   \sqrt{
\braket{\Psi ^{(\text{i})} _{x _{2}} (\tau) | [\hat a ^{\text{(i)}\dag } _{x _{1}}  (\tau ) \hat a ^{\text{(i)}} _{x _{1}}  (\tau )  ]^{2}  | \Psi ^{(\text{i})}_{x _{2}} (\tau) }
}
+ B _{0}  \sqrt{
\braket{\Psi ^{(\text{i})} _{x _{1}} (\tau) | [\hat a ^{\text{(i)}\dag } _{x _{2}}  (\tau ) \hat a ^{\text{(i)}} _{x _{2}}  (\tau )  ]^{2}  | \Psi ^{(\text{i})}_{x _{1}} (\tau) } 
} 
.
\end{align}
Using the aforementioned assumption that the probe state stays Gaussian throughout the entire search protocol and Wick's theorem, we can bound the expectation values of the squared boson number operators as (see Appendix~\ref{app.sec:bound.n2.gaus} for detail)
\begin{align}
\forall x_1, x_2: \quad 
& \braket{\Psi ^{(\text{i})} _{x _{2}} (\tau) | [\hat a ^{\text{(i)}\dag } _{x _{1}}  (\tau ) \hat a ^{\text{(i)}} _{x _{1}}  (\tau )  ]^{2}  | \Psi ^{(\text{i})}_{x _{2}} (\tau) }
\nonumber \\
\label{neq:bound.n2.gaus}
\le {}& 3 \braket{\Psi ^{(\text{i})} _{x _{2}} (\tau) | \hat n ^{\text{(i)}  } _{x _{1}}  (\tau )    | \Psi ^{(\text{i})}_{x _{2}} (\tau) } ^{2} 
+ 2 \braket{\Psi ^{(\text{i})} _{x _{2}} (\tau) | \hat n ^{\text{(i)}  } _{x _{1}}  (\tau )    | \Psi ^{(\text{i})}_{x _{2}} (\tau) }
.
\end{align}

Substituting Eq.~\eqref{neq:bound.n2.gaus} into Eq.~\eqref{neq:tderiv.overlap.4m}, we can now bound the norm of the time derivative of the state overlap in~\eqref{neq:tderiv.overlap.4m} as
\begin{align}
|{\partial _{\tau }  \braket{\Psi ^{(\text{i})}_{x_{1}} (\tau ) | \Psi ^{(\text{i})}_{x_{2}}  (\tau ) }} |
\le {} & B _{0} \sqrt{3 \braket{\Psi ^{(\text{i})} _{x_{2}} (\tau) | \hat n ^{\text{(i)}  } _{x _{1}}  (\tau )    | \Psi ^{(\text{i})}_{x_{2}} (\tau) } ^{2} + 2 \braket{\Psi ^{(\text{i})} _{x_{2}} (\tau) | \hat n ^{\text{(i)}  } _{x _{1}}  (\tau )    | \Psi ^{(\text{i})}_{x_{2}} (\tau) } }
\nonumber \\
& + B _{0} \sqrt{3 \braket{\Psi ^{(\text{i})} _{x_{1}} (\tau) | \hat n ^{\text{(i)}  } _{x _{2}}  (\tau )    | \Psi ^{(\text{i})}_{x_{1}} (\tau) } ^{2} + 2 \braket{\Psi ^{(\text{i})} _{x_{1}} (\tau) | \hat n ^{\text{(i)}  } _{x _{2}}  (\tau )    | \Psi ^{(\text{i})}_{x_{1}} (\tau) } }
.
\end{align}
Substituting the above inequality into Eq.~\eqref{neq:gen.bound.coh.deriv.1}, we have
\begin{align}
p ^{(0)}_{\text{th}} & < \frac{2}{N (N-1) }  \sum _{x _{1} \ne x_{2} } \sqrt{2 B _{0}  \int ^{T_{\text{tot}}} _{0}
\sqrt{3 \braket{\Psi ^{(\text{i})} _{x_{2}} (\tau) | \hat n ^{\text{(i)}  } _{x _{1}}  (\tau )    | \Psi ^{(\text{i})}_{x_{2}} (\tau) } ^{2} + 2 \braket{\Psi ^{(\text{i})} _{x_{2}} (\tau) | \hat n ^{\text{(i)}  } _{x _{1}}  (\tau )    | \Psi ^{(\text{i})}_{x_{2}} (\tau) } } \, d \tau  }
\\
& \le \frac{2}{N (N-1) }  \sum _{x _{1} \ne x_{2} } \sqrt{2 B _{0}  \int ^{T_{\text{tot}}} _{0}
\left ( \sqrt{3} \braket{\Psi ^{(\text{i})} _{x _{2}} (\tau) | \hat n ^{\text{(i)}  } _{x _{1}}  (\tau )    | \Psi ^{(\text{i})}_{x _{2}} (\tau) }
+ \sqrt{2  \braket{\Psi ^{(\text{i})} _{x _{2}} (\tau) | \hat n ^{\text{(i)}  } _{x _{1}}  (\tau )    | \Psi ^{(\text{i})}_{x _{2}} (\tau) } } \right ) d \tau  }
.
\end{align}
Using the Cauchy-Schwarz inequality and the upper bound on the time-averaged boson number in Eq.~\eqref{neq:tavg.nph}, we obtain 
\begin{align}
p ^{(0)}_{\text{th}} & < 2 \sqrt{\frac{2 B _{0} }{N (N-1) }  \sum _{x _{1} \ne x_{2} }  \int ^{T_{\text{tot}}} _{0}
\left ( \sqrt{3} \braket{\Psi ^{(\text{i})} _{x_{2}} (\tau) | \hat n ^{\text{(i)}  } _{x _{1}}  (\tau )    | \Psi ^{(\text{i})}_{x_{2}} (\tau) }
+ \sqrt{2  \braket{\Psi ^{(\text{i})} _{x_{2}} (\tau) | \hat n ^{\text{(i)}  } _{x _{1}}  (\tau )    | \Psi ^{(\text{i})}_{x_{2}} (\tau) } } \right ) d \tau 
}
\nonumber \\
& \le 2\sqrt{2 B _{0} T_{\text{tot}} \left ( \sqrt{3} \frac{n _{\budgetlb} }{N-1} + \sqrt{2} \sqrt{\frac{n _{\budgetlb} }{N-1}} \right ) }
.
\label{eq:bosonic_gaussian_final_ineq}
\end{align}
In the above equation, we used the Cauchy-Schwarz inequality and Eq.~\eqref{neq:tavg.nph} to derive the following inequality on the total time-integrated square-root of the boson number:
\begin{align}
& \frac{ 1 }{N (N-1) }  \sum _{x _{1} \ne x_{2} }  \int ^{T_{\text{tot}}} _{0}
\sqrt{ \braket{\Psi ^{(\text{i})} _{x_{2}} (\tau) | \hat n ^{\text{(i)}  } _{x _{1}}  (\tau )    | \Psi ^{(\text{i})}_{x_{2}} (\tau) } } d \tau 
\nonumber \\
\le {} & \frac{ 1 }{N (N-1) }  \sum _{x _{1} \ne x_{2} } 
\sqrt{ T_{\text{tot}}  \int ^{T_{\text{tot}}} _{0}
\braket{\Psi ^{(\text{i})} _{x_{2}} (\tau) | \hat n ^{\text{(i)}  } _{x _{1}}  (\tau )    | \Psi ^{(\text{i})}_{x_{2}} (\tau) }
d \tau }
\nonumber \\
\le {} & \sqrt{\frac{ T_{\text{tot}} }{N (N-1) }  \sum _{x _{1} \ne x_{2} }  \int ^{T_{\text{tot}}} _{0}
\braket{\Psi ^{(\text{i})} _{x_{2}} (\tau) | \hat n ^{\text{(i)}  } _{x _{1}}  (\tau )    | \Psi ^{(\text{i})}_{x_{2}} (\tau) }
d \tau }
\nonumber \\
\le {}& \sqrt{\frac{T_{\text{tot}}}{N-1} \cdot \max _{x_{2}}
\sum _{x_{1}} \int ^{T_{\text{tot}}} _{0}
\braket{\Psi ^{(\text{i})} _{x_{2}} (\tau) | \hat n ^{\text{(i)}  } _{x _{1}}  (\tau )    | \Psi ^{(\text{i})}_{x_{2}} (\tau) }
d \tau }
\nonumber \\
\le {}& T_{\text{tot}} \sqrt{\frac{n _{\budgetlb}}{N-1} }
.
\end{align}
Thus, we have proved the following lower bound for purely Gaussian strategies (where we change the subscript to emphasize the setting):
\begin{align}
\label{neq:Ttot.Gaus.gen}
T_{\mathrm{Gaussian}} > \frac{(p ^{(0)}_{\text{th}})^{2}}{8} 
\frac{\sqrt{N-1}}{B _{0}  \sqrt{n _{\budgetlb}} \left (\sqrt{2} +  \sqrt{ 3 \frac{ n _{\budgetlb} }{N-1}} \right) }
.
\end{align}
The total query time for any purely Gaussian strategy therefore satisfies
\begin{align}
\label{neq:Ttot.lb.Gaus.coh}
T_{\text{Gaussian}}  = \Omega \left (\min \left \{ 
\frac{\sqrt{N}}{ \sqrt{n _{\budgetlb}} B _{0} }, \frac{N}{n _{\budgetlb}B _{0}} \right\} \right )
.
\end{align}

Note that in the low boson-number regime $n _{\budgetlb} \ll 1 \ll N$, Eq.~\eqref{neq:Ttot.lb.Gaus.coh} is weaker than the general lower bound in Eq.~\eqref{neq:Ttot.lb.gen.coh}. In this regime, the coherent-state strategy attains $T_{\text{tot}}  = \Theta \left ( \sqrt{N} / { n _{\budgetlb} B _{0} }\right )$, saturating Eq.~\eqref{neq:Ttot.lb.gen.coh}. Hence~\eqref{neq:Ttot.lb.Gaus.coh} is not tight for $n _{\budgetlb} \ll 1$, and non-Gaussianity offers no advantage in this regime.

For $n_{\budgetlb}\gg N$, Eq.~\eqref{neq:Ttot.lb.Gaus.coh} instead gives $T_{\mathrm{Gaussian}}=\Omega(N/n_{\budgetlb}B_0)$, which can be attained by probing all modes once with each mode in a squeezed vacuum state, i.e.,
\begin{align}
\ket{\psi _{\text{sq}}} = \bigotimes_{x=1}^{N}  \ket{\psi _{\text{sq},r}}_{x} 
, \quad 
\ket{\psi _{\text{sq},r}}_{x} 
= e ^{i \frac{r}{2} \hat a _{x}  ^ {\dag  2} - \text{H.c.} }
\ket{0}_{x}
,
\end{align}
and undoing the squeezing before photon counting. Here $r$ denotes the squeezing strength, which is chosen to satisfy $\sinh^2r=n_{\budgetlb}/N$ to satisfy the boson-number constraint. After the unitary evolution, unmarked modes return exactly to vacuum, while the marked mode remains in vacuum with probability $[1+\sinh^2(2r)\sin^2(B_0t)]^{-1/2}$, so choosing $t=\Theta(N/(n_{\budgetlb}B_0))$ allows identifying the marked mode. Non-Gaussian probes nevertheless remain necessary for all $n_{\budgetlb}\gg1$ to attain the optimal joint query scaling in
$N$ and $n_{\budgetlb}$; in particular, for $n_{\budgetlb}\gg N$ the general bound~\eqref{neq:Ttot.lb.gen.coh} is smaller by a factor $\sqrt{N}$.

We note that coherent-state probes cannot reach the scaling in Eq.~\eqref{neq:Ttot.lb.Gaus.coh} when $n _{\budgetlb} \gg N$. 
In Appendix~\ref{app.sec:bound.Ttot.lb.coh}, we prove the following bound  [Eq.~\eqref{seq:Ttot.lb.coh}] on the complexity of boson unstructured search using any protocol for which the
unconditional reduced probe state remains coherent throughout (assuming $N\gg 1$):
\begin{align}
\label{neq:Ttot.lb.coh}
T_{\text{coh}} = \Omega \left ( \frac{1}{B_0}\sqrt{\frac{N}{n_{\budgetlb}}} \right )
.
\end{align}
The same bound holds in the presence of boson loss, since pure loss maps coherent states to coherent states and the environment can be absorbed into the oracle-independent ancillas.
In particular, this bound is strictly stronger than the bound~\eqref{neq:Ttot.lb.Gaus.coh} for Gaussian protocols when $n_{\budgetlb} \gg N$.

\subsection{Proof for lower bound on query complexity for bosonic search with linear boson loss}

\label{nsec:loss.bound}

In this section, we derive a query lower bound for bosonic unstructured search in the case where the oracle is a quantum operation that involves both coherent unitary evolution and uniform photon loss at rate $\gamma$. Specifically, we
consider a continuous-time oracle generated by one of the $x_{0}$-dependent Lindbladians 
\begin{align}
\label{neq:time.oracle.gamma.def}
\mathcal{L} _{x _{0}} \hat \rho_{\text{S}} = -i [ \hat H  _{x _{0}} , \hat \rho_{\text{S}} ] + \gamma \sum _{x} \mathcal{D} [ \hat a _{x}] \hat \rho_{\text{S}}
, 
\end{align}
where $\mathcal{D}  [\hat O] \cdot \coloneqq \hat O \cdot \hat O ^{\dag} - \frac{1}{2} \{\hat O ^{\dag} \hat O  ,  \cdot \} $ denotes the standard Lindbladian dissipator for jump operator $\hat O$.
In what follows, we omit the time dependence of density matrices when it does not cause confusion.
Similarly to the previous sections, we are interested in understanding the  query complexity of identifying $x_{0}$ under the time-averaged total boson-number constraint in Eq.~\eqref{neq:def.nS.avg.var}, $T_{\text{tot}}^{-1} \int ^{T_{\text{tot}}} _{0}
\braket{ \hat n _{\text{S}} (t) } dt \le n _{\budgetlb} $. We also allow arbitrary ancillas and probe-ancilla quantum control, 
as long as the probe state satisfies the aforementioned boson-number constraint~\eqref{neq:def.nS.avg.var} and the regularity condition~\eqref{neq:nS.regular.cond}. As in the lossless case [see discussion above Eq.~\eqref{neq:Psi.xo.def}], this framework automatically covers strategies involving mid-circuit measurements and feedback.

Our proof technique adapts the argument in~\cite{Regev2008} for a faulty oracle to the case with continuous oracle evolution under an energy constraint. The final result, the lower bound~\eqref{neq:Ttot.lb.gen.loss}, also exhibits similar scaling in $N $ as previous query lower bounds for faulty and/or noisy unstructured search using qubits or qudits~\cite{Regev2008,Temme2014,Rosmanis2023}.

The query lower bound proved in this section can also be viewed as the decision-problem analog of a related result in metrology known as the \textit{Hamiltonian-Not-in-Lindblad-Span} (HNLS) condition. Previous works~\cite{Guta2012,Jiang2018} prove that the quantum Fisher information (QFI) for estimating a Hamiltonian's strength in the presence of Markovian noise scales at best as the standard quantum limit (i.e., the classical limit) in sensing time, when the target Hamiltonian does not satisfy the HNLS condition. For bosonic systems, Ref.~\cite{Maccone2025} derives energy-dependent noisy metrological bounds using related techniques. These results hold when allowing arbitrary ancillas, quantum control, and measurement during the sensing process. Our final result can thus be viewed as an analog of the QFI bounds in metrology in the case of finite oracle strength. However, we use different proof techniques (not involving HNLS or any metrological bounds) to connect the boson-number constraint to a query lower bound. In contrast, a na\"{i}ve extension of HNLS to the regime with finite oracle strength would require additional assumptions on the time-averaged boson number over trajectories with varying oracle strengths, which is not needed in our setting [see the discussion below Eq.~\eqref{neq:loss.bound.Bures.dil.0}].

We begin by considering a general search protocol specified by a time-dependent, $x _{0}$-independent control Hamiltonian, $\hat H _{\text{c}} (t)$. As mentioned, the control generally acts on both the probe and the ancillas and can create entanglement between them, as long as Eq.~\eqref{neq:def.nS.avg.var} is satisfied. The $x _{0}$-dependent oracle and the control together generate the following probe-ancilla evolution:
\begin{align}
\partial _{t}\hat \rho_{x _{0}} & = -i [ \hat H  _{x _{0}}  + \hat H _{\text{c}} (t), \hat \rho_{x _{0}} ] + \gamma \sum _{x} \mathcal{D} [ \hat a _{x}] \hat \rho_{x _{0}} 
\nonumber \\
& \eqqcolon (\mathcal{L} _{x _{0} } + \mathcal{L} _{ \text{c}} (t)) \hat \rho_{x _{0}} 
.
\end{align}
For a protocol to successfully identify the value of $x_{0}$ after time $T_{\text{tot}}$, a necessary condition is that the probability of successfully distinguishing between arbitrary pairs of oracles $x _{1} \ne x _{2}$, averaged over all possible pairs, is greater than a threshold value:
\begin{align}
\label{neq:cond.dist.gen.noisy}
p_{\text{suc}} & \coloneqq \frac{1}{N (N-1)} \sum _{x _{1} \ne x _{2}} 
\frac{1+\frac{1}{2} \| \hat \rho _{x_{1}} (T_{\text{tot}}) -  \hat \rho _{x_{2}} (T_{\text{tot}})  \| _{1}}{2}  
> \frac{1+ p ^{(1)}_{\text{th}} }{2}
,
\end{align}
where $p ^{(1)}_{\text{th}} \in (0,1)$ is again a threshold constant. Note that this necessary condition automatically includes multi-shot protocols, since we allow arbitrary ancillas and probe-ancilla control operations, which can simulate multi-shot protocols via delayed measurements. Subtracting $1/2$ from both sides of Eq.~\eqref{neq:cond.dist.gen.noisy} and multiplying the remaining terms by a factor of $2$, we can rewrite the necessary condition as 
\begin{align}
\label{neq:loss.gen.1norm}
\frac{1}{N (N-1)} \sum _{x _{1} \ne x _{2}} \frac{1}{2} \| \hat \rho _{x_{1}} (T_{\text{tot}}) -  \hat \rho _{x_{2}} (T_{\text{tot}})  \| _{1}
> p ^{(1)}_{\text{th}}
.
\end{align}

To derive an upper bound for the left-hand side of Eq.~\eqref{neq:loss.gen.1norm}, we consider an $x$-independent unitary dilation of the probe's dissipative dynamics, where a fresh environmental mode is supplied at each infinitesimal time interval. This dilation is also known as a quantum collision model~\cite{Attal2006,Palma2022}. The probe-ancilla-environmental dynamics corresponding to the $x$th oracle can be described by a stochastic Schr{\"{o}}dinger equation in the It{\^{o}} form~\cite{Gardiner2004}:
\begin{align}
\partial _{t} \ket{\Phi_{x _{0}} (t)} & = -i \left [\hat H  _{x _{0} } + \hat H _{\text{c}} (t) - \frac{i\gamma }{2} \sum_{x} \hat a _{x}^{\dag} \hat a _{x} \right  ] 
\ket{\Phi_{x_{0}} (t)} 
+ \sqrt{\gamma } \sum_{x} [ \hat a _{x} \hat B ^{\dag} _{x} (t) - \text{H.c.} ] \ket{\Phi_{x_{0}} (t)} 
\nonumber \\
& = -i \left [\hat H  _{x _{0} } + \hat H _{\text{c}} (t) - \frac{i\gamma }{2} \sum_{x} \hat a _{x}^{\dag} \hat a _{x} \right  ] 
\ket{\Phi_{x_{0}} (t)} 
+ \sqrt{\gamma } \sum_{x} \hat a _{x} \hat B ^{\dag} _{x} (t) \ket{\Phi_{x_{0}} (t)} 
\label{neq:def.Phi.SAE}
,
\end{align}
where $\hat B _{x} (t)$ denotes the lowering operators for the environmental mode coupled to the $x$th probe at time $t$. 
These operators are known as quantum white-noise operators in the open quantum systems literature (see Ch.~5 of~\cite{Gardiner2004} for details). 
They satisfy the canonical commutation relations 
\begin{align}
\label{neq:q.noise.op.ccr}
[\hat B _{x} (t), \hat B ^{\dag}_{x'} (t')]  = \delta _{xx'} \delta (t-t')
, \quad 
[\hat B _{x} (t), \hat B _{x'} (t')]  = 0
,
\end{align}
where the time dependence $\delta (t-t')$ reflects Markovianity of the dissipative dynamics (due to coupling to a memoryless environment). Further, since the dissipation involves pure loss, the environmental modes are at zero temperature, so
\begin{subequations}
\label{neq:q.noise.op.def}
\begin{align}
\braket{\hat B^{\dag} _{x} (t) \hat B _{x'} (t')} _{\text{E}} &= 0, \qquad\qquad\qquad
 \braket{\hat B_{x} (t) \hat B _{x'} (t')} _{\text{E}} = 0
, \\
\braket{\hat B _{x} (t) \hat B ^{\dag}_{x'} (t')} _{\text{E}} &= \delta_{xx'} \delta (t-t')
,
\end{align}
\end{subequations}
where $\braket{\cdot}_{\text{E}}$ denotes expectation with respect to the environmental modes. Since $\braket{\hat B^{\dag} _{x} (t) \hat B _{x'} (t')} _{\text{E}} = 0$ implies that every $\hat B _{x}(t)$ annihilates the environmental state, Eq.~\eqref{neq:q.noise.op.def} is equivalent to the condition that all environmental modes are initialized in vacuum~\cite{Parthasarathy1984}. As the vacuum is a Gaussian state, Eqs.~\eqref{neq:q.noise.op.ccr} and~\eqref{neq:q.noise.op.def} then fix all environmental correlation functions via Wick's theorem.

Using the fact that the trace distance between two quantum states is contractive under partial trace, we can upper bound the trace distance between the probe-ancilla mixed states under any pair of oracles by the trace distance between the corresponding dilated pure-state dynamics:
\begin{align}
\label{neq:bound.qdp}
& \| \hat \rho _{x_{1}} (T_{\text{tot}}) -  \hat \rho _{x_{2}}  (T_{\text{tot}}) \| _{1}
\le \| \ketbra{\Phi_{x_{1}}} -  \ketbra{\Phi_{x_{2}}} \| _{1}
,
\end{align}
where we define the final probe-ancilla-environment state $\ket{\Phi_{x_{1}}} $ corresponding to any $x$-dependent oracle as
\begin{align}
\ket{\Phi_{x_{1}}} & \coloneqq \ket{\Phi_{x_{1} }  (T_{\text{tot}})} 
.
\end{align}
Substituting Eq.~\eqref{neq:bound.qdp} into Eq.~\eqref{neq:loss.gen.1norm}, we derive the following lower bound for the averaged state distinguishability corresponding to any given family of dilated unitaries:  
\begin{align}
\label{neq:loss.bound.1norm.0}
p ^{(1)}_{\text{th}} & < \frac{1}{2 N  (N-1) } \sum _{x _{1} \ne x _{2}} \| \hat \rho _{x_{1}} (T_{\text{tot}}) -  \hat \rho _{x_{2}} (T_{\text{tot}})  \| _{1}
\\
& \le \frac{1}{2N (N-1) }    \sum _{x _{1} \ne x _{2}}  \| \ketbra{\Phi_{x_{1}}} -  \ketbra{\Phi_{x_{2}}} \| _{1}
\nonumber \\
& = \frac{1}{N (N-1) }   \sum _{x _{1} \ne x _{2}} \sqrt{1- |\braket{ \Phi_{x_{1}} | \Phi_{x_{2}} } |^{2} }
\nonumber \\
\label{neq:loss.bound.Bures.dil.0}
& \le  \frac{1}{N (N-1) }  \sum _{x _{1}\ne x _{2}} \sqrt{ 2 (1- |\braket{ \Phi_{x_{1}} | \Phi_{x_{2}} } |) }
.
\end{align}
In the $B _{0} \to 0$ limit, the term on the last line of the right-hand side of the form $\sqrt{ 2 (1- |\braket{ \Phi_{x_{1}} | \Phi_{x_{2}} }  |) }$   
can be related to the quantum Fisher information (QFI) $\mathcal{F} _{Q}$ with respect to $B _{0}$. This suggests a connection between Eq.~\eqref{neq:loss.bound.Bures.dil.0} and the HNLS-based metrological bounds of Ref.~\cite{Jiang2018}, which limit the achievable QFI minimized over all possible purifications satisfying Eq.~\eqref{neq:def.Phi.SAE} when the Hamiltonian lies in the Lindblad span. Here, the oracle Hamiltonian $\hat H_{x} = B_{0} \hat n _{x}$ lies in the Lindblad span, $\text{span}\{\hat{\mathbb{I}}, \hat L_{x}, \hat L_{x}^{\dag}, \hat L_{x}^{\dag}\hat L_{y} \}$ with $\hat L_{x} =\hat a _{x}$, since $\hat a_{x}^{\dag} \hat a _{x} =\hat n_{x} $.
For finite $B_{0}$, the quantity in the sum on the right-hand side is equal to the Bures distance between $\ket{\Phi_{x_{1}}}$ and $\ket{\Phi_{x_{2}}}$.

To derive a query lower bound in the finite-$B _{0}$ regime, one may attempt to continuously vary the oracle strength from $0$ to $B _{0}$ and apply the QFI bounds in~\cite{Jiang2018} along the parametric path. However, such a procedure would require extra assumptions about the boson number budget on fictitious trajectories with different oracle strengths. In what follows, we provide an alternative proof that directly bounds the trace distance in Eq.~\eqref{neq:loss.bound.1norm.0}. Our proof uses a ``split-dissipation" technique inspired by the decomposition method of~\cite{Regev2008}, and we adapt it to the case with a boson-number constraint.

It is convenient to introduce an artificial reference evolution $\ket{\Phi ^{\text{(r)}}_{x_{1},x_{2}} (t)}$ for any given pair of query trajectories $\ket{\Phi_{x_{1}} (t)}$ and $\ket{\Phi_{x_{2}} (t)}$, which originates from the same initial state $\ket{\Phi ^{\text{(r)}}_{x_{1},x_{2}} (0)}   = \ket{\Phi_{\text{in}}}$ and undergoes evolution generated by the oracle Hamiltonians for both $x_{1}$ and $x_{2}$, but evolves under different oracle-independent probe-environment interactions after dilation. In particular, we split the dissipative dynamics for the $x_{2}$th mode of the reference state $\ket{\Phi ^{\text{(r)}}_{x_{1},x_{2}} (t)} $ into two parts. 
The first part is due to loss to the same environmental modes that are coupled to $\ket{\Phi_{x_{1}} }$
and contributes $\gamma |z|^{2} $ to the total decay rate, where $z$ is a complex constant coefficient satisfying $|z|<1$ that we will specify later. In contrast, the second part is coupled to a set of record modes with quantum white-noise operators $\hat D  _{x_{2}} (t) $, which also satisfies Eq.~\eqref{neq:q.noise.op.def} and is completely decoupled from the probe-ancilla-environment system associated with $\ket{\Phi_{x_{1}} }$. Concretely, the dilated reference state dynamics satisfy the equation of motion
\begin{align}
\partial _{t} \ket{\Phi ^{\text{(r)}}_{x_{1},x_{2}} (t)} 
=  {}& -i \left [ \hat H  _{x_{1} }  + \hat H  _{x_{2} } +  \hat H _{\text{c}} (t) - \frac{i\gamma }{2} \sum_{x} \hat a _{x}^{\dag} \hat a _{x} \right  ]  \ket{\Phi ^{\text{(r)}}_{x_{1},x_{2}} (t)} 
\nonumber \\
\label{neq:def.Phi.ref.SAE}
& + \sqrt{\gamma } \hat a _{x_{2}} [ z \hat B^{\dag} _{x_{2}} (t) + \sqrt{1-|z|^{2}} \hat D ^{\dag} _{x_{2}} (t) ]  \ket{\Phi ^{\text{(r)}}_{x_{1},x_{2}} (t)} 
+ \sqrt{\gamma }  \sum_{x\ne x _{2}} \hat a _{x} \hat B^{\dag} _{x} (t)  \ket{\Phi ^{\text{(r)}}_{x_{1},x_{2}} (t)}.
\end{align}
Equation~\eqref{neq:def.Phi.ref.SAE} differs from Eq.~\eqref{neq:def.Phi.SAE} in two crucial aspects: first, the oracle Hamiltonian $\hat H  _{x _{0} } $ in~\eqref{neq:def.Phi.SAE} is replaced by $\hat H  _{x _{1} }  + \hat H  _{x _{2} } $; second, the dissipation of mode $x_{2}$ is dilated differently, where $\hat B^{\dag} _{x_{2}} $ in~\eqref{neq:def.Phi.SAE} is replaced by $z \hat B^{\dag} _{x_{2}} + \sqrt{1-|z|^{2}} \hat D ^{\dag} _{x_{2}}$.
We stress that this split of the dissipative environment is solely for convenience, and does not impact the probe-ancilla evolution (after tracing out the environmental modes).
Similarly to the query trajectories, we denote the final reference state by
\begin{align}
& \ket{\Phi ^{\text{(r)}}_{x_{1},x_{2}} }
\coloneqq \ket{\Phi ^{\text{(r)}}_{x_{1},x_{2} } (T_{\text{tot}} ) }.
\end{align}
Thus, Eq.~\eqref{neq:loss.bound.1norm.0} can be upper bounded as
\begin{align}
p ^{(1)}_{\text{th}} & < \frac{1}{N (N-1)}  \sum _{x _{1} \ne x _{2}} \left ( \| \hat \rho _{x_{1}} (T_{\text{tot}}) -  \hat \rho ^{\text{(r)}}_{x_{1},x_{2}}  (T_{\text{tot}})  \| _{1} 
+  \| \hat \rho _{x_{2}} (T_{\text{tot}}) -  \hat \rho ^{\text{(r)}}_{x_{1},x_{2}}  (T_{\text{tot}})  \| _{1}  \right )
,
\end{align}
where we define $ \hat \rho ^{\text{(r)}}_{x_{1},x_{2}}  (T_{\text{tot}}) \coloneqq \text{Tr}_{\text{E}}\ket{\Phi ^{\text{(r)}}_{x_{1},x_{2}} }\!\!\bra{\Phi ^{\text{(r)}}_{x_{1},x_{2}} }$.
Noting that the sum on the right-hand side of the above inequality is symmetric under the exchange of $x_{1}$ and $x_{2}$, we can rewrite the above upper bound as
\begin{align}
\label{neq:loss.bound.1norm.dil.1}
p ^{(1)}_{\text{th}} & < \frac{2 }{N (N-1)}  \sum _{x _{1} \ne x _{2}}  \! \| \hat \rho _{x_{1}} (T_{\text{tot}}) -  \hat \rho ^{\text{(r)}}_{x_{1},x_{2}}  (T_{\text{tot}})  \| _{1} 
.
\end{align}

We next prove a general upper bound on the trace distance between the oracle trajectory $\hat \rho _{x_{1}} (T_{\text{tot}}) $ and the corresponding reference state $\hat \rho ^{\text{(r)}}_{x_{1},x_{2}}  (T_{\text{tot}}) $:
\begin{align}
\| \hat \rho _{x_{1}} (T_{\text{tot}}) -  \hat \rho ^{\text{(r)}}_{x_{1},x_{2}}  (T_{\text{tot}})  \| _{1} 
\le {} & 
4 B_{0} 
\sqrt{\frac{1}{ \gamma }\int ^{T_{\text{tot}}} _{0} \braket{ \Phi  _{ x_{1}  }  ( t )| \hat n _{x_{2}} | \Phi  _{ x_{1}  }  ( t ) }  dt }
\nonumber \\
\label{neq:bound.Bures.sq}
= {} & 
4 B_{0} 
\sqrt{\frac{1}{ \gamma }\int ^{T_{\text{tot}}} _{0} \text{Tr}[ \hat n _{x_{2}} \hat \rho _{ x_{1}  }  ( t ) ] \, dt }
.
\end{align}
To prove the above inequality, we first rewrite the reference state $\ket{\Phi ^{\text{(r)}}_{x_{1},x_{2}} (t)} $ in terms of a contribution $\ket{v(t)}$ where no jumps occur in the record modes, and a remainder term $\ket{\Phi ^{\text{(r)}}_{x_{1},x_{2}} (t)} - \ket{v(t)}$ where at least one jump occurs. By definition, the two terms are orthogonal to each other:
\begin{align}
\bra{v (t)} (\ket{\Phi ^{\text{(r)}}_{x_{1},x_{2}} (t)} - \ket{v (t)}) = 0
.
\end{align}
The norm of $\ket{v (t)}$ can change during the search protocol, and its dynamics are generated by a non-Hermitian operator:
\begin{align}
\partial _{t}\ket{v (t)} 
= {}&  -i \! \left [\! \hat H  _{x_{1} }  + \hat H  _{x_{2} } +  \hat H _{\text{c}} (t) - \frac{i\gamma }{2} \sum_{x} \hat a _{x}^{\dag} \hat a _{x} \!\right  ]  \ket{v (t)} 
\nonumber \\& 
+  \sqrt{\gamma } \sum_{x\ne x_{2}} \hat a _{x} \hat B^{\dag} _{x} (t)  \ket{v (t)} 
\label{neq:def.vref.SAE}
+ z \sqrt{\gamma } \hat a _{x_{2}} \hat B^{\dag} _{x_{2}} (t)  \ket{v (t)} 
.
\end{align}
The stochastic equation above does not conserve probability, and the norm of $\ket{v (t)} $ satisfies
\begin{align}
\partial _{t}\braket{v (t)|v (t)} = - \gamma (1-|z|^{2}) \braket{v (t)|\hat a _{x_{2}}^{\dag} \hat a _{x_{2}}|v (t)}
,
\end{align}
which we integrate over the time interval $[0,t]$ to obtain
\begin{align}
1 - \braket{v (t)|v (t)}  
= {}&- \int ^{t}_{0} \partial _{\tau}\braket{v (\tau)|v (\tau)} d \tau
\\
\label{neq:vnorm.diff}
={}&\gamma (1-|z|^{2}) \int ^{t}_{0}\braket{v (\tau)|\hat a _{x_{2}}^{\dag} \hat a _{x_{2}}|v (\tau)}d \tau
.
\end{align}
Further, $\ket{\Phi ^{\text{(r)}}_{x_{1},x_{2}} (t)} - \ket{v(t)}$ is also orthogonal to the oracle evolution $\ket{\Phi_{x_{1}}(t)} $ $\forall t$, because the oracle trajectory is not coupled to the record modes. From the triangle inequality, we thus have
\begin{align}
\| \hat \rho _{x_{1}} (T_{\text{tot}}) -  \hat \rho ^{\text{(r)}}_{x_{1},x_{2}}  (T_{\text{tot}})  \| _{1} 
\le {} & \| \ketbra {\Phi_{x_{1}} }  - \ketbra{v}  \|_{1} + 1- \|  \ket{v}\| ^{2}
\nonumber \\
\le {} & 2 ( \| \ket{\Phi_{x_{1}} }  - \ket{v}\| + 1- \|  \ket{v}\| )
\nonumber \\
\label{neq:bound.Bures.sq.1}
\le {} & 4 \| \ket{\Phi_{x_{1}} }  - \ket{v}\| 
,
\end{align}
where $\| \cdot \| $ denotes the Euclidean norm of state vectors, and we use the triangle inequality $1- \|  \ket{v}\| \le \| \ket{\Phi_{x_{1}} }  - \ket{v}\| $ to derive the last inequality.

We now derive an upper bound for the vector norm of the difference between the oracle trajectory and the no-record-jump reference state $\| \ket{\Phi_{x_{1}} }  - \ket{v}\| $, which can be rewritten in terms of the norm of $\ket{v}$ and the overlap between the two states:
\begin{align}
\| \ket{\Phi_{x_{1}} }  - \ket{v}\| ^{2} 
= 1+ \braket{v|v} - 2 \text{Re} \braket{\Phi_{x_{1}} |v}
.
\end{align}
Taking the time derivative on both sides, we have
\begin{align}
\partial _{t} \| \ket{\Phi_{x_{1}} (t) }  - \ket{v (t)}\| ^{2} 
= {} & - \gamma (1-|z|^{2}) \braket{v (t)|\hat a _{x_{2}}^{\dag} \hat a _{x_{2}}|v (t)} 
\label{neq:td.Phi.v.diff}
- 2  \text{Re} ( \partial _{t} \braket{\Phi_{x_{1}}  (t)|v (t)} )
.
\end{align}
Using the It{\^{o}} product rule, we can write the overlap increment $d (\braket{\Phi_{x_{1}}  (t)|v (t)} )$ as 
\begin{align}
d (\braket{\Phi_{x_{1}}  (t)|v (t)} )
= {} & \braket{ d \Phi_{x_{1}}  (t)|v (t)} + \braket{\Phi_{x_{1}}  (t)| d v (t)} + \braket{d \Phi_{x_{1}}  (t)|d v (t)}
,
\end{align}
so that substituting Eqs.~\eqref{neq:def.Phi.SAE} and~\eqref{neq:def.vref.SAE} into the above equation, we can compute the time derivative of the overlap term $ \braket{\Phi_{x_{1}} (t) |v (t)}$:
\begin{align}
\frac{ d }{dt } (\braket{\Phi_{x_{1}}  (t)|v (t)} )
= {} & \braket{\Phi_{x_{1}} (t)| ( -i \hat H  _{x_{2} } - \gamma \sum_{x} \hat a _{x}^{\dag} \hat a _{x}  ) |v (t)} 
+ \frac{\braket{d \Phi_{x_{1}}  (t)|d v (t)}}{dt}
&& \text{(It{\^{o}} rule)} 
\nonumber \\
= {} & \braket{\Phi_{x_{1}} (t)| ( -i \hat H  _{x_{2} } - \gamma (1-z) \hat a _{x_{2}}^{\dag} \hat a _{x_{2}}  ) |v (t)} 
\nonumber \\
\label{neq:td.Phi.v.prod}
= {} & - [i B_{0} + \gamma (1-z) ] \braket{\Phi_{x_{1}} (t)| \hat a _{x_{2}}^{\dag} \hat a _{x_{2}} |v (t)} 
,
\end{align}
where the overlap term $\braket{ d \Phi_{x_{1}} | d v} $ originates from the It{\^{o}} product rule. Substituting Eq.~\eqref{neq:td.Phi.v.prod} into Eq.~\eqref{neq:td.Phi.v.diff} and applying the Cauchy-Schwarz inequality, we have
\begin{align}
\partial _{t} \| \ket{\Phi_{x_{1}} (t) }  - \ket{v (t)}\| ^{2} 
= {} & - \gamma (1-|z|^{2}) \braket{v (t)|\hat a _{x_{2}}^{\dag} \hat a _{x_{2}}|v (t)} 
+ 2  \text{Re} \left (  [i B_{0} + \gamma (1-z) ] \braket{\Phi_{x_{1}} (t)| \hat a _{x_{2}}^{\dag} \hat a _{x_{2}} |v (t)} \right )
\nonumber \\
\le {} & \frac{| i B_{0} + \gamma (1-z) | ^{2} }{\gamma (1-|z|^{2}) }
\braket{\Phi_{x_{1}} (t)| \hat a _{x_{2}}^{\dag} \hat a _{x_{2}} |\Phi_{x_{1}} (t) } 
=  \frac{| i B_{0} + \gamma (1-z) | ^{2} }{\gamma (1-|z|^{2}) }
\braket{\Phi_{x_{1}} (t)| \hat n _{x_{2}} |\Phi_{x_{1}}  (t)} 
\label{neq:td.Phi.v.diff.1}
.
\end{align}
The above equation relates the change in the distance between an oracle trajectory and a conditional reference trajectory to the boson-number budget that only depends on the oracle trajectory.
We stress that our proof does not require any additional assumption about the boson-number dynamics of the reference state $\ket{\Phi ^{\text{(r)}}_{x_{1},x_{2}} (t)} $.
By choosing 
\begin{align}
z = \frac{\gamma }{ -i B_{0} + \gamma }
,
\end{align}
we can simplify Eq.~\eqref{neq:td.Phi.v.diff.1} as 
\begin{align}
&\partial _{t} \| \ket{\Phi_{x_{1}} (t) }  - \ket{v (t)}\| ^{2} 
\le \frac{ B_{0} ^{2} }{\gamma  }
\braket{\Phi_{x_{1}} (t)| \hat n _{x_{2}} |\Phi_{x_{1}}  (t)} 
.
\end{align}
At the initial time, the oracle-trajectory state coincides with the no-record-jump branch of the reference state
\begin{align}
\ket{\Phi_{x_{1}} (t =0) }  = \ket{v (t=0)} = \ket{\Phi_{\text{in}}}
,
\end{align}
so that the final squared two-norm of the difference between the two states can be upper bounded by
\begin{align}
& \| \ket{\Phi_{x_{1}} (T_{\text{tot}}) }  - \ket{v (T_{\text{tot}})}\| ^{2} 
\le  \frac{ B_{0} ^{2} }{\gamma  }
\int ^{T_{\text{tot}}} _{0} \braket{ \Phi  _{ x_{1}  }  ( t )| \hat n _{x_{2}} | \Phi  _{ x_{1}  }  ( t ) }  dt 
.
\end{align}
Substituting the above equation into Eq.~\eqref{neq:bound.Bures.sq.1}, we have
\begin{align}
\| \hat \rho _{x_{1}} (T_{\text{tot}}) -  \hat \rho ^{\text{(r)}}_{x_{1},x_{2}}  (T_{\text{tot}})  \| _{1} 
\le {} & 4 \| \ket{\Phi_{x_{1}} (T_{\text{tot}}) }  - \ket{v (T_{\text{tot}})}\| 
\nonumber \\
\le {} & 4 B_{0} 
\sqrt{\frac{1 }{ \gamma  }\int ^{T_{\text{tot}}} _{0} \braket{ \Phi  _{ x_{1}  }  ( t )| \hat n _{x_{2}} | \Phi  _{ x_{1}  }  ( t ) }  dt }
,
\end{align}
which proves Eq.~\eqref{neq:bound.Bures.sq}.

Substituting Eq.~\eqref{neq:bound.Bures.sq} into Eq.~\eqref{neq:loss.bound.1norm.dil.1}, we can relate the threshold probability to other protocol parameters as 
\begin{align}
\frac{p ^{(1)}_{\text{th}}\sqrt{\gamma }}{8 B_{0} } & < \frac{\sum _{x _{1} \ne x _{2}}  \! \sqrt{ 
\int ^{T_{\text{tot}}} _{0} \braket{ \Phi  _{ x_{1}  }  ( t )| \hat n _{x_{2}} | \Phi  _{ x_{1}  }  ( t ) }  dt  } }{N (N-1) }. 
\end{align}
Using the Cauchy-Schwarz inequality and the boson-number constraint in Eq.~\eqref{neq:def.nS.avg.var}, we can further upper bound the above equation as 
\begin{align}
\frac{p ^{(1)}_{\text{th}} \sqrt{\gamma }}{8 B_0}
& <  \sqrt{ \frac{\sum _{x _{1} \ne x _{2}}  \int ^{T_{\text{tot}}} _{0} \braket{ \Phi  _{ x_{1}  }  ( t )| \hat n _{x_{2}} | \Phi  _{ x_{1}  }  ( t ) }  dt }{N (N-1)} 
 } 
\nonumber \\
& \le  \sqrt{ \frac{n _{\budgetlb} T_{\text{tot}} }{N-1}    } 
,
\end{align}
establishing the lower bound
\begin{align}
T_{\text{tot}} > \frac{[p ^{(1)}_{\text{th}} ]^{2} }{64} \frac{(N-1) \gamma }{B_0^{2} n _{\budgetlb}}
.
\end{align}

In the $N\gg 1$ regime, this establishes a lower bound for general bosonic unstructured search under photon loss:
\begin{align}
\label{neq:Ttot.lb.gen.loss}
T_{\text{tot}}  = \Omega \left ( \frac{N \gamma  }{n _{\budgetlb} B _{0} ^{2}  }\right )
.
\end{align}
The bound in Eq.~\eqref{neq:Ttot.lb.gen.coh} still applies in this case, as photon loss can be viewed as coupling to certain ancillas that the user does not have access to, so that we obtain the final lower bound
\begin{align}
\label{neq:Ttot.lb.gen.comb}
T_{\text{tot}}  = \Omega \left (\max \left \{ \frac{\sqrt{N}}{ n _{\budgetlb} B _{0} }
, \frac{N \gamma }{ n _{\budgetlb}   B _{0} ^{2}  } 
\right \} 
\right )
,
\end{align}
which proves Eq.~\eqref{neq:Ttot.lower.bound.0}.
Under constant-rate photon loss and for $1\ll n_{\budgetlb}=O(N)$, the lower bound is again achievable with Gaussian probes, so the non-Gaussian advantage disappears [see the discussion below Eq.~\eqref{neq:Ttot.lb.gen.loss.1} for detail].

\subsection{Optimal protocols achieving the lower bounds \texorpdfstring{in Eqs.~\eqref{neq:Ttot.lb.Gaus.coh} and~\eqref{neq:Ttot.lb.gen.comb}}{} }

\label{nsec:opt.protocol.ent}

The bounds in Eqs.~\eqref{neq:Ttot.lb.Gaus.coh} and~\eqref{neq:Ttot.lb.gen.comb} are tight when\footnote{As noted in footnote~\ref{footnote:small_nb}, a coherent-state strategy can saturate the matching lower bound for $n_{\budgetlb} \ll 1$, so we omit that case here.} $n_{\budgetlb} \gg 1$. In this section, we explain the protocols introduced in Sec.~\ref{nsec:summary} (for $1\ll n_{\budgetlb} \ll N$) and additional strategies (for $n_{\budgetlb} \gg N$) that collectively achieve these lower bounds across all parameter regimes.
We first note that the scaling of the final bound in Eq.~\eqref{neq:Ttot.lb.gen.comb} depends on the value of $\gamma$. For sufficiently small photon-loss rate such that 
\begin{align}
\frac{N \gamma }{ n _{\budgetlb}   B _{0} ^{2}  } \ll \frac{\sqrt{N}}{ n _{\budgetlb} B _{0} } 
\quad \text{i.e.} \quad
\gamma \ll \frac{   B _{0} }{\sqrt{N}  },
\end{align}
the first term of Eq.~\eqref{neq:Ttot.lb.gen.comb} dominates, and the bound scales as
\begin{align}
\label{neq:Ttot.Grover}
T_{\text{tot}}  = \Omega \left (\frac{\sqrt{N}}{ n _{\budgetlb} B _{0} }\right )
.
\end{align}
We can use either the entangled [with initial state~\eqref{neq:nph.Wstate}] or unentangled [with initial state~\eqref{neq:nph.FockSum.prod}] non-Gaussian strategy to achieve Eq.~\eqref{neq:Ttot.Grover}, although non-Gaussianity is necessary for saturating the optimal joint scaling in $N$ and $n _{\budgetlb}$. We prove that purely Gaussian probes cannot achieve  Eq.~\eqref{neq:Ttot.Grover} in Sec.~\ref{nsec:Gaus.bound}.

We now provide more details on the non-Gaussian protocols discussed in Sec.~\ref{nsec:summary}.
First, if one prepares the probe in the $W$-Fock state
\begin{align}
&\text{Entangled, non-Gaussian strategy:  }
\label{neq:psi.WFock}
\ket{\psi _{\text{in}}} = \frac{1}{\sqrt{N}} \sum _{x=1}^{N}    \ket{ M }  _{x}
= \frac{1}{\sqrt{N}} \sum _{x=1}^{N} \frac{1}{\sqrt{ M !}}\hat a _{x}  ^\dag {}^{ M } \ket{\text{vac}}
\end{align}
with total boson number $M = \lfloor n _{\budgetlb} \rfloor $ [see Eq.~\eqref{neq:nph.Wstate}],
then a Grover-style search within the subspace spanned by Fock states $\{ \ket{ M } _{x} \}$ achieves the lower bound in Eq.~\eqref{neq:Ttot.Grover}. Specifically, the iterate $U _{\text{G}} e ^{-i B _{0} \Delta t \hat a _{x_{0}} ^\dag \hat a _{x_{0}} }$ performs dynamics analogous to discrete Grover search, giving the final state
\begin{align}
\left ( \hat U _{\text{G}} e ^{-i B _{0} \Delta t \hat a _{x_{0}} ^\dag \hat a _{x_{0}} } \right ) ^{K} \ket{\psi _{\text{in}}} 
\end{align}
[see Eq.~\eqref{neq:Wstate.Grover}].
For $\Delta t = \frac{\pi}{M B _{0}} $ and $K = O(\sqrt{N})$, the final state acquires $\Omega (1)$ amplitude in $\{ \ket{ M } _{x_{0}} \}$. Direct photodetection then enables identifying the $x_{0}$ mode as the mode with highest boson number.
Note that the state $ \ket{\psi _{\text{in}}} $ in \eqref{neq:psi.WFock} is both non-Gaussian and entangled. In fact, one can show that it contains non-Gaussian entanglement for $M>2$ and $N>1$~\cite{Zhuang2026}, i.e., the non-Gaussian state~\eqref{neq:psi.WFock} with $M>2$ cannot be prepared by applying Gaussian operations on a non-Gaussian product state.

Intriguingly, entanglement is not necessary to achieve the lower bound in Eq.~\eqref{neq:Ttot.Grover}, as shown by the unentangled non-Gaussian protocol discussed in Sec.~\ref{nsec:summary} [see Eq.~\eqref{neq:prod.Grover}]. We now discuss the boson-number budget and performance of the protocol in more detail. Recall that the initial probe state of the protocol is a uniform unentangled state 
\begin{align}
&\text{Unentangled, non-Gaussian strategy: }
\ket{\psi _{\text{prod}}} =  \bigotimes  _{x=1}^{N}  \ket{\psi _{M';N}}_{x}  
\end{align}
[see Eq.~\eqref{neq:nph.FockSum.prod}], where $M' = \lfloor n _{\budgetlb}/2 \rfloor $, and the $x$th mode superposition Fock state $\ket{\psi _{M';N}}_{x} $ is defined as
\begin{align}
\ket{\psi _{M';N}}_{x} &\coloneqq \sqrt{1-\frac{1}{N}} \ket{0}_{x}
+ \frac{1}{\sqrt{N}} \frac{1}{\sqrt{ M' !}}\hat a _{x}  ^\dag {}^{ M' } \ket{0}_{x}
\nonumber \\
&= \sqrt{1-\frac{1}{N}} \ket{0}_{x}
+ \frac{1}{\sqrt{N}} \ket{ M' }  _{x} 
.
\end{align}
For this product state $\ket{\psi _{\text{prod}}} $, each individual bosonic mode has average occupation
\begin{align}
\braket{\hat n _{x}} = M'/N,
\end{align}
so the expectation 
of the total boson number in the entire $N$-mode state is $\sum _{x} \braket{\hat n _{x}} = N \times (M'/N) = M'$. The algorithm then evolves each mode individually under a Grover-like amplitude amplification process, giving
\begin{align}
& \left ( e ^{- i \pi \sum _{x=1}^{N} \ketbra{\psi _{M' ;N} }_{x} }
e ^{-i B _{0} \Delta t \hat a _{x_{0}} ^\dag \hat a _{x_{0}} }
\right ) ^{K} \ket{\psi _{\text{prod}}}
\nonumber\\
\label{neq:nph.FockSum.prod.1}
&= \prod _{x=1}^{N} \left [ 
\left ( e ^{- i \pi \ketbra{\psi _{M' ;N} }_{x} }
e ^{-i B _{0} \delta _{x,x_{0}} \Delta t \hat a _{x_{0}} ^\dag \hat a _{x_{0}} }
\right ) ^{K} \ket{\psi _{M' ;N}}_{x} 
\right ] 
.
\end{align}
After $K = O(\sqrt{N})$ time steps with $\Delta t = \frac{\pi}{M' B _{0}} $, a unitary $\hat U _{\text{det}}$ is applied to each bosonic mode, performing the basis change  
\begin{align}
\label{neq:psiM.Udet.def.}
& \hat U _{\text{det}} \ket{\psi _{M';N}}_{x}  = \ket{0}_{x}
, \\
& \hat U _{\text{det}} \ket{\psi ^{\perp} _{M';N}}_{x}  = \ket{M'}_{x}
,
\end{align}
where $\ket{\psi ^{\perp} _{M';N}}_{x} = \frac{1}{\sqrt{N}} \ket{0}_{x}
- \sqrt{1-\frac{1}{N}} \ket{ M' }  _{x}  $ denotes the state in $\text{span} \{ \ket{0}_{x} , \ket{M'}_{x} \}$ that is orthogonal to $\ket{\psi _{M';N}}_{x}$. Photon counting then identifies the target mode with constant success probability, which can be amplified to $1-\epsilon$ for any $\epsilon>0$ with $O(\log(1/\epsilon))$ repetitions.
Alternatively, without applying $\hat U _{\text{det}} $, one could directly measure each bosonic mode via photodetection and repeat to achieve a high success probability.

Throughout the product-state search process~\eqref{neq:nph.FockSum.prod.1}, only the target mode's boson number changes during the time evolution: without the target phase shift $e ^{-i B _{0} \delta _{x,x_{0}} \Delta t \hat a _{x_{0}} ^\dag \hat a _{x_{0}} }$, the Grover iteration $e ^{- i \pi \ketbra{\psi _{M' ;N} }_{x} }$ only repeatedly induces $\pi$ overall phase shifts on each mode. In the target mode, however, the Grover iteration in Eq.~\eqref{neq:nph.FockSum.prod.1} rotates the initial state to a final state $\simeq \ket {M' }$, which can be easily distinguished from all the other modes whose boson number is $M' /N$ using the measurement procedure discussed above [see Eq.~\eqref{neq:psiM.Udet.def.}]. Thus, this protocol identifies the value of $x_{0}$ in query time $K \Delta t = O(\frac{\sqrt{N}}{M' B _{0}}) $, achieving Eq.~\eqref{neq:Ttot.Grover} using a maximal instantaneous mean boson number $< 2M' \le n _{\budgetlb} $. Furthermore, the protocol only involves unentangled probe states throughout the search process. We note that none of the control operations in this unentangled non-Gaussian strategy have any entangling power, i.e., they cannot create any entanglement from any unentangled input, in contrast to the beamsplitters used in the coherent-state protocol discussed below. Nevertheless, the probe is still highly non-Gaussian.

The protocol used in Ref.~\cite{Spreeuw2002} demonstrates that non-Gaussianity is not necessary to achieve the optimal $\Theta(\sqrt{N})$ scaling. More specifically, initializing the probe in the coherent state
\begin{align}
&\text{Coherent state + passive linear optics: }
\ket{\psi _{\text{coh}}} = \ket{\alpha} ^{\otimes N}
\end{align}
with $\alpha$ chosen to satisfy the boson-number constraint, $|\alpha | ^{2} = n _{\budgetlb} /{N}$,
and then performing Grover search using  beamsplitters, also achieves the $\Theta (\sqrt{N})$ speedup~\cite{Spreeuw2002}. However, Ref.~\cite{Spreeuw2002} does not consider boson-number scaling when using Gaussian strategies, which, as we show in Eq.~\eqref{neq:Ttot.lb.Gaus.coh}, must be suboptimal in its joint dependence on $N$ and $n_{\budgetlb}$. In fact, for $1\ll n_{\budgetlb}=O(N) $, measuring the coherent-state output after $O(\sqrt{N / n _{\budgetlb}})$ Grover iterations with displaced photodetection achieves
\begin{align}
T_{\text{Gaussian}}  = \Theta \left (\frac{\sqrt{N}}{ \sqrt{n _{\budgetlb}} B _{0} }\right )
,
\end{align} 
which is also optimal for any Gaussian probe in the absence of loss in that regime [see Eq.~\eqref{neq:Ttot.lb.Gaus.coh}].

In contrast, when $\gamma=\omega(B_0/\sqrt{N})$, the $\gamma$-dependent lower bound~\eqref{neq:Ttot.lb.gen.loss} dominates:
\begin{align}
\label{neq:Ttot.lb.gen.loss.1}
\gamma=\omega(B_0/\sqrt{N})  \quad \Longrightarrow \quad
T_{\text{tot}}  = \Omega \left (\frac{N \gamma }{ n _{\budgetlb} B _{0} ^{2}  } \right ) 
.
\end{align}
Moreover, for $\gamma \gg \sqrt{n _{\budgetlb}} B _{0}  / {\sqrt{N}}$, this bound is achievable by applying the coherent-state strategy~\eqref{seq:boson.Grover}, with a modified two-stage readout procedure (described below), to clusters of $\Theta (\max\{ B _{0}^{2} /\gamma ^{2}, 1\} )$ modes using $\Theta ( \max\{ \gamma ^{2}/B _{0}^{2} , 1\} )$ bosons per cluster.\footnote{For $\gamma\gtrsim B_{0}$, each cluster is a single mode, and its search reduces to a phase measurement of that mode over time $O(\gamma^{-1})$, without Grover iterations. In this regime, parallel search of $\Theta(\min\{n _{\budgetlb},N\})$ single-mode coherent states with $\Theta(\max\{1,n _{\budgetlb}/N\})$ bosons per mode, each mode being independently probed $\Theta(\gamma ^{2}/[B_{0}^{2}\max\{1,n _{\budgetlb}/N\}])$ times, achieves the lower bound~\eqref{neq:Ttot.lb.gen.loss.1}.} This parameter regime includes the typical case with constant $\gamma = \Theta (B_{0}) $ and finite boson number $n _{\budgetlb} = O(N)$. In this case, search within each cluster costs query time $O( \gamma ^{-1}  )$, and the boson-number constraint enables searching $\Theta (\min \{n _{\budgetlb}, n _{\budgetlb} B _{0}^{2}/ \gamma ^{2}\} )$ clusters in parallel. Since we assume $n _{\budgetlb}\gg 1$ and $\gamma =\Omega( \sqrt{n _{\budgetlb}} B _{0}  / {\sqrt{N}})$, the boson-number constraint is satisfied by construction. These parameter choices keep the accumulated loss during each cluster interrogation bounded by a constant. For clusters containing more than one mode, the following two-stage readout applies throughout this regime.
\begin{enumerate}
    \item [(i)] In the first stage, every mode is displaced by the negative of its known no-mark amplitude (including the deterministic loss damping), so that all modes of an unmarked cluster are exactly vacuum and never click, while the marked cluster retains a constant residual mean boson number and hence clicks with constant probability. During this stage, clusters are interrogated in sequential batches, with the budget-allowed number of clusters searched in parallel within each batch. Each batch is tested a fixed number of times, and any click identifies the cluster containing the marked element. With the parameter choices above, this stage costs the query time stated above.
    \item [(ii)] The coherent-state search~\eqref{seq:boson.Grover} is then repeated on the identified cluster only, with the displaced photon-counting readout [see discussion below Eq.~\eqref{neq:coh.nx0.fin}] and the displacement adjusted for loss during the Grover iterations.
\end{enumerate}
For (ii), the preceding success probability analysis [see discussion below Eq.~\eqref{neq:coh.K.pop.sum}] continues to provide a constant success probability, and a constant number of repetitions suffice to reach the desired threshold. Stage (ii) thus adds $O(1)$ runs of duration $O(\gamma^{-1})$, which does not exceed $N\gamma/(n_{\budgetlb}B_0^2)$ when $\gamma=\Omega(B_0\sqrt{n_{\budgetlb}/N})$. 
Thus, the coherent-state strategy combined with a two-stage readout achieves the query lower bound~\eqref{neq:Ttot.lb.gen.loss.1} using the given boson-number budget. We note that distributing the boson-number budget over many clusters is essential to achieve $O(n _{\budgetlb}^{-1})$ scaling. In contrast, concentrating all bosons into a single cluster is suboptimal in the lossy regime by a factor $\Theta (\sqrt{n _{\budgetlb}})$. This can be understood intuitively as due to the fact that the trace distance between the oracle-on and -off coherent-state trajectories (quantifying state distinguishability) grows as $\sqrt{n _{\budgetlb}} B_{0} t $, which favors long probe times with few bosons per pixel.

For the intermediate parameter regime $B_{0}/\sqrt{N}\ll \gamma \ll B_{0}\sqrt{n _{\budgetlb}/N}$ with $n_{\budgetlb}=O(N)$, the lower bound~\eqref{neq:Ttot.lb.Gaus.coh} associated with Gaussian probes is strictly tighter than the lossy query bound~\eqref{neq:Ttot.lb.gen.loss}:
\begin{align}
\label{neq:cond.nonG.lossy}
 \frac{ B _{0} }{\sqrt{N}  } \ll \gamma \ll \frac{ \sqrt{n _{\budgetlb}} B _{0} }{\sqrt{N}}
\quad
\Longrightarrow \quad \frac{N \gamma }{ n _{\budgetlb} B _{0} ^{2}  } \ll \frac{\sqrt{N}}{\sqrt{n _{\budgetlb}} B _{0}}
.
\end{align}
In this regime, the lossy-search lower bound~\eqref{neq:Ttot.lb.gen.loss} is achieved by applying the entangled strategy~\eqref{neq:Wstate.Grover} with $M=\lfloor n_{\budgetlb}\rfloor$ bosons to blocks of $L=\Theta(B_0^2/\gamma^2)$ modes, one block at a time, where $4\le L\ll N$. Each block search uses $\Theta(\sqrt{L})$ Grover iterations of duration $\pi/(MB_0)$, i.e., interrogation time $\Theta(1/(M\gamma))$, so that the no-loss probability $e^{-\gamma M t}$ remains $\Omega(1)$. Since all controls conserve the total boson number, a total-number measurement rejects every trajectory in which a loss occurred, and on the no-loss branch the decay term in Eq.~\eqref{neq:def.Phi.SAE} is a scalar within the fixed-$M$ sector, so the lossless Grover dynamics are unchanged. After the Grover iterations, readout on each block proceeds with three steps: 
\begin{enumerate}
    \item [(i)]{measure the total boson number and discard outcomes below $M$; }
    \item [(ii)]{within the surviving $M$-boson sector, apply the POVM
    $\{\ket{\psi _{\text{in},L}}\!\bra{\psi _{\text{in},L}},\,\hat {\mathbb{I}}-\ket{\psi _{\text{in},L}}\!\bra{\psi _{\text{in},L}} \}$ nondestructively, where $\ket{\psi _{\text{in},L}}$ is the initial W-Fock state of the block [Eq.~\eqref{neq:nph.Wstate}], and discard the $\ket{\psi _{\text{in},L}}$ outcome;}
    \item [(iii)] {conditioned on obtaining the second outcome (i.e., projection onto $\hat {\mathbb{I}}-\ket{\psi _{\text{in},L}}\!\bra{\psi _{\text{in},L}}$), perform boson counting to obtain a candidate mode (the mode containing $M$ bosons).}
\end{enumerate}
The first step excludes cases undergoing photon loss during the evolution. On the no-loss branch, an unmarked block returns exactly to $\ket{\psi _{\text{in},L}}$ up to a phase: since the oracle acts trivially on it and the decay term is a scalar in the fixed-$M$ sector, it never reaches step (iii). For a marked block, writing $\ket{\psi _{\text{in},L}}=\sin\varphi_L\ket{M}_{x_{0}}+\cos\varphi_L\ket{\psi _{\perp,L}}$ with $\sin\varphi_L=L^{-1/2}$ and $\ket{\psi _{\perp,L}} = (L-1)^{-\frac{1}{2}}\sum_{x\ne x_{0}} \ket{M}_{x}$, step (ii) succeeds with probability $\sin^2(2K\varphi_L)$, after which step (iii) returns $x_0$ with probability $1-1/L$. We choose $K\approx\pi/(4\varphi_L)$ and integer block sizes $L=\lfloor B_0^2/\gamma^2\rfloor\ge4$, merging a smaller final block into its predecessor; blocks of one or two modes are excluded, since there the $\pi$-phase oracles differ only by a global phase. Thus, with the stated choice of $K$, a constant number of repetitions suffices.
Searching the $\lceil N/L\rceil$ blocks sequentially, each repeated $O(1)$ times, gives total query time $\Theta\big((N/L)\,(M\gamma)^{-1}\big)=\Theta(N\gamma/n_{\budgetlb}B_0^2)$, which saturates Eq.~\eqref{neq:Ttot.lb.gen.loss}.

For $n_{\budgetlb}\gg N$, the Gaussian lower bound becomes $T_{\text{Gaussian}}=\Omega(\max\{N/n_{\budgetlb}B_{0},\,N\gamma/n_{\budgetlb}B_{0}^{2}\})$ [Eqs.~\eqref{neq:Ttot.lb.Gaus.coh} and~\eqref{neq:Ttot.lb.gen.loss}]. For $B_{0}/\sqrt{N} \ll \gamma \le B_{0}/2 $, the block protocol above attains $\Theta(N\gamma/n_{\budgetlb}B_{0}^{2})$, while Gaussian strategies are slower by a factor $B_{0}/\gamma$. The remaining window $B_{0}/2\le\gamma\ll B_{0}\sqrt{n_{\budgetlb}/N}$ is covered by an unentangled non-Gaussian strategy, which starts in a similar initial state as~\eqref{neq:prod.Grover} but does not involve Grover diffusion. Specifically, the modes are probed in sequential batches of $P=\lfloor n_{\budgetlb}/M\rfloor\le N$ modes, where $M=\Theta(\max\{n_{\budgetlb}/N,\log N\})$, with each mode initialized in $(\ket{0}+\ket{M})/\sqrt{2}$. Each mode is interrogated for time $1/(M\gamma)$, acquiring relative phase $B_{0}/\gamma$ with constant survival probability, and is then tested by projection onto the state in $\operatorname{span}\{\ket{0},\ket{M}\}$ orthogonal to $\ket{0}+e^{-1/2}\ket{M}$, the no-loss branch of the initial state without the oracle. Loss events that end up in $\operatorname{span}\{\ket{1},\ket{2},\ldots,\ket{M-1}\}$ exit this subspace and never pass the test; only the branch in which all $M$ bosons are lost returns to $\ket{0}$, so an unmarked mode passes with probability $O(M^{-M})$, whereas the marked mode passes with probability\footnote{Explicitly, $p_{\text{false}}=(1-e^{-1/M})^M/[2(e+1)]\le M^{-M}/[2(e+1)]$ and $p_{\text{marked}}=p_{\text{false}}+\frac{2}{e+1}\sin^2(B_0/2\gamma)$.} $\Theta(B_{0}^{2}/\gamma^{2})$. After $\Theta(\gamma^{2}/B_{0}^{2})$ repetitions per batch, the marked mode passes with constant probability, while the expected number of false passes over all modes is $O(N M^{1-M})=o(1)$, where we use $\gamma^{2}/B_{0}^{2}\ll n_{\budgetlb}/N\le M$ in this window and $M\gtrsim\log N$. The total query time is $\Theta\big((N/P)(\gamma^{2}/B_{0}^{2})(M\gamma)^{-1}\big)=\Theta(N\gamma/n_{\budgetlb}B_{0}^{2})$. We note that this query complexity outperforms any protocol using fully coherent states: comparing Eq.~\eqref{neq:Ttot.lb.coh}, which still holds in the presence of boson loss (Sec.~\ref{nsec:Gaus.bound}), with Eq.~\eqref{neq:Ttot.lb.gen.loss} shows that coherent probes reach the lossy bound only for $\gamma=\Omega( B_0\sqrt{n_{\budgetlb}/N})$. Thus, non-Gaussian probes outperform coherent probes in this window, while entanglement is still unnecessary.

We summarize the tight lower bound for the lossy bosonic unstructured search over all search strategies with $n_{\budgetlb} \gg 1$ as
\begin{align}
T_{\text{tot}}  & = \Theta \left ( 
\max \left \{ \frac{\sqrt{N}}{ n _{\budgetlb} B _{0} }, \frac{N \gamma }{ n _{\budgetlb}   B _{0} ^{2}  } \right \} 
\right )
= \begin{cases}
\Theta\left (\frac{\sqrt{N}}{ n _{\budgetlb} B _{0} }\right ), & \gamma =O(\frac{ B _{0} }{\sqrt{N}  }), \\
\Theta \left (\frac{N \gamma }{ n _{\budgetlb}   B _{0} ^{2}  } \right ), & \gamma =\Omega (\frac{ B _{0} }{\sqrt{N}  }).
\end{cases}
\end{align}
Specifically, when $\gamma =O(B_{0}/\sqrt{N})$, running one of the non-Gaussian protocols and repeating them $O(1)$ times achieves the Grover speedup and the optimal $n _{\budgetlb} $ scaling.

\subsection{Query models with alternative energy constraints}
\label{nsec:query.alt.limit}

We now discuss how our results change with energy constraints other than Eq.~\eqref{neq:def.nS.avg.var}. Our lower bound in Eq.~\eqref{neq:Ttot.lower.bound.0} directly applies to other energy constraints, but the protocols achieving such bounds can be drastically different. For example, if we constrain the average boson number per pixel instead of the total boson number,
\begin{align}
\forall x, \quad 
\frac{1}{T_{\text{tot}}}\int ^{T_{\text{tot}}} _{0}
\braket{ \hat n _{x} (t) } dt \le n _{\text{sing}} 
,
\end{align}
then the marked element can be identified in $O(1)$ queries for $n _{\text{sing}} = \Theta(1)$ (see the discussion above Eq.~\eqref{neq:def.nS.avg.var}). Alternatively, we could constrain the time-integrated total boson number
\begin{align}
B _{0} \int ^{T_{\text{tot}}} _{0}
\braket{ \hat n _{\text{S}} (t) } dt \le n _{\text{sum}} 
,
\end{align}
where we include $B _{0} $ as a dimensional constant. Unlike~\eqref{neq:def.nS.avg.var}, which fixes the time-averaged boson number irrespective of $T_{\text{tot}}$, the above equation only constrains the total boson budget, so a protocol can trade query time for instantaneous boson-number budget. 
To use the lower bound in Eq.~\eqref{neq:Ttot.lower.bound.0} in this case, we introduce the time-averaged boson number $n ^{(0)} _{\budgetlb} $ for any given protocol as an additional parameter that satisfies the condition
\begin{align}
n ^{(0)} _{\budgetlb} \coloneqq \frac{1}{T_{\text{tot}}}\int ^{T_{\text{tot}}} _{0}
\braket{ \hat n _{\text{S}} (t) } dt 
\le \frac{n _{\text{sum}} }{B _{0} T_{\text{tot}}}
,
\end{align}
so that the lower bound in Eq.~\eqref{neq:Ttot.lower.bound.0} can be rewritten in terms of the new constraint parameter $n _{\text{sum}} $ as 
\begin{align}
T_{\text{tot}} & = \Omega \left (\max \left \{ \frac{\sqrt{N}}{ n ^{(0)} _{\budgetlb}  B _{0} }
, 
\frac{N \gamma }{ n ^{(0)} _{\budgetlb}    B _{0} ^{2}  } 
\right \} 
\right )
\quad \Longrightarrow \quad
n _{\text{sum}} = \Omega \left ( \max \left \{ 
\sqrt{N} , 
\frac{N \gamma }{   B _{0}  }
\right \} \right )
\label{neq:nsum.lower.bound}
,
\end{align}
where the total query time $T_{\text{tot}}  $ drops out from the final result, and we instead derive a lower bound on $n _{\text{sum}} $. 
Choosing $1\ll n_{\budgetlb}^{(0)}$ and applying the
matching protocols of Sec.~\ref{nsec:opt.protocol.ent} shows that the above lower bound on $n_{\text{sum}}$
is also tight. This means that when we constrain the time-integrated boson number $n _{\text{sum}} $ instead of the time-averaged value, we can only successfully accomplish bosonic unstructured search when $n _{\text{sum}} $ satisfies Eq.~\eqref{neq:nsum.lower.bound}. Moreover, as long as Eq.~\eqref{neq:nsum.lower.bound} is satisfied, the search can be completed in arbitrarily short time by choosing an arbitrarily large $n ^{(0)} _{\budgetlb} $ and implementing one of the protocols achieving Eq.~\eqref{neq:Ttot.lower.bound.0} described in Sec.~\ref{nsec:opt.protocol.ent}.

Finally, all the query lower bounds in Secs.~\ref{nsec:Ttot.gen.bound}, \ref{nsec:Gaus.bound}, and~\ref{nsec:loss.bound} hold equally well when the peak instead of the time-averaged boson number is constrained, i.e.,
\begin{align}
\label{neq:ns.peak.cond}
\forall t \in [0, T_{\text{tot}} ], \quad 
\braket{ \hat n _{\text{S}} (t) } \le n _{\budgetlb} 
,
\end{align}
because a properly constrained peak boson number is a sufficient condition for the time-averaged budget to hold. Further, since all protocols in Sec.~\ref{nsec:opt.protocol.ent} keep the instantaneous mean boson number at or below $n _{\budgetlb} $ by construction, they naturally satisfy a peak-boson-number constraint. Thus,  for $1\ll n_{\budgetlb}$, the aforementioned query bounds are also tight under a peak-boson-number constraint~\eqref{neq:ns.peak.cond}.

\subsection{Query complexity for discrete-phase oracle models}

\label{nsec:oracle.phi0}

In this section, we prove a query lower bound for the alternative query model~\eqref{neq:phi.oracle.def}, where the oracle is a fixed-angle unitary acting on an unknown target mode,
\begin{align}
\hat U _{x_{0} } (\phi_{0}) = e^{-i \phi_{0} \hat a _{x_{0}} ^\dag \hat a _{x_{0}} } 
= e^{-i \phi_{0} \hat n _{x_{0}}  } 
,
\end{align}
and $\phi_{0}$ is a known constant fixed throughout the protocol, $\phi_{0} \ne 0  \pmod {2\pi}$. The goal is to find the value of $x_{0} $ using as few queries as possible. We consider protocols with arbitrary probe-ancilla control operations subject to a discretized boson-number constraint. Specifically, we adapt the time-averaged boson-number constraint in
Eq.~\eqref{neq:def.nS.avg.var} to the discrete oracle model as
\begin{align}
\label{neq:nS.avg.cond.0}
\frac{1}{K} \sum _{j=1} ^{K}
\braket{ \hat n _{\text{S}, j}  } \le n _{\budgetlb} 
, 
\end{align}
where $\braket{ \hat n _{\text{S}, j}  }$ denotes the total boson number during the application of the $j$th oracle.

To prove a lower bound, we use an argument that adapts Eqs.~\eqref{neq:Psi.xo.def} and~\eqref{neq:pth.1norm.coh} to discrete queries. In particular, we prove two query lower bounds that can be written compactly as
\begin{align}
\label{neq:Ktot.bound.phi0.0}
K_{\text{tot}}  = \Omega \left (\frac{\sqrt{N}}{ \min \left \{ n _{\budgetlb} \left | \sin  \frac{\phi_{0}}{2}  \right |  , \sqrt{ n _{\budgetlb} \left | \sin  \frac{\phi_{0}}{2}  \right | } 
\right \}  }\right )
.
\end{align}
While the $\Omega (\sqrt{N} /n _{\budgetlb}  | \sin  \frac{\phi_{0}}{2}   |  )$ lower-bound scaling can be viewed as a discrete version of the continuous-time result in Eq.~\eqref{neq:Ttot.lb.gen.coh}, the $\Omega (\sqrt{N} / \sqrt{n _{\budgetlb}  | \sin  \frac{\phi_{0}}{2} |})$ lower bound exhibits a qualitatively different scaling in boson number. The proof for this second bound also differs from the continuous-time case in one technical aspect: in Eq.~\eqref{neq:bound.evol.phi0}, the proof explicitly uses the discrete spectrum of the oracle unitary to constrain the amount of change in state distinguishability (as measured by the trace distance between states evolving under two different oracles) per query throughout the search protocol. In this sense, the proof for the $\Omega (\sqrt{N} / \sqrt{n _{\budgetlb}  | \sin  \frac{\phi_{0}}{2} |})$ scaling more closely resembles finite-dimensional query lower bounds~\cite{Ambainis2000}.
We note that this argument would not produce a nontrivial lower bound in the continuous-time setting, because in the latter case, the time step can be made arbitrarily small, and a corresponding lower bound on the total query time would vanish as  $\phi_{0} ^{\frac{1}{2}} $ [consider the $\phi_{0}\to 0$ limit of Eq.~\eqref{neq:K.lower.bound.phi0}].

Similarly to Sec.~\ref{nsec:Ttot.gen.bound}, we begin by writing the probe-ancilla state after $K$ queries for any search protocol in terms of the oracle unitaries interleaved with arbitrary probe-ancilla control unitaries $\hat V _{j}$ as
\begin{align}
\label{neq:Psi.xo.phi0.def}
\ket{\Psi _{x _{0}}} & = \hat V _{K+1} \mathcal{T} \prod _{j=1} ^{K} \left( e^{-i \phi_{0} \hat n _{x_{0}}  }  \hat V _{j} \right) (\ket{\psi _{0}} \otimes \ket{\text{vac}}_{\text{A}} )
.
\end{align}
For convenience, we transform to the toggling frame defined by the control unitaries $\hat V _{j} $ as 
\begin{align}
\label{neq:Psi.xo.phi0.int}
\ket{\Psi ^{\text{(t)}} _{x _{0}}} & = \tilde{\mathcal{T}} \prod _{j=1} ^{K+1} \hat V ^{\dag} _{j}  \ket{\Psi  _{x _{0}}} 
\\
& = \mathcal{T} \prod _{j=1} ^{K} e^{-i \phi_{0} \hat n ^{\text{(t)}}_{x_{0},j}  }  (\ket{\psi _{0}} \otimes \ket{\text{vac}}_{\text{A}} )
, 
\end{align}
where $\tilde{\mathcal{T}} $ denotes anti-time-ordering, and the toggling-frame number operator $\hat n ^{\text{(t)}}_{x_{0},j} $ at time-step $j$ is defined as
\begin{align}
\hat n ^{\text{(t)}}_{x_{0},j} &\coloneqq \hat U ^{\dag}_{j}
\hat n  _{x_{0}} \hat U_{j}
, \quad 
\hat U_{j} \coloneqq \mathcal{T} \prod _{\ell=1} ^{j} \hat V  _{\ell} 
.
\end{align}
The toggling frame can be viewed as a discrete version of the interaction picture with respect to the probe-ancilla control unitaries. Writing the probe-ancilla state after the $j$th query as
\begin{align}
\label{neq:Psi.tog.j.def}
\ket{\Psi ^{\text{(t)}} _{x _{0};j}} & = \mathcal{T} \prod _{\ell=1} ^{j} e^{-i \phi_{0} \hat n ^{\text{(t)}}_{x_{0},\ell} }  (\ket{\psi _{0}} \otimes \ket{\text{vac}}_{\text{A}} )
,
\end{align}
the boson-number constraint~\eqref{neq:nS.avg.cond.0} can be rewritten in terms of the toggling-frame operators as 
\begin{align}
\label{neq:ntot.bound.phi0}
\forall x_{0}:\quad 
\frac{1}{K} \sum _{j=1} ^{K} \sum _{x_{1}=1}^{N} \braket{ \Psi ^{\text{(t)}} _{x _{0};j-1}| \hat n ^{\text{(t)}}_{x_{1},j} |\Psi ^{\text{(t)}} _{x _{0};j-1} } \le n _{\budgetlb} 
.
\end{align}

We can rewrite Eq.~\eqref{neq:pth.1norm.coh} in terms of the overlap between the toggling-frame states as 
\begin{align}
p ^{(0)}_{\text{th}} & < \frac{1}{N (N-1)} \sum _{x _{1} \ne x _{2}} \frac{1}{2} \| \ket{\Psi _{x _{1}}}\!\bra{\Psi _{x _{1}}}  -  \ket{\Psi _{x _{2}}}\!\bra{\Psi _{x _{2}}}   \| _{1} 
\nonumber \\
& =  \frac{1}{N (N-1)} \sum _{x _{1} \ne x _{2}} \frac{1}{2} \| \ket{\Psi ^{\text{(t)}}_{x _{1}}}\!\bra{\Psi ^{\text{(t)}} _{x _{1}}}  -  \ket{\Psi ^{\text{(t)}} _{x _{2}}}\!\bra{\Psi ^{\text{(t)}} _{x _{2}}}   \| _{1} 
\nonumber \\
& =  \frac{1}{N (N-1)} \sum _{x _{1} \ne x _{2}} \sqrt{1- |\! \bra{\Psi ^{\text{(t)}}_{x _{1} }} \Psi ^{\text{(t)}}_{x _ {2} } \rangle | ^{2} } 
. \end{align}
Using the fact that the state overlap satisfies $|\! \braket{\Psi ^{\text{(t)}}_{x _{1} } | \Psi ^{\text{(t)}}_{x _ {2} } }  \!| \le 1 $, we obtain
\begin{align}
\label{neq:pth.1norm.tog}
p ^{(0)}_{\text{th}} <   \frac{\sqrt{2}}{N (N-1)} \sum _{x _{1} \ne x _{2}} \sqrt{1- |\! \braket{\Psi ^{\text{(t)}}_{x _{1} } | \Psi ^{\text{(t)}}_{x _ {2} } }  \!| } 
.
\end{align}

We now derive an upper bound on the change in the overlap between different query-trajectory states 
using an argument similar to the derivation in Eq.~\eqref{neq:gen.qsl.bound} but adapted to discrete oracles. We first rewrite the overlap as a sum over accumulated changes in each time step:
\begin{align}
& 1- \braket{\Psi ^{\text{(t)}}_{x _{1} } | \Psi ^{\text{(t)}}_{x _ {2} } } 
\label{neq:Psi.overlap.evol}
=  \sum _{j = 1}^{K} \left ( \braket{\Psi ^{\text{(t)}}_{x _{1} ;j-1  } | \Psi ^{\text{(t)}}_{x _ {2} ;j-1 } } - \braket{\Psi ^{\text{(t)}}_{x _{1} ;j} | \Psi ^{\text{(t)}}_{x _ {2} ;j } } \right )
,
\end{align}
where $\ket{\Psi ^{\text{(t)}} _{x _{1};j}} $ and $\ket{\Psi ^{\text{(t)}} _{x _{2};j}} $ denote the probe-ancilla state after $j$th query as defined in Eq.~\eqref{neq:Psi.tog.j.def}, and we use the relations
\begin{align}
& \ket{\Psi ^{\text{(t)}}_{x _{1} ; K } } = \ket{\Psi ^{\text{(t)}} _{x _{1}}} 
, \quad 
\ket{\Psi ^{\text{(t)}}_{x _{2} ; K } } = \ket{\Psi ^{\text{(t)}} _{x _{2}}} 
,\\
& \ket{\Psi ^{\text{(t)}}_{x _{1} ; 0  } } = \ket{\Psi ^{\text{(t)}}_{x _ {2} ;0 } }= \ket{\psi _{0}} \otimes \ket{\text{vac}}_{\text{A}} 
.
\end{align}
From Eq.~\eqref{neq:Psi.overlap.evol} and the triangle inequality, we thus have
\begin{align}
| 1- \braket{\Psi ^{\text{(t)}}_{x _{1} } | \Psi ^{\text{(t)}}_{x _ {2} } }  |
\label{neq:query.evol.step}
\le {} & \sum _{j = 1}^{K} \left | \braket{\Psi ^{\text{(t)}}_{x _{1} ;j-1  } | \Psi ^{\text{(t)}}_{x _ {2} ;j-1 } } - \braket{\Psi ^{\text{(t)}}_{x _{1} ;j} | \Psi ^{\text{(t)}}_{x _ {2} ;j } } \right |
.
\end{align}
From Eq.~\eqref{neq:Psi.tog.j.def}, we can relate the query evolution in two consecutive time steps as 
\begin{align}
\label{neq:Psi.phi0.evol.tog}
\ket{\Psi ^{\text{(t)}}_{x _ {1} ;j } } = e^{-i \phi_{0} \hat n ^{\text{(t)}}_{x_{1},j} }  \ket{\Psi ^{\text{(t)}}_{x _ {1} ;j-1 } }
,
\end{align}
so that the right-hand side of Eq.~\eqref{neq:query.evol.step} can be rewritten as 
\begin{align}
| 1- \braket{\Psi ^{\text{(t)}}_{x _{1} } | \Psi ^{\text{(t)}}_{x _ {2} } }  |
\le {}& \sum _{j = 1}^{K} \left | \braket{\Psi ^{\text{(t)}}_{x _{1} ;j-1  } | \Psi ^{\text{(t)}}_{x _ {2} ;j-1 } } - \braket{\Psi ^{\text{(t)}}_{x _{1} ;j} | \Psi ^{\text{(t)}}_{x _ {2} ;j } } \right |
\nonumber \\
\label{neq:Psi.overlap.evol.1}
= {} & \sum _{j = 1}^{K} \left | \braket{\Psi ^{\text{(t)}}_{x _{1} ;j-1  } | (e^{i \phi_{0} \hat n ^{\text{(t)}}_{x_{1},j} } e^{-i \phi_{0} \hat n ^{\text{(t)}}_{x_{2},j} } -\mathbb{I} ) | \Psi ^{\text{(t)}}_{x _ {2} ;j-1 } } \right |
.
\end{align}
Substituting the relation 
\begin{align}
e^{i \phi_{0} \hat n ^{\text{(t)}}_{x_{1},j} } e^{-i \phi_{0} \hat n ^{\text{(t)}}_{x_{2},j} } -\mathbb{I} 
= {} & e^{i \phi_{0} \hat n ^{\text{(t)}}_{x_{1},j} }  (e^{-i \phi_{0} \hat n ^{\text{(t)}}_{x_{2},j} } -\mathbb{I} )+  
e^{i \phi_{0} \hat n ^{\text{(t)}}_{x_{1},j} } - \mathbb{I} 
\end{align}
into the right-hand side of Eq.~\eqref{neq:Psi.overlap.evol.1} and again invoking the triangle inequality, we have
\begin{align}
| 1- \braket{\Psi ^{\text{(t)}}_{x _{1} } | \Psi ^{\text{(t)}}_{x _ {2} } }  |
\le {}& \sum _{j = 1}^{K} \left | \braket{\Psi ^{\text{(t)}}_{x _{1} ;j-1  } | \Psi ^{\text{(t)}}_{x _ {2} ;j-1 } } - \braket{\Psi ^{\text{(t)}}_{x _{1} ;j} | \Psi ^{\text{(t)}}_{x _ {2} ;j } } \right |
\\
\le {}& \sum _{j = 1}^{K} \left | \braket{\Psi ^{\text{(t)}}_{x _{1} ;j-1  } | e^{i \phi_{0} \hat n ^{\text{(t)}}_{x_{1},j} }  (e^{-i \phi_{0} \hat n ^{\text{(t)}}_{x_{2},j} } -\mathbb{I} ) | \Psi ^{\text{(t)}}_{x _ {2} ;j-1 } } \right |
\nonumber \\& 
+ \sum _{j = 1}^{K} \left | \braket{\Psi ^{\text{(t)}}_{x _{1} ;j-1  } | ( e^{i \phi_{0} \hat n ^{\text{(t)}}_{x_{1},j} } -\mathbb{I} ) | \Psi ^{\text{(t)}}_{x _ {2} ;j-1 } } \right |
.
\end{align}
Substituting the above inequality into Eq.~\eqref{neq:pth.1norm.tog} and applying the Cauchy-Schwarz inequality, we derive that $p ^{(0)}_{\text{th}}$ is at most
\begin{align}
& \frac{\sqrt{2}}{N (N-1)} \!\sum _{x _{1} \ne x _{2}} \!\!\sqrt{\sum _{j = 1}^{K} \left (  \left | {\langle\Psi ^{\text{(t)}}_{x _{1} ;j-1  } | e^{i \phi_{0} \hat n ^{\text{(t)}}_{x_{1},j} }  (e^{-i \phi_{0} \hat n ^{\text{(t)}}_{x_{2},j} } -\mathbb{I} ) | \Psi ^{\text{(t)}}_{x _ {2} ;j-1 } \rangle } \right |  + \left | {\langle\Psi ^{\text{(t)}}_{x _{1} ;j-1  } |  ( e^{i \phi_{0} \hat n ^{\text{(t)}}_{x_{1},j} } -\mathbb{I} ) | \Psi ^{\text{(t)}}_{x _ {2} ;j-1 } \rangle } \right | \right )} 
\nonumber\\
& \le  \sqrt{ \frac{ 2 }{N (N-1)} \!\sum _{x _{1} \ne x _{2}}  \sum _{j = 1}^{K} \left (  \left | {\langle\Psi ^{\text{(t)}}_{x _{1} ;j-1  } | e^{i \phi_{0} \hat n ^{\text{(t)}}_{x_{1},j} }  (e^{-i \phi_{0} \hat n ^{\text{(t)}}_{x_{2},j} } -\mathbb{I} ) | \Psi ^{\text{(t)}}_{x _ {2} ;j-1 } \rangle} \right |  + \left | {\langle\Psi ^{\text{(t)}}_{x _{1} ;j-1  } |  ( e^{i \phi_{0} \hat n ^{\text{(t)}}_{x_{1},j} } -\mathbb{I} ) | \Psi ^{\text{(t)}}_{x _ {2} ;j-1 } \rangle} \right | \right )} 
\label{neq:Psi.phi0.bound}
.
\end{align}

In order to derive a nontrivial query lower bound, we bound the terms contributing to the sum on the right-hand side of Eq.~\eqref{neq:Psi.phi0.bound}. 
In what follows, we provide two separate proofs that constrain the terms differently, which lead to different scaling with $n _{\budgetlb} $ in the final query lower bounds. Combining the two bounds produces  the final result in Eq.~\eqref{neq:Ktot.bound.phi0.0}, 
which provides a tight query lower bound for the fixed-phase oracle model.

We first prove a query lower bound that can be viewed as a direct generalization of the lower bound in Sec.~\ref{nsec:Ttot.gen.bound} for the discrete oracle~\eqref{neq:phi.oracle.def}. Using the Cauchy-Schwarz inequality and the fact that $[ \hat n ^{\text{(t)}}_{x_{2},j}, \hat n ^{\text{(t)}}_{x_{1},j} ] =0 $ $\forall x_{1}, x_{2}, j$, we derive the following upper bounds:
\begin{align}
& \left | \braket{\Psi ^{\text{(t)}}_{x _{1} ;j-1  } | e^{i \phi_{0} \hat n ^{\text{(t)}}_{x_{1},j} }  (e^{-i \phi_{0} \hat n ^{\text{(t)}}_{x_{2},j} } -\mathbb{I} ) | \Psi ^{\text{(t)}}_{x _ {2} ;j-1 } } \right | 
\nonumber\\
\le {}& 2 \sqrt{\braket{ \Psi ^{\text{(t)}}_{x _{1} ;j-1  }  | \sqrt{\mathbb{I} - \cos (\phi_{0} \hat n ^{\text{(t)}}_{x_{2},j}) } | \Psi ^{\text{(t)}}_{x _{1} ;j-1  } } 
\braket{ \Psi ^{\text{(t)}}_{x _{2} ;j-1  }  | \sqrt{\mathbb{I} - \cos (\phi_{0} \hat n ^{\text{(t)}}_{x_{2},j}) } | \Psi ^{\text{(t)}}_{x _{2} ;j-1  } } 
}
\\
= {}& 2 \sqrt{\braket{ \Psi ^{\text{(t)}}_{x _{1} ;j  }  | \sqrt{\mathbb{I} - \cos (\phi_{0} \hat n ^{\text{(t)}}_{x_{2},j}) } | \Psi ^{\text{(t)}}_{x _{1} ;j  } }  
\braket{ \Psi ^{\text{(t)}}_{x _{2} ;j  }  | \sqrt{\mathbb{I} - \cos (\phi_{0} \hat n ^{\text{(t)}}_{x_{2},j}) } | \Psi ^{\text{(t)}}_{x _{2} ;j  } } 
}
\label{neq:Psi.bound.nx2.0}
,
\end{align}
where we use Eq.~\eqref{neq:Psi.phi0.evol.tog} and the relations
\begin{align}
(e^{-i \phi_{0} \hat n ^{\text{(t)}}_{x_{2},j} } -\mathbb{I} ) (e^{i \phi_{0} \hat n ^{\text{(t)}}_{x_{2},j} } -\mathbb{I} ) 
={} & (e^{i \phi_{0} \hat n ^{\text{(t)}}_{x_{2},j} } -\mathbb{I} ) (e^{-i \phi_{0} \hat n ^{\text{(t)}}_{x_{2},j} } -\mathbb{I} ) 
\nonumber\\
= {}& 2[\mathbb{I} - \cos (\phi_{0} \hat n ^{\text{(t)}}_{x_{2},j})  ]
.
\end{align}
Similarly, for the second term in the sum in Eq.~\eqref{neq:Psi.phi0.bound}, we have
\begin{align}
&\left | \braket{\Psi ^{\text{(t)}}_{x _{1} ;j-1  } |  ( e^{i \phi_{0} \hat n ^{\text{(t)}}_{x_{1},j} } -\mathbb{I} ) | \Psi ^{\text{(t)}}_{x _ {2} ;j-1 } } \right | 
\nonumber\\
\le {}& 2 \sqrt{\braket{ \Psi ^{\text{(t)}}_{x _{1} ;j-1  }  | \sqrt{\mathbb{I} - \cos (\phi_{0} \hat n ^{\text{(t)}}_{x_{1},j}) } | \Psi ^{\text{(t)}}_{x _{1} ;j-1  } } 
\braket{ \Psi ^{\text{(t)}}_{x _{2} ;j-1  }  | \sqrt{\mathbb{I} - \cos (\phi_{0} \hat n ^{\text{(t)}}_{x_{1},j}) } | \Psi ^{\text{(t)}}_{x _{2} ;j-1  } } 
}
\\
= 2 {}& \sqrt{\braket{ \Psi ^{\text{(t)}}_{x _{1} ;j  }  | \sqrt{\mathbb{I} - \cos (\phi_{0} \hat n ^{\text{(t)}}_{x_{1},j}) } | \Psi ^{\text{(t)}}_{x _{1} ;j  } } 
\braket{ \Psi ^{\text{(t)}}_{x _{2} ;j  }  | \sqrt{\mathbb{I} - \cos (\phi_{0} \hat n ^{\text{(t)}}_{x_{1},j}) } | \Psi ^{\text{(t)}}_{x _{2} ;j } }  
}
\label{neq:Psi.bound.nx1.0}
,
\end{align}
where we again use Eq.~\eqref{neq:Psi.phi0.evol.tog} and the fact that $[\hat n ^{\text{(t)}}_{x_{1},j}, \hat n ^{\text{(t)}}_{x_{2},j}]=0$ to evolve the state from the $(j-1)$st time step to the $j$th.
Substituting Eqs.~\eqref{neq:Psi.bound.nx1.0} and~\eqref{neq:Psi.bound.nx2.0} into Eq.~\eqref{neq:Psi.phi0.bound}, we obtain
\begin{align}
[p ^{(0)}_{\text{th}} ]^{2} < {}&  \frac{ 8 }{N (N-1)}  \sum _{j = 1}^{K}  
\sum _{x _{1} \ne x _{2}} 
\sqrt{ \braket{\Psi ^{\text{(t)}}_{x _{1} ;j  } | \sqrt{\mathbb{I} - \cos (\phi_{0} \hat n ^{\text{(t)}}_{x_{1},j}) } | \Psi ^{\text{(t)}}_{x _ {1} ;j } } 
\braket{\Psi ^{\text{(t)}}_{x _{2} ;j  } |  \sqrt{\mathbb{I} - \cos (\phi_{0} \hat n ^{\text{(t)}}_{x_{1},j}) } | \Psi ^{\text{(t)}}_{x _ {2} ;j } }
}
\label{neq:Psi.phi0.bound.expt.0}
,
\end{align}
where we use the fact that the sum on the right-hand side is symmetric under exchanging $x_{1}$ and $x_{2}$.

Noting that the eigenvalues of the toggling-frame number operators are nonnegative integers, we obtain the following upper bounds for arbitrary states $\ket{\Psi}$:
\begin{align}
\braket{\Psi | \sqrt{\mathbb{I} - \cos (\phi_{0} \hat n ^{\text{(t)}}_{x_{1},j}) }  | \Psi}
&\le \sqrt{2} \left ( \sup _{n_0 \in \mathbb{N}^{+}} \frac{| \sin \frac{\phi_{0}  n_0 }{2}| }{n_0} \right ) \braket{\Psi | \hat n ^{\text{(t)}}_{x_{1},j}|\Psi  }
\\
\label{neq:phi0.evol.bound.n}
&\le \sqrt{2} \left | \sin  \frac{\phi_{0}}{2}  \right | \braket{\Psi | \hat n ^{\text{(t)}}_{x_{1},j}|\Psi  }
.
\end{align}
Substituting the above upper bound into Eq.~\eqref{neq:Psi.phi0.bound.expt.0}, we obtain 
\begin{align}
[p ^{(0)}_{\text{th}} ]^{2} &  < \frac{ 16 }{N (N-1)} \left | \sin  \frac{\phi_{0}}{2}  \right | 
\sum _{x _{1} \ne x _{2}}\sum _{j = 1}^{K} 
\sqrt{\braket{\Psi ^{\text{(t)}}_{x _{1} ;j } |  \hat n ^{\text{(t)}}_{x_{1},j}| \Psi ^{\text{(t)}}_{x _ {1} ;j } } } 
\sqrt{\braket{\Psi ^{\text{(t)}}_{x _{2} ;j } |  \hat n ^{\text{(t)}}_{x_{1},j}| \Psi ^{\text{(t)}}_{x _ {2} ;j } } } 
,
\end{align}
so that using the Cauchy-Schwarz inequality, we can bound the right-hand side as 
\begin{align}
[p ^{(0)}_{\text{th}} ]^{2} &  < \frac{ 16 \left | \sin  \frac{\phi_{0}}{2}  \right |  }{N (N-1)} \sum _{x _{1} \ne x _{2}}\! 
\sqrt{\sum _{j = 1}^{K} \braket{\Psi ^{\text{(t)}}_{x _{1} ;j } |  \hat n ^{\text{(t)}}_{x_{1},j}| \Psi ^{\text{(t)}}_{x _ {1} ;j } } \sum _{j' = 1}^{K} \braket{\Psi ^{\text{(t)}}_{x _{2} ;j' } |  \hat n ^{\text{(t)}}_{x_{1},j'}| \Psi ^{\text{(t)}}_{x _ {2} ;j' } }} 
\nonumber \\
& \le \frac{ 16 \left | \sin  \frac{\phi_{0}}{2}  \right |  }{N (N-1)} \sqrt{K n _{\budgetlb} }
\label{neq:Ktot.bound.mid}
\sum _{x _{1} \ne x _{2}} 
\sqrt{\sum _{j' = 1}^{K} \braket{\Psi ^{\text{(t)}}_{x _{2} ;j' } |  \hat n ^{\text{(t)}}_{x_{1},j'}| \Psi ^{\text{(t)}}_{x _ {2} ;j' } } } 
,
\end{align}
where we use the discretized boson-number constraint~\eqref{neq:ntot.bound.phi0} to bound the first sum under the square root as 
\begin{align}
\sum _{j = 1}^{K} \braket{\Psi ^{\text{(t)}}_{x _{1} ;j } |  \hat n ^{\text{(t)}}_{x_{1},j}| \Psi ^{\text{(t)}}_{x _ {1} ;j } } 
\le K n _{\budgetlb} 
.
\end{align}
Using the Cauchy-Schwarz inequality and substituting Eq.~\eqref{neq:ntot.bound.phi0} into Eq.~\eqref{neq:Ktot.bound.mid}, we finally have
\begin{align}
[p ^{(0)}_{\text{th}} ]^{2} &<\frac{ 16 \left | \sin  \frac{\phi_{0}}{2}  \right |  }{N } \sum _{x _{2}} 
\sqrt{\frac{K n _{\budgetlb} }{N-1}
\sum _{x _{1} , x _{1} \ne x _{2}} \sum _{j' = 1}^{K} \braket{\Psi ^{\text{(t)}}_{x _{2} ;j' } |  \hat n ^{\text{(t)}}_{x_{1},j'}| \Psi ^{\text{(t)}}_{x _ {2} ;j' } } 
}
\nonumber \\
& \le \frac{ 16 \left | \sin  \frac{\phi_{0}}{2}  \right | }{\sqrt{ N -1}} K n _{\budgetlb} 
,
\end{align}
which can be rewritten as a query lower bound for the discrete oracle model:
\begin{align}
\label{neq:K.lower.bound.phi0.0}
K > \frac{[p ^{(0)}_{\text{th}} ]^{2} }{16} \frac{\sqrt{N-1}}{ \left | \sin  \frac{\phi_{0}}{2}  \right | n _{\budgetlb} }
.
\end{align}
More specifically, in the asymptotically small query-phase limit, we have
\begin{align}
\label{neq:Ktot.lower.bound.phi0.0}
K _{\text{tot}} = \Omega \left ( \frac{\sqrt{N}}{ \phi_{0} n _{\budgetlb} } \right )
,
\end{align}
which is consistent with Eq.~\eqref{neq:Ktot.phi0.bound.inT} and generalizes the continuous-time version of the query lower bound to the discrete setting.

Next, we use a different way to bound the terms contributing to the sum on the right-hand side of Eq.~\eqref{neq:Psi.phi0.bound}. This leads to a qualitatively different query lower bound that can be tighter for finite query phase $\phi _{0}$.
Using the Cauchy-Schwarz inequality and the fact that $[ \hat n ^{\text{(t)}}_{x_{2},j}, \hat n ^{\text{(t)}}_{x_{1},j} ] =0 $ $\forall x_{1}, x_{2}, j$, we can bound the overlaps on the right-hand side of Eq.~\eqref{neq:Psi.phi0.bound} as
\begin{align}
\left | \braket{\Psi ^{\text{(t)}}_{x _{1} ;j-1  } | e^{i \phi_{0} \hat n ^{\text{(t)}}_{x_{1},j} }  (e^{-i \phi_{0} \hat n ^{\text{(t)}}_{x_{2},j} } -\mathbb{I} ) | \Psi ^{\text{(t)}}_{x _ {2} ;j-1 } } \right | 
\label{neq:Psi.bound.nx2}
\le {}& \sqrt{2 \braket{ \Psi ^{\text{(t)}}_{x _{1} ;j-1  }  | \left [ \mathbb{I} - \cos (\phi_{0} \hat n ^{\text{(t)}}_{x_{2},j}) \right ]  | \Psi ^{\text{(t)}}_{x _{1} ;j-1  } } 
 }
,
\end{align}
where we use the relations $(e^{-i \phi_{0} \hat n ^{\text{(t)}}_{x_{2},j} } -\mathbb{I} ) (e^{i \phi_{0} \hat n ^{\text{(t)}}_{x_{2},j} } -\mathbb{I} )  = 2[\mathbb{I} - \cos (\phi_{0} \hat n ^{\text{(t)}}_{x_{2},j})  ]$.
Similarly, for the second term in the sum in Eq.~\eqref{neq:Psi.phi0.bound}, we have
\begin{align}
\left | \braket{\Psi ^{\text{(t)}}_{x _{1} ;j-1  } |  ( e^{i \phi_{0} \hat n ^{\text{(t)}}_{x_{1},j} } -\mathbb{I} ) | \Psi ^{\text{(t)}}_{x _ {2} ;j-1 } } \right |
\le {}& \sqrt{2 \braket{\Psi ^{\text{(t)}}_{x _{2} ;j-1  } |  \left [\mathbb{I} - \cos (\phi_{0} \hat n ^{\text{(t)}}_{x_{1},j}) \right ] | \Psi ^{\text{(t)}}_{x _ {2} ;j-1 } } }
\label{neq:Psi.bound.nx1}
.
\end{align}
Substituting Eqs.~\eqref{neq:Psi.bound.nx1} and~\eqref{neq:Psi.bound.nx2} into Eq.~\eqref{neq:Psi.phi0.bound} and again using Eq.~\eqref{neq:Psi.phi0.evol.tog}, we obtain the following upper bound relating properties of the instantaneous probe states to the threshold probability: 
\begin{align}
[p ^{(0)}_{\text{th}} ]^{2} < {}&  \frac{ 4 }{N (N-1)}  
\sum _{j = 1}^{K}  \sum _{x _{1} \ne x _{2}}  \left (  
\sqrt{ 2 \braket{\Psi ^{\text{(t)}}_{x _{2} ;j  } |  \left [ \mathbb{I} - \cos (\phi_{0} \hat n ^{\text{(t)}}_{x_{1},j}) \right ] | \Psi ^{\text{(t)}}_{x _ {2} ;j } } }\right )
\label{neq:Psi.phi0.bound.expt}
,
\end{align}
where we again use the fact that the sum on the right-hand side is symmetric under exchanging $x_{1}$ and $x_{2}$.

Once again, using the fact that the eigenvalues of the toggling-frame number operators are nonnegative integers, we obtain the following upper bounds for an arbitrary state $\ket{\Psi}$:
\begin{align}
\braket{\Psi | \left [ \mathbb{I} - \cos (\phi_{0} \hat n ^{\text{(t)}}_{x_{1},j}) \right ]  | \Psi}
&\le  2 \braket{\Psi | \left | \sin  (\phi_{0} \hat n ^{\text{(t)}}_{x_{1},j}/2)  \right |
| \Psi}
\nonumber\\
&\le 2\left ( \sup _{n_0 \in \mathbb{N}^{+}} \frac{| \sin \frac{\phi_{0}  n_0 }{2}| }{n_0} \right ) \braket{\Psi | \hat n ^{\text{(t)}}_{x_{1},j}|\Psi  }
\\
\label{neq:bound.evol.phi0}
&\le 2 \left | \sin  \frac{\phi_{0}}{2}  \right | \braket{\Psi | \hat n ^{\text{(t)}}_{x_{1},j}|\Psi  }
.
\end{align}
Substituting the above upper bound into Eq.~\eqref{neq:Psi.phi0.bound.expt}, we obtain 
\begin{align}
[p ^{(0)}_{\text{th}} ]^{2} &  < \frac{ 8 }{N (N-1)} \sqrt{\left | \sin  \frac{\phi_{0}}{2}  \right | } 
\sum _{j = 1}^{K} \sum _{x _{1} \ne x _{2}} 
\sqrt{\braket{\Psi ^{\text{(t)}}_{x _{2} ;j } |  \hat n ^{\text{(t)}}_{x_{1},j}| \Psi ^{\text{(t)}}_{x _ {2} ;j } } } 
,
\end{align}
so that, using the Cauchy-Schwarz inequality, we can bound the right-hand side as 
\begin{align}
[p ^{(0)}_{\text{th}} ]^{2} &  < \frac{ 8 }{N } \sqrt{\left | \sin  \frac{\phi_{0}}{2}  \right | } \sum _{x _{2}} 
\sqrt{\frac{K}{N-1} \sum _{j = 1}^{K} \sum _{x _{1},x _{1} \ne x _{2}} \braket{\Psi ^{\text{(t)}}_{x _{2} ;j } |  \hat n ^{\text{(t)}}_{x_{1},j}| \Psi ^{\text{(t)}}_{x _ {2} ;j } } } 
\nonumber \\
& \le 8 \max _{x_{2}}
\sqrt{\frac{K \left | \sin  \frac{\phi_{0}}{2}  \right | }{N-1} \sum _{j = 1}^{K} \sum _{x _{1},x _{1} \ne x _{2}} \braket{\Psi ^{\text{(t)}}_{x _{2} ;j } |  \hat n ^{\text{(t)}}_{x_{1},j}| \Psi ^{\text{(t)}}_{x _ {2} ;j } } } 
.
\end{align}
Substituting the discretized boson-number constraint~\eqref{neq:ntot.bound.phi0} into the above equation, we finally have
\begin{align}
[p ^{(0)}_{\text{th}} ]^{2} <  8 K  \sqrt{ \left | \sin  \frac{\phi_{0}}{2}  \right | \frac{ n _{\budgetlb} }{N-1}}
,
\end{align}
which can be rewritten as a query lower bound for the discrete-time oracle model:
\begin{align}
\label{neq:K.lower.bound.phi0}
K > \frac{[p ^{(0)}_{\text{th}} ]^{2} }{8} \sqrt{\frac{N-1}{ \left | \sin  \frac{\phi_{0}}{2}  \right | n _{\budgetlb} }}
.
\end{align}
Note that the above bound exhibits a distinct scaling in $n _{\budgetlb} $ compared with the continuous-time bound in Sec.~\ref{nsec:Ttot.gen.bound} and Eq.~\eqref{neq:K.lower.bound.phi0.0}. 
In particular, in the asymptotically small query-phase ($\phi_{0} \to 0$) limit, Eq.~\eqref{neq:K.lower.bound.phi0} gives $K = \Omega (\phi_{0}^{-\frac{1}{2}})$, which is looser than the  scaling in Eq.~\eqref{neq:K.lower.bound.phi0.0} (the discrete generalization of the bound in Sec.~\ref{nsec:Ttot.gen.bound}), $K = \Omega (\phi_{0}^{-1})$.

Combining Eqs.~\eqref{neq:K.lower.bound.phi0.0} and ~\eqref{neq:K.lower.bound.phi0}, we obtain the query lower bound for fixed-phase oracles in Eq.~\eqref{neq:Ktot.bound.phi0.0}.
We now show that the lower bound~\eqref{neq:Ktot.bound.phi0.0} is achievable in all parameter regimes, using the entangled strategy of Sec.~\ref{nsec:opt.protocol.ent} with a phase-matched Grover iteration~\cite{Hoyer2000,Long2001}. With the W-Fock state~\eqref{neq:nph.Wstate} of $M$ bosons, each query imprints the phase $\theta\coloneqq M\phi_0$ on $\ket{M}_{x_0}$. Replacing the Grover diffusion by $\hat I+(e^{-i\theta}-1)\ketbra{\psi_{\text{in}}}$ [constructed as in Sec.~\ref{nsec:nonG.ent.FockR}, with the phase $\pi$ replaced by $\theta$], each iteration rotates the state by an angle $\Theta(|\sin(\theta/2)|/\sqrt{N})$ within $\operatorname{span}\{\ket{\psi_{\text{in}}},\ket{M}_{x_0}\}$, so that $\Theta(\sqrt{N}/|\sin(\theta/2)|)$ queries identify $x_0$ with constant probability. We choose $M$ such that $\theta\in(0,\pi]$, which avoids values $\theta\in2\pi\mathbb{Z}$ at which the oracle acts trivially. For $n_{\budgetlb}\phi_0<\pi$, which up to constant factors corresponds to $n_{\budgetlb}|\sin(\phi_0/2)|=O(1)$, we take $M=\lfloor n_{\budgetlb}\rfloor$, so that $|\sin(\theta/2)|=\Theta(n_{\budgetlb}|\sin(\phi_0/2)|)$ and $K_{\text{tot}}=\Theta(\sqrt{N}/n_{\budgetlb}|\sin(\phi_0/2)|)$. For $n_{\budgetlb}\phi_0\ge\pi$, we instead take $M=\lfloor\pi/\phi_0\rfloor=\Theta(1/|\sin(\phi_0/2)|)$, so that $|\sin(\theta/2)|=\Theta(1)$, and prepare $P=\min\{N,\lfloor n_{\budgetlb}/M\rfloor\}=\Theta(\min\{N,n_{\budgetlb}|\sin(\phi_0/2)|\})$ such W-Fock states on disjoint clusters whose sizes differ by at most one,
searched in parallel. After $\Theta(\sqrt{N/P})$ queries, measuring every cluster yields $P$ candidates, one of which is $x_0$ with constant probability; these are checked with $O(1)$ further queries by preparing each candidate in $(\ket{0}+\ket{M})/\sqrt{2}$, which uses at most $n_{\budgetlb}/2$ bosons. After the query, each candidate is tested by projection onto $(\ket{0}-\ket{M})/\sqrt{2}$. Unmarked candidates are rejected with certainty, whereas the marked candidate is detected with probability $\sin^2(\theta/2)\ge1/2$, since $\theta\in[\pi/2,\pi]$ in this regime. A constant number of repetitions of the search and
verification procedure reaches the required success probability. If $N<4P$, the W-Fock searches are skipped, and the candidate test is applied directly to all modes in at most four sequential batches of at most $P$ modes each, which uses at most $n_{\budgetlb}/2$ bosons and a constant number of queries; this branch realizes the one-query floor in Eq.~\eqref{neq:Ktot.bound.phi0.fin}.
In all cases, $K_{\text{tot}}=O(\max\{1,\sqrt{N}/\min\{n_{\budgetlb}|\sin\frac{\phi_0}{2}|,\sqrt{n_{\budgetlb}|\sin\frac{\phi_0}{2}|}\}\})$, so the lower bound is tight:
\begin{align}
\label{neq:Ktot.bound.phi0.fin}
K_{\text{tot}}  = \Theta \left (\max\left\{1,
\frac{\sqrt{N}}{ \min \{ n _{\budgetlb} \left | \sin  \frac{\phi_{0}}{2}  \right |  , \sqrt{ n _{\budgetlb} \left | \sin  \frac{\phi_{0}}{2}  \right |  } \}  }
\right\}\right )
.
\end{align}

\section{Experimental proposals for implementing the unentangled non-Gaussian unstructured search}

\label{nsec:nonG.prod.expt}

In this section, we discuss three methods for implementing the unentangled non-Gaussian unstructured  search algorithm [Eq.~\eqref{neq:prod.Grover}], using single-mode primitives established and experimentally demonstrated in prior literature. 
Recall that the non-Gaussian, unentangled strategy [Eqs.~\eqref{neq:nph.FockSum.prod} and~\eqref{neq:nph.FockSum.prod.1}] only involves two non-Gaussian ingredients:\footnote{In what follows, to simplify notation, we use $M$ to denote a boson number that should be chosen as either $M = \lfloor n _{\budgetlb}/2 \rfloor $ for the unentangled protocol, or $M = \lfloor n _{\budgetlb} \rfloor $ for the entangled non-Gaussian protocol. However, the experimental proposals in Secs.~\ref{nsec:nonG.prod.expt} and~\ref{nsec:nonG.ent.expt} do not rely on fine tuning the value of $M$.}
\begin{enumerate}
    \item {ability to prepare each probe mode $x$ in a non-Gaussian state $\ket{\psi _{M;N}} _{x} =  \sqrt{1-\frac{1}{N}} \ket{0}_{x} + \frac{1}{\sqrt{N}} \ket{ M } _{x}  $;}
    \item {ability to implement the single-mode non-Gaussian Grover diffusion unitary $e ^{- i \pi \ketbra{\psi _{M;N} }_{x} }$ on each mode $x$.}
\end{enumerate}
As the search operations are completely decoupled across the $N$ modes, it suffices to consider the implementation for a single mode $x$, which can then be replicated across the remaining probe modes to complete the search. Therefore, we focus on single-mode operations and drop the mode index $x$ in the rest of this section, whenever it does not cause confusion.

Before describing the detailed protocols, we first note that both of the above non-Gaussian tasks can be accomplished with a single-mode fixed-angle non-Gaussian rotation gate between Fock states, $\hat U _{\text{sing};M} (\theta_{1} ) $, defined via its operation on the vacuum and $M$-boson Fock state:
\begin{align}
\label{neq:0toFockS.sing}
\hat U _{\text{sing};M} (\theta_{1} )  \ket{0}
&= \cos \theta_{1}  \ket{0} + \sin \theta_{1}  \ket{M}
, \\
\hat U _{\text{sing};M} (\theta_{1} )  \ket{M}
& = - \sin \theta_{1}  \ket{0} + \cos \theta_{1}  \ket{M}
,
\end{align}
where $\theta_{1} = \arcsin \frac{1}{\sqrt{N}}$,
and its inverse, $\hat U _{\text{sing};M} ^{\dag} ( \theta_{1}  )$.
With access to $\hat U _{\text{sing};M} ( \theta_{1}  ) $, we immediately obtain the required non-Gaussian initial states, since $\hat U _{\text{sing};M} ( \theta_{1} )  \ket{0} = \ket{\psi _{M;N}} $. Further, we can also construct the Grover diffusion (up to a constant phase factor) by conjugating a Gaussian phase shift with the non-Gaussian rotation: 
\begin{align}
- \hat P _{\{ \ket{0}, \ket{M}\} }
e ^{- i \pi \ketbra{\psi _{M;N} } }
\hat P _{\{ \ket{0}, \ket{M}\} }
&= \hat P _{\{ \ket{0}, \ket{M}\} }
\hat U _{\text{sing};M} ( \theta_{1}  ) e ^{- i \frac{\pi}{M} \hat a  ^\dag \hat a } \hat U _{\text{sing};M} ^{\dag} ( \theta_{1}  )
\hat P _{\{ \ket{0}, \ket{M}\} }
,
\end{align}
where $\hat P _{\{ \ket{0}, \ket{M}\} }$ denotes the projector onto the linear span of $\{ \ket{0}, \ket{M}\}$. 
Thus, we can implement the entire unentangled search algorithm with access to a pair of non-Gaussian rotations, $\hat U _{\text{sing};M} ( \theta_{1}  )$ and its inverse.

Implementation of such non-Gaussian gates that create (or undo) superpositions of Fock states is a well-studied topic in the literature. Three main classes of protocols have been considered, based on the specific non-Gaussian couplings they utilize: 
\begin{enumerate}
    \item[a)] {The \textit{Jaynes-Cummings} strategy. The non-Gaussian gate $\hat U _{\text{sing};M} ( \theta  )$ and its inverse can be approximated using an ancilla qubit and tunable Jaynes-Cummings interaction between the ancilla and the probe mode~\cite{Molmer2013}. (Related state-preparation techniques are described in \cite{Eberly1996}.)
    At the level of ideal selective Fock-state operations, let $\hat F$ be a sequence of adjacent swaps satisfying $\hat F\ket{0}=\ket{0}$ and $\hat F\ket{1}=\ket{M}$. Then the desired rotation is $\hat U_{\text{sing};M}(\theta) =\hat F\hat R_{01}(\theta)\hat F^\dagger$ on $\operatorname{span}\{\ket{0},\ket{M}\}$, where $\hat R_{01}(\theta)$ is the corresponding rotation on $\operatorname{span}\{\ket{0},\ket{1}\}$.
    Related state-preparation and qubit-to-oscillator encoding and decoding operations have been demonstrated using superconducting circuits~\cite{Cleland2009,Chakram2026}.} 
    \item[b)] {The \textit{dispersive} strategy. Alternatively, the non-Gaussian gates can be implemented by concatenating displacements in the probe mode and the selective number-dependent arbitrary phase (SNAP) gates~\cite{Jiang2015}. The target unitary can be approximated to arbitrary precision on a chosen finite Fock subspace using SNAP gates and displacements (see Eq.~(8) of~\cite{Jiang2015}).}
    \item[c)] {The \textit{NOON} strategy. If we attach an ancilla bosonic mode or an $M$-dimensional qudit to each probe mode, we can extend the single-mode unitary to a two-mode non-Gaussian rotation between the ancilla Fock state and a NOON-like state between each pair of probe and ancilla:
    \begin{align}
\label{neq:0toFockS.NOON}
&\hat U _{\text{NOON};M} (\theta_{1} )  \ket{0_{\text{p}} M _{\text{A}} }
= \cos \theta_{1}  \ket{0_{\text{p}} M _{\text{A}} } + \sin \theta_{1}  \ket{M_{\text{p}} 0 _{\text{A}} }
, 
\end{align}
where $\theta_{1} = \arcsin \frac{1}{\sqrt{N}}$ denotes the hybridization phase. The two-mode probe-ancilla state performs equally well in the unstructured search problem as the desired single-mode state, as long as we replace $\hat U _{\text{det}}$ in Eq.~\eqref{neq:psiM.Udet.def.} with its two-mode counterpart.\footnote{This realization is separable across probe--ancilla pairs but uses entanglement within each pair; the abstract protocol in Sec.~\ref{nsec:summary} remains ancilla-free and fully separable.} The unitary  $\hat U _{\text{NOON};M} (\theta_{1} )$ in Eq.~\eqref{neq:0toFockS.NOON} can be synthesized using Jaynes-Cummings interactions~\cite{Chakram2026}.
Further, Sections~\ref{nsec:ent.Kerr} and~\ref{nsec:ent.expSWAP} describe two approaches to realizing this unitary, using Kerr and ancilla-controlled beamsplitters as the non-Gaussian resource, respectively. Alternatively, Ref.~\cite{Austin2026} proposes a protocol that uses cat states and SNAP and displacement unitaries to generate NOON states, which would also enable the unentangled unstructured search algorithm via a mechanism similar to Eq.~\eqref{neq:0toFockS.NOON}. 
    }
\end{enumerate}

For all three methods described above, the ancilla-probe nonlinearity is directly parallelizable if each probe mode is coupled to a separate ancilla. 
In this case, the time cost of each Grover-style iteration is independent of $N$. However, if one has access to only $O(1)$ ancilla qubits in total, then sequentially transforming each probe mode would take $O(N)$ steps, potentially introducing a significant architecture-induced overhead on the overall search time. In contrast, the entangled Grover diffusion operation can be implemented in $O(\log N)$ time using the Kerr strategy of Sec.~\ref{nsec:ent.Kerr}, which requires no ancilla. 
The nonlinear elements in such an implementation (Kerr phase shifts) are directly parallelizable. In this sense, an unentangled strategy with access to only $O(1)$ ancillas can be less efficient than the entangled strategy, even though the times querying the oracle exhibit the same scaling for the two strategies.

\section{Experimental proposals for entangled non-Gaussian unstructured search}

\label{nsec:nonG.ent.expt}

In this section, we provide two strategies for implementing the entangled
non-Gaussian unstructured search algorithm [see Eq.~\eqref{neq:Wstate.Grover}]. The first strategy constructs the entire search process using a Fock-state input, a fixed Kerr-phase-shift unitary gate, and fully Gaussian operations. The second strategy uses an exponential-SWAP (E-swap) gate~\cite{LauPlenio2016} and constructs the Grover diffusion unitary using an ancilla-qubit-assisted gate. The only non-Gaussian operation in the second strategy, an entangling gate between the probe and the ancilla qubit, can be achieved with either a qubit-resonator dispersive coupling, or a qubit-controlled beamsplitter interaction.

All of the non-Gaussian control ingredients involved in these protocols have been demonstrated experimentally in various platforms, including superconducting circuits~\cite{Schoelkopf2019,Schoelkopf2023}, trapped ions~\cite{Matsukevich2020}, and high-Q acoustic resonators coupled to superconducting qubits~\cite{Schoelkopf2018}.

\subsection{Non-Gaussian bosonic unstructured search via a parallelized multimode Fock-state rotation gadget}

\label{nsec:nonG.ent.FockR}

In this section, we describe a general procedure that implements an entangled
non-Gaussian Grover algorithm [see Eq.~\eqref{neq:Wstate.Grover}], which provides the backbone for the first two strategies. 
We begin with observing that the non-Gaussian algorithm only requires two non-Gaussian ingredients:
\begin{itemize}
    \item {Preparation of the non-Gaussian W-Fock initial state:
    \begin{align}
    \ket{\psi _{\text{in}}} = \frac{1}{\sqrt{N}} \sum _{x=1}^{N}    \ket{ M }  _{x}
= \frac{1}{\sqrt{N}} \sum _{x=1}^{N} \frac{1}{\sqrt{ M !}}\hat a _{x}  ^\dag {}^{ M } \ket{\text{vac}}
,    
\end{align}
    where $M = \lfloor n _{\budgetlb} \rfloor$.
    }
    \item {The Grover diffusion unitary with respect to $\ket{\psi _{\text{in}}}$:
    \begin{align}
    \hat U _{\text{G}} = \hat {\mathbb{I} } - 2\ketbra{\psi _{\text{in}}} 
    .
    \end{align}
    }
\end{itemize}
Both non-Gaussian resources can be deterministically created with perfect fidelity (ignoring practical imperfections) using a single non-Gaussian unitary rotation $\hat W_{x_{1}   \rightarrow \psi _{\text{in}}} $ and its inverse, satisfying
\begin{align}
\label{neq:Uswap.j0toW.def}
\hat W _{x_{1}  \rightarrow \psi _{\text{in}} } \ket{ M }  _{x_{1}} = \ket{ \psi _{\text{in}} } 
,
\end{align}
where $x_{1} $ 
denotes a fixed mode index. By definition, Eq.~\eqref{neq:Uswap.j0toW.def} rotates the corresponding Fock state $\ket{ M }  _{x_{1}}$ into the desired initial state.
Further, we can implement the Grover diffusion unitary $\hat U _{\text{G}} $ by conjugating an appropriately chosen Gaussian phase-shift unitary on the $x_{1}$ mode with the non-Gaussian swap gate (within $\text{span}\{ \ket{M}_{x} \}$, which the dynamics never leaves):
\begin{align}
\hat P _{\{ \ket{M}_{x} \} } \hat U _{\text{G}} \hat P _{\{ \ket{M}_{x} \} }  
= {}&\hat P _{\{ \ket{M}_{x} \} } e ^{-i \pi \ketbra{\psi _{\text{in}}}  }\hat P _{\{ \ket{M}_{x} \} } 
\\
= {}&\hat P _{\{ \ket{M}_{x} \} } \hat W _{x_{1}  \rightarrow \psi _{\text{in}} }
e ^{-i \frac{\pi}{M} \hat a _{x_{1}} ^\dag \hat a _{x_{1}} }
\hat W ^{\dag} _{x_{1}  \rightarrow \psi _{\text{in}} } \hat P _{\{ \ket{M}_{x} \} } 
.
\end{align}
Thus, to enable the non-Gaussian unstructured search, it suffices to construct the rotation $\hat W _{x_{1}  \rightarrow \psi _{\text{in}} } $. 
As we show below, this unitary rotation can be realized using only Gaussian transformations and a naturally present non-Gaussian unitary control, which can be either a Kerr phase-shift unitary (Sec.~\ref{nsec:ent.Kerr}) or a probe-ancilla-qubit-controlled beamsplitter unitary (Sec.~\ref{nsec:ent.expSWAP}).

Before discussing detailed strategies for constructing $\hat W _{x_{1}  \rightarrow \psi _{\text{in}} } $, we first note that the unitary rotation for an arbitrary mode index $x_{2}$ can be generated via  $\hat W _{x_{1}  \rightarrow \psi _{\text{in}} }$ with an arbitrarily fixed $x_{1}$. Specifically, this can be achieved by conjugating $\hat W _{x_{1}  \rightarrow \psi _{\text{in}} }$ with a swap operation $\hat O _{x_{1} \leftrightarrow x_{2}} $ between modes $x_{1}$ and $x_{2}$: 
\begin{align}
\hat W _{x_{2}  \rightarrow \psi _{\text{in}} }
= \hat O _{x_{1} \leftrightarrow x_{2}} \hat W _{x_{1}  \rightarrow \psi _{\text{in}} } 
\hat O  ^{\dag}_{x_{1} \leftrightarrow x_{2}} 
.
\end{align}
The swap unitary is Gaussian, and can be implemented as
\begin{align}
\hat O _{x_{1} \leftrightarrow x_{2}}
\coloneqq e ^{i \frac{\pi}{2} (\hat a _{x_{2}} ^\dag \hat a _{x_{2}} + \hat a _{x_{1}} ^\dag \hat a _{x_{1}})} e ^{-i \frac{\pi}{2} (\hat a _{x_{2}} ^\dag \hat a _{x_{1}} + \hat a_{x_1}^\dag \hat a_{x_2}) }
.
\end{align}
Within the subspace $\text{span} \{ \ket{ M }  _{x}  \} $ spanned by single-mode $M$-boson Fock states, $\hat O _{x_{1} \leftrightarrow x_{2}}$ swaps the two corresponding Fock states, while keeping all the other basis states invariant:
\begin{align}
\hat O _{x_{1} \leftrightarrow x_{2}} \ket{ M }  _{x } =
\begin{cases}
\ket{ M }  _{x_{2}} , & x = x_{1}, \\
\ket{ M }  _{x_{1}} , & x = x_{2}, \\
\ket{ M }  _{x} , & x \ne x_{1}, x_{2}. 
\end{cases}
\end{align}
Thus, it suffices to synthesize the unitary rotation $\hat W _{x_{1}  \rightarrow \psi _{\text{in}} } $ for a single fixed $x_{1}$, and all other similar rotations $\hat W _{x _{2}   \rightarrow \psi _{\text{in}} }  $ with a different $x_{2}$ can be directly synthesized by Gaussian unitary conjugation. In the rest of this section, without loss of generality, we assume $x_{1} = 1$.

The multimode non-Gaussian unitary $\hat W _{x_{2}  \rightarrow \psi _{\text{in}} } $ can be implemented as a sequence of two-mode unitary rotations between neighboring Fock basis states:
\begin{align}
\label{neq:j0toW.TMrots}
\hat W _{x_{1}  \rightarrow \psi _{\text{in}} } 
&= \hat R _{N-1,N} (\theta _{N-1 }) \cdots \hat R _{j,j+1} (\theta _{j }) \cdots  \hat R _{2,3} (\theta _{2}) 
\hat R _{1,2} (\theta _{1}) 
,
\end{align}
where the two-mode non-Gaussian rotations are defined via the relation
\begin{align}
\hat R _{x,x+1} (\theta ) \ket{ M }  _{\ell }  
&\coloneqq
\begin{cases}
\ket{ M }  _{x } \cos \theta   + \ket{ M }  _{x+1 } \sin \theta   , & \ell  = x , \\
\ket{ M }  _{x+1 } \cos \theta - \ket{ M }  _{x } \sin \theta  , & \ell  = x+1, \\
\ket{ M }  _{\ell } , & \ell \ne x , x +1, 
\end{cases}
\end{align}
and the rotation angles are given by
\begin{align}
\theta _{x } \coloneqq \arccos\frac{1}{ \sqrt{ N-x+1 }}
.
\end{align}
The sequence of two-mode unitary rotations on the right-hand side of Eq.~\eqref{neq:j0toW.TMrots} implements the following transformation between the bosonic states (recalling that $x_{1}=1$):
\begin{align}
\ket{ M }  _{1 } & \xrightarrow{\hat R _{1,2} (\theta _{1}) }
\frac{1}{\sqrt{N}} \ket{ M }  _{1 } + \sqrt{1-\frac{1}{N}} \ket{ M }  _{2} 
\nonumber \\
& \xrightarrow{\hat R _{2,3} (\theta _{2}) }
\frac{1}{\sqrt{N}} \ket{ M }  _{1 } + \frac{1}{\sqrt{N}} \ket{ M }  _{2 } + \sqrt{1-\frac{2}{N}} \ket{ M }  _{3} 
\nonumber \\
& \xrightarrow{\hat R _{3,4} (\theta _{3}) }
\cdots 
\xrightarrow{\hat R _{N-1,N} (\theta _{N-1}) }
\ket{\psi _{\text{in}}} = \frac{1}{\sqrt{N}} \sum _{x=1}^{N}    \ket{ M }  _{x}
.
\end{align}
For an arbitrary angle $\theta$, the unitary rotation $\hat R _{x,x+1} (\theta ) $ between the two Fock states can be realized using a fixed-angle rotation between the two Fock states as
\begin{align}
\label{neq:TMFock.piover2.theta}
& \hat R _{x,x+1} (\theta ) = e ^{ i \frac{\pi }{2M} \hat n _{x+1} } \hat R ^{\dag} _{x,x+1} (\tfrac{\pi}{4}) 
e ^{ -i \frac{\theta }{M}(\hat n _{x+1} - \hat n _{x}) }
\hat R _{x,x+1} (\tfrac{\pi}{4}) e ^{ - i \frac{\pi }{2M} \hat n _{x+1} } 
.
\end{align}
Thus, to implement the entangled 
non-Gaussian unstructured search, it suffices to construct a family of fixed-angle Fock-state rotations $\hat R _{x,x+1} (\frac{\pi}{4}) $. Note also that Eq.~\eqref{neq:TMFock.piover2.theta} can be viewed as a realization of the E-swap gate within an encoding space $\text{span} \{ \ket{ M }  _{x} , \ket{ M }  _{x+1}  \} $~\cite{LauPlenio2016}.

The decomposition in Eq.~\eqref{neq:j0toW.TMrots} is sequential, but an alternative parallel decomposition implements an equally valid $\hat W _{x_{1}  \rightarrow \psi _{\text{in}} } $ in depth $O(\log N)$, at the cost of requiring $O(N)$ nonlinear elements simultaneously. To illustrate the idea, we assume $N = 2 ^{m}$ for $m \in \mathbb{N}$. However, the decomposition applies to arbitrary $N\in \mathbb{N} ^{+}$ by assigning the $N$ modes to the leaves of a balanced binary tree and applying
a similar construction as that described below. 
For $N = 2 ^{m}$, to simplify notation, we label mode $x$ by the $m$-bit binary representation of $x-1$, $x-1 = [x^{(m)} x^{(m-1)}\ldots x^{(1)}]$. Algorithm~\ref{alg:parallel.nonG.Grover} implements $\hat W _{x_{1} } $ using only $m$ layers of commuting non-Gaussian two-mode rotations.
\begin{algorithm}[H]
\caption{Parallel implementation of $\hat W _{x_{1}  \rightarrow \psi _{\text{in}} } $ in depth $m= O(\log N) $ using $O(N) $ simultaneous applications of non-Gaussian elements}
\label{alg:parallel.nonG.Grover}
\begin{algorithmic}[1]
\For {$\ell =1$ \textbf{to} $m$}
	\State Find all pairs of $(x,x')$ such that $x^{(\ell)} =0$, ${x} ^{(k)}  = {x'} ^{(k)} $ for $k>\ell$, and $x ^{(k)}  = \neg {x'} ^{(k)} $ for $k\le \ell$
    \State Apply $\hat R _{x ,x'} (\frac{\pi}{4}) $ for all $(x ,x')$ pairs identified above. All those unitaries can be applied in parallel, because they all act on mutually disjoint pairs.
\EndFor
\end{algorithmic}
\end{algorithm}

\subsection{Strategy 1: The Fock state + Kerr + Gaussian rotations strategy}

\label{nsec:ent.Kerr}

In this section, we construct $\hat R _{j,j+1} (\frac{\pi}{4}) $ by combining fixed-strength Kerr unitaries and two-mode beamsplitters. Specifically, we note that 
\begin{align}
&\hat R _{j,j+1} (\tfrac{\pi}{4}) =  
e ^{i \frac{\pi}{4M } (\hat n _{j } -\hat n _{j+1 } )   }
e ^{ - \frac{\pi}{4} (\hat a _{j+1} ^\dag \hat a _{j} - \text{H.c.} ) } 
e ^{- i \frac{\pi}{2} \hat n _{j } ^{2} } e ^{ \frac{\pi}{4} (\hat a _{j+1} ^\dag \hat a _{j} - \text{H.c.} ) }  
e ^{ i \frac{\pi}{2M}  \hat n _{j+1 }   }
,
\end{align}
which can be proved either by explicitly computing its action, or by first mapping the fixed total $M$-boson-number sector of a two-mode bosonic system to a spin-$M/2$ system (also known as the Schwinger boson representation~\cite{Schwinger1952}) and then applying the corresponding construction in spins~\cite{Wineland2004}.
Substituting the above equation into Eq.~\eqref{neq:TMFock.piover2.theta}, we obtain
\begin{align}
\hat R _{j,j+1} ( \theta  ) 
&=  e ^{ - \frac{\pi}{4} (\hat a _{j+1} ^\dag \hat a _{j} - \text{H.c.} ) } 
e ^{ i \frac{\pi}{2} \hat n _{j } ^{2} } e ^{ \frac{\pi}{4} (\hat a _{j+1} ^\dag \hat a _{j} - \text{H.c.} ) }
e ^{-i \frac{\theta }{M}(\hat n _{j+1} - \hat n _{j}) }
e ^{ - \frac{\pi}{4} (\hat a _{j+1} ^\dag \hat a _{j} - \text{H.c.} ) } 
e ^{- i \frac{\pi}{2} \hat n _{j } ^{2} } e ^{ \frac{\pi}{4} (\hat a _{j+1} ^\dag \hat a _{j} - \text{H.c.} ) }
\nonumber \\
\label{neq:Eswap.Kerr}
&= e ^{ - \frac{\pi}{4} (\hat a _{j+1} ^\dag \hat a _{j} - \text{H.c.} ) } 
e ^{ i \frac{\pi}{2} \hat n _{j } ^{2} }  
e ^{i \frac{\theta }{M}(\hat a _{j+1} ^\dag \hat a _{j} + \text{H.c.} ) }
e ^{- i \frac{\pi}{2} \hat n _{j } ^{2} } e ^{ \frac{\pi}{4} (\hat a _{j+1} ^\dag \hat a _{j} - \text{H.c.} ) }
.
\end{align}
Thus, we have proved that Grover diffusion can be implemented using fixed-angle Kerr-phase-shift and Gaussian rotations.

\subsection{Strategy 2: The ancilla-qubit-probe E-swap strategy}

\label{nsec:ent.expSWAP}

The Kerr construction in the above section uses $O(N)$ fine-tuned nonlinear Kerr operations, applied in $O(\log N)$ parallel layers (Algorithm~\ref{alg:parallel.nonG.Grover}) or $O(N)$ sequential layers. For better controllability, we propose an alternative approach using a single ancilla qubit and the exponential-swap operations [see Eq.~\eqref{neq:TMFock.piover2.theta}], 
which are proposed in Ref.~\cite{LauPlenio2016} and experimentally demonstrated in~\cite{Schoelkopf2019,Matsukevich2020,Schoelkopf2023}. The qubit-assisted strategy implements a unitary $\hat R _{x_{1},x_{2}} (\frac{\pi}{4})$ for an arbitrary pair of modes using a controlled-beamsplitter unitary between $x_{1}$ and $x_{2}$ modes, 
\begin{align}
\hat U _{\text{CBS}} (\theta; x_{1},x_{2} ) = e ^{-i \theta \hat \sigma_{z} (\hat a _{x_{1}} ^\dag \hat a _{x_{2}} + \text{H.c.}  ) }
,
\end{align}
and Gaussian rotations. 
The qubit-controlled beamsplitter interaction is compatible with quantum platforms such as trapped ions~\cite{Matsukevich2020} and superconducting circuits~\cite{Schoelkopf2023}.
Alternatively, it can also be exactly synthesized using tunable  dispersive couplings between an auxiliary qubit and the probe modes as follows:
\begin{align}
& e ^{-i \theta \hat \sigma_{z} (\hat a _{x_{1}} ^\dag \hat a _{x_{2}} + \text{H.c.}  ) }
= e ^{ - \frac{\pi}{4} (\hat a _{x_{1}} ^\dag \hat a _{x_{2}} - \text{H.c.} ) }
e ^{-i \theta \hat \sigma_{z} (\hat a _{x_{1}} ^\dag \hat a _{x_{1}} - \hat a _{x_{2}} ^\dag \hat a _{x_{2}} ) }
e ^{ \frac{\pi}{4} (\hat a _{x_{1}} ^\dag \hat a _{x_{2}} - \text{H.c.} ) }
.
\end{align}
The interactions above can also be used to implement the E-swap gate through the ancilla-assisted construction in Eqs.~(11) and (13) of Ref.~\cite{LauPlenio2016}.

As pointed out in Ref.~\cite{LauPlenio2016}, the E-swap operation enables encoding from an ancilla qubit into the logical qubit formed by the corresponding bosonic Fock state basis $\{ \ket{0M}, \ket{M0}\}$, so that replacing $\hat R _{x ,x'} (\frac{\pi}{4}) $ in Algorithm~\ref{alg:parallel.nonG.Grover} by the ancilla-qubit-controlled operation can directly prepare the $W$-Fock state. 
Further, the E-swap operation also enables decoding from the Fock-basis states into an ancilla qubit, which can be used to implement the Grover iteration. Combining both ingredients, we have an experimentally feasible protocol for implementing the entangled bosonic unstructured search. Note that a single ancilla implements these gadgets sequentially; parallel layers require one ancilla per pair. Alternatively, the Kerr strategy~\eqref{neq:Eswap.Kerr} can also be viewed as an implementation of the E-swap operation using a Kerr nonlinearity instead of an ancilla qubit.

\section{Discussion and outlook}

In this work, we prove tight query complexity lower bounds for bosonic unstructured search and show that, in the lossless regime, optimal joint query scaling in the mode number and the total boson number is only achievable using non-Gaussian probe states. In contrast, for sufficiently strong photon loss, the optimal scaling is already attained by classical coherent-state probes. We also show that entanglement is not necessary to achieve the optimal query complexity in this problem. This reveals that quantum resources such as entanglement and non-Gaussianity can play very different roles in an unstructured search problem, as compared with conventional metrological tasks.

For future work, it would be interesting to explore whether the Grover-style speedup can survive in energy-constrained settings in the presence of other structured dissipation.  It would also be interesting to identify scenarios where entanglement can provide a speedup. In the temporal domain, Ref.~\cite{Choi2025} considers a related problem of identifying an AC signal using a qubit sensor.
It is also of practical interest to consider the case where the signal is encoded in temporal rather than spatial degrees of freedom and explore a bosonic, energy-constrained version of the Grover-Heisenberg scaling discussed in~\cite{Choi2025}.

\section*{Acknowledgments}

We thank Wenhua He, Chaohan Cui, Saikat Guha, Joseph Carolan, Chinmay Nirkhe, Sisi Zhou, and Amir H.~Safavi-Naeini for helpful discussions.
Y.-X.W.~acknowledges support from a QuICS Hartree Postdoctoral Fellowship. 
M.H.~is supported by the NSF QLCI grant (award No.~OMA-2120757).
E.A., S.A., and A.V.G.\ were supported in part by ARL (W911NF-24-2-0107), ONR MURI, AFOSR MURI, NSF STAQ program,  and NQVL:QSTD:Design:FTL. E.A., S.A., A.M.C., and A.V.G.\ were supported in part by NSF QLCI (awards No.~OMA-2120757 and No.~OSI-2553574) and DoE ASCR Quantum Testbed Pathfinder program (award No.~DE-SC0024220). E.A., S.A., A.M.C., and A.V.G. also acknowledge support from  the U.S.~Department of Energy, Office of Science, National Quantum Information Science Research Centers, Quantum Systems Accelerator (award No.~DE-SCL0000121) and the U.S.\ Department of Energy, Office of Science, Accelerated Research in Quantum Computing, Fundamental Algorithmic Research toward Quantum Utility (FAR-Qu).

Claude Sonnet 5, accessed through Claude in August--September 2026, was used to brainstorm ideas for the proof in Sec.~\ref{nsec:Ttot.gen.bound} and proofread the manuscript. 
OpenAI GPT-6, accessed through ChatGPT in September--October 2026, was used to brainstorm ideas for the proof in Sec.~\ref{nsec:loss.bound}, assist literature search on experimental strategies to implement the non-Gaussian protocols, and proofread the manuscript.

\onecolumngrid

\appendix

\section{Derivation of \texorpdfstring{Eqs.~\eqref{neq:coh.K.pop.sum}: Time}{time} evolution under the coherent-state protocol}

\label{app.sec:coh.npop.K}

In the main text, we state that $K$ time steps of Gaussian Grover-style operations and a coherent input, as  
\begin{align}
\label{app.eq:boson.Grover}
\ket{\psi _{\text{out},K}}
&= \left ( e ^{-i  \delta \Delta t \hat n _{ k = {0}}  } e ^{-i B _{0}  \Delta t  \hat n _{x_{0}}   }
\right ) ^{K}
\ket{\psi _{\text{coh}}}
, \quad 
T _{\text{tot} } = K \Delta t
,
\end{align}
transform the target and non-target mode boson populations as [see Eq.~\eqref{neq:coh.K.pop.sum}]
\begin{subequations}
\label{app.eq:coh.K.pop.sum}
\begin{align}
\label{app.eq:nx0.K.coh.approx}
\braket{\hat n _{x_{0}} ( K \Delta t ) } & = n _{\budgetlb} \sin ^{2} [ (2 {K} + 1) \varphi]
, \\
\label{app.eq:nxne.K.coh.approx}
\braket{\hat n _{x\ne x_{0}} ( K \Delta t ) }  & = \frac{n _{\budgetlb} }{N-1} \cos ^{2} [ (2 {K} + 1) \varphi]
.
\end{align}
\end{subequations}
In this Appendix, we provide a detailed derivation of Eqs.~\eqref{app.eq:coh.K.pop.sum}.

More specifically, we can rewrite the initial state in terms of the displacement operator as 
\begin{align}
\ket{\psi _{\text{coh}}} &= e ^{ \sqrt{N} \alpha (\hat a _{ k = {0}} ^{\dag} - \text{H.c.})  } \ket{\text{vac}},
\end{align}
where
\begin{align}
\hat a _{ k = {0}} &= \frac{1}{\sqrt{N}} \hat a _{x_{0}} + \sqrt{1-\frac{1}{N}} \hat a _{\perp}
\end{align}
and $\hat a _{\perp} \coloneqq  \frac{1}{\sqrt{N-1}} \sum _{x\ne x_{0} } \hat a _{x}$ denotes a uniformly delocalized mode that is orthogonal to $\hat a _{x_{0}}$. For convenience, we also introduce $\hat a _{k\perp}= \sqrt{1-\frac{1}{N}} \hat a _{x_{0}} -\frac{1}{\sqrt{N}}  \hat a _{\perp} $ as a collective mode in $\text{span}\{ \hat a_{x_{0}} ,\hat a _{\perp} \}$ orthogonal to $\hat {a} _{k=0}$, so that 
\begin{subequations}
\label{neq:a.x0.conv.k}
\begin{align}
\hat a _{x_0} &= \hat a _{ k = {0}} \sin \varphi  + \hat a _{k\perp} \cos\varphi 
, \\
\hat a _{\perp} &= \hat a _{ k = {0}} \cos \varphi  - \hat a _{k\perp} \sin\varphi 
,
\end{align}
\end{subequations}
where $\varphi = \arcsin {(} N ^{-\frac{1}{2}} )$ denotes the phase of the beamsplitter operation that transforms the momentum mode basis $\hat a _{ k = {0}} $ and $\hat a _{k\perp} $ to $\hat a _{x_0} $ and $\hat a _{\perp} $.

We can rewrite the probe state after $K$ Grover iterations as
\begin{align}
\ket{\psi _{\text{out},K}} = e ^{(-) ^{K} \sqrt{N} \alpha (\hat {a} _{k = {0},K} ^{\dag}   - \text{H.c.})  } \ket{\text{vac}}
,
\end{align}
where $\hat {a} _{k=0,K} ^{\dag}  $ denotes the rotated Schr{\"{o}}dinger-picture raising operator for the displaced mode at time $K \Delta t $:
\begin{align}
\label{neq:intframe.a.com.def}
\hat {a} _{k = {0},K} ^{\dag}  & \coloneqq (-) ^{K} \hat U_{\text{G}, K}  \hat {a} _{k = {0}} ^{\dag}  \hat U_{\text{G}, K}  ^{\dag} 
\end{align}
where
\begin{align}
\hat U_{\text{G}, K} &  \coloneqq \left ( e ^{-i  \delta \Delta t \hat a _{ k = {0}} ^\dag \hat a _{ k = {0} } } e ^{-i B _{0}  \Delta t  \hat a _{x_{0}} ^\dag \hat a _{x_{0}} }
\right ) ^{K}
\\
& = \left ( e ^{-i  \pi  \hat a _{ k = {0}} ^\dag \hat a _{ k = {0} } } e ^{-i \pi \hat a _{x_{0}} ^\dag \hat a _{x_{0}} }
\right ) ^{K}
.
\end{align}
Similarly, we define a rotated operator $\hat a _{k\perp,K} \coloneqq (-) ^{K} \hat U_{\text{G}, K}  \hat {a} _{k\perp} \hat U_{\text{G}, K} ^{\dag} $, which is a collective mode orthogonal to $\hat {a} _{k = {0},K} $.
Since the unitary in Eq.~\eqref{seq:boson.Grover} is generated by a fully quadratic bosonic Hamiltonian, we can solve the dynamics analytically and derive $\hat {a} _{k=0,K}   $ and $\hat a _{k\perp,K}  $ as
\begin{align}
\label{neq:aint.com.K}
\hat {a} _{k=0,K}  & = \hat a _{k=0} \cos \theta _{K} + \hat a _{k\perp} \sin \theta _{K} 
, \\
\hat a _{k\perp,K}  & = -\hat a _{k=0} \sin \theta _{K} + \hat a _{k\perp} \cos \theta _{K} 
,
\end{align}
where the $K$-dependent rotation angle is
\begin{align}
\label{neq:theta.K.coh.approx}
\theta _{K} & = 2K \varphi = 2K \arcsin \frac{1}{\sqrt{N}}  
.
\end{align}
Thus, the expectated boson number in each probe mode can be evaluated as
\begin{align}
\braket{\psi _{\text{out},K}| \hat {a} _{x } ^{\dag} \hat {a} _{x } | \psi _{\text{out},K}}
= 
\braket{\psi _{\text{coh}}| \hat {a} _{x,K} ^{\text{(i)} \dag} \hat {a} _{x,K} ^{\text{(i)}}| \psi _{\text{coh}} } 
,
\end{align}
where we introduce the interaction-picture operator $\hat {a} _{x ,K} ^{\text{(i)}} \coloneqq \hat U_{\text{G}, K} ^{\dag}  \hat {a} _{x } \hat U_{\text{G}, K}  $ via the inverse unitary transformation of that in Eq.~\eqref{neq:intframe.a.com.def}. Using the definition of the $k=0$ mode lowering operator in Eq.~\eqref{neq:a.x0.conv.k}, the action of the unitary transformation $\hat U_{\text{G}, K} $ as specified in Eq.~\eqref{neq:aint.com.K}, and the permutation symmetry of all the modes $a_{x}$ with $x \ne x_{0}$ during the search process, we can compute the target-mode boson-number expectation value as
\begin{align}
\braket{\psi _{\text{coh}}| \hat {a} _{x_{0},K} ^{\text{(i)} \dag} \hat {a} _{x_{0},K} ^{\text{(i)}}| \psi _{\text{coh}} }
& = \braket{\psi _{\text{coh}}| \hat {a} _{k=0} ^{\dag} \hat {a} _{k=0} | \psi _{\text{coh}} } \sin ^{2} (\theta _{K} + \varphi)
\nonumber \\
& = N |\alpha | ^{2} \sin ^{2} (\theta _{K} + \varphi)
\nonumber \\
& = n _{\budgetlb} \sin ^{2} (\theta _{K} + \varphi) = n _{\budgetlb} \sin ^{2} [ (2 {K} + 1) \varphi]
, 
\end{align}
which produces Eq.~\eqref{app.eq:nx0.K.coh.approx}. Note that the second line in the above equation uses the fact the Heisenberg-picture transformation $\hat {a} _{x ,K} ^{\text{(i)}} \coloneqq \hat U_{\text{G}, K} ^{\dag}  \hat {a} _{x } \hat U_{\text{G}, K}  $ is related to the Schr{\"{o}}dinger-picture transformation, as specified in Eq.~\eqref{neq:aint.com.K}, via a unitary inversion, which transforms $\theta _{K} \to - \theta _{K} $.
Similarly, the expectation values of the other probe modes are 
\begin{align}
\braket{\psi _{\text{coh}}| \hat {a} _{x \ne x_{0},K} ^{\text{(i)} \dag} \hat {a} _{x \ne x_{0},K} ^{\text{(i)}}| \psi _{\text{coh}} }
& = \frac{1}{N-1}\braket{\psi _{\text{coh}}| \hat {a} ^{\dag}_{k=0}  \hat {a} _{k=0}  | \psi _{\text{coh}} } \cos ^{2} (\theta _{K} + \varphi)
\nonumber \\
& = \frac{N}{N-1} |\alpha | ^{2} \cos ^{2} (\theta _{K} + \varphi) 
=  \frac{n _{\budgetlb} }{N-1} \cos ^{2} (\theta _{K} + \varphi) 
\\
& = \frac{n _{\budgetlb} }{N-1} \cos ^{2} [ (2 {K} + 1) \varphi]
,
\end{align}
which proves Eq.~\eqref{app.eq:nxne.K.coh.approx}.

\section{Derivation of \texorpdfstring{Eq.~\eqref{neq:bound.n2.gaus}: Upper}{upper} bounds on second-order boson number moments for Gaussian states}

\label{app.sec:bound.n2.gaus}

In Sec.~\ref{nsec:Gaus.bound} in the main text, we use an upper bound [Eq.~\eqref{neq:bound.n2.gaus}] on the second-order boson number moments for an arbitrary Gaussian state,
\begin{align}
\braket{\Psi ^{(\text{i})} _{x _{2}} (\tau) | [ \hat n ^{\text{(i)} } _{x _{1}}  (\tau )  ] ^{2}  | \Psi ^{(\text{i})}_{x _{2}} (\tau) }& = \braket{\Psi ^{(\text{i})} _{x _{2}} (\tau) | [\hat a ^{\text{(i)}\dag } _{x _{1}}  (\tau ) \hat a ^{\text{(i)}} _{x _{1}}  (\tau )  ]^{2}  | \Psi ^{(\text{i})}_{x _{2}} (\tau) }
\nonumber \\
&\le 3 \braket{\Psi ^{(\text{i})} _{x _{2}} (\tau) | \hat n ^{\text{(i)}  } _{x _{1}}  (\tau )    | \Psi ^{(\text{i})}_{x _{2}} (\tau) } ^{2} + 2 \braket{\Psi ^{(\text{i})} _{x _{2}} (\tau) | \hat n ^{\text{(i)}  } _{x _{1}}  (\tau )    | \Psi ^{(\text{i})}_{x _{2}} (\tau) },
\label{app.eq:bound.n2.gaus}
\end{align}
to derive a tight query-time lower bound for all Gaussian protocols.
In this Appendix, we provide a detailed derivation of Eq.~\eqref{app.eq:bound.n2.gaus}.
To bound the expectation values on the right-hand side of the above equation, we rewrite all bosonic raising and lowering operators in terms of their means $\braket{\Psi ^{(\text{i})} _{x _{2}} (\tau) | \hat a ^{\text{(i)}} _{x _{1}}  (\tau )  | \Psi ^{(\text{i})}_{x _{2}} (\tau) }$ and fluctuations
\begin{align}
\label{app.eq:def.da}
\delta \hat a ^{\text{(i)}} _{x _{1}}  (\tau ) \coloneqq \hat a ^{\text{(i)}} _{x _{1}}  (\tau )- \braket{\Psi ^{(\text{i})} _{x _{2}} (\tau) | \hat a ^{\text{(i)}} _{x _{1}}  (\tau )  | \Psi ^{(\text{i})}_{x _{2}} (\tau) } 
,
\end{align}
so that we have
\begin{align}
\braket{\Psi ^{(\text{i})} _{x _{2}} | [\hat a ^{\text{(i)}\dag } _{x _{1}}  \hat a ^{\text{(i)}} _{x _{1}}   ]^{2} | \Psi ^{(\text{i})}_{x _{2}} }
&= \left | \braket{\Psi ^{(\text{i})} _{x _{2}}   | \hat a ^{\text{(i)}} _{x _{1}}    | \Psi ^{(\text{i})}_{x _{2}}  }\right | ^{4}
+ \braket{\Psi ^{(\text{i})} _{x _{2}}   | \hat a ^{\text{(i)}} _{x _{1}}    | \Psi ^{(\text{i})}_{x _{2}}  } ^{2} \braket{\Psi ^{(\text{i})} _{x _{2}} | [\delta \hat a ^{\text{(i)}\dag } _{x _{1}}   ]^{2} | \Psi ^{(\text{i})}_{x _{2}}  }
\nonumber \\
&\quad + \braket{\Psi ^{(\text{i})} _{x _{2}}   | \hat a ^{\text{(i)}\dag } _{x _{1}}    | \Psi ^{(\text{i})}_{x _{2}}  } ^{2} \braket{\Psi ^{(\text{i})} _{x _{2}} | [\delta \hat a ^{\text{(i)}} _{x _{1}}   ]^{2} | \Psi ^{(\text{i})}_{x _{2}}  } + \braket{\Psi ^{(\text{i})} _{x _{2}} | [\delta \hat a ^{\text{(i)}\dag } _{x _{1}}  \delta \hat a ^{\text{(i)}} _{x _{1}}   ]^{2} | \Psi ^{(\text{i})}_{x _{2}} } 
\nonumber \\
\label{neq:expt.n2.gen}
&\quad + \left | \braket{\Psi ^{(\text{i})} _{x _{2}}   | \hat a ^{\text{(i)}} _{x _{1}} | \Psi ^{(\text{i})}_{x _{2}}  }\right | ^{2} \left ( 4 \braket{\Psi ^{(\text{i})} _{x _{2}} | \delta \hat a ^{\text{(i)}\dag } _{x _{1}}  \delta \hat a ^{\text{(i)}} _{x _{1}}   | \Psi ^{(\text{i})}_{x _{2}} } +1 \right ) 
.
\end{align}
In the above equation and the following derivations,  to simplify notations, we omit the $\tau$ dependence of the interaction-picture operators and the probe states when it does not cause confusion.
Since the reduced probe state stays Gaussian throughout the search process, the fourth moment $\braket{\Psi ^{(\text{i})} _{x _{2}} | [\delta \hat a ^{\text{(i)}\dag } _{x _{1}}  \delta \hat a ^{\text{(i)}} _{x _{1}}   ]^{2} | \Psi ^{(\text{i})}_{x _{2}} } $ satisfies Wick's theorem in terms of Gaussian fluctuation operators $\delta \hat a ^{\text{(i)}} _{x _{1}}  (\tau )$ and $\delta \hat a ^{\text{(i)} \dag} _{x _{1}}  (\tau )$, giving
\begin{align}
\braket{\Psi ^{(\text{i})} _{x _{2}} | [\delta \hat a ^{\text{(i)}\dag } _{x _{1}}  \delta \hat a ^{\text{(i)}} _{x _{1}}   ]^{2} | \Psi ^{(\text{i})}_{x _{2}} } 
&= \braket{\Psi ^{(\text{i})} _{x _{2}} | [\delta \hat a ^{\text{(i)}\dag } _{x _{1}}   ]^{2} | \Psi ^{(\text{i})}_{x _{2}}  }\braket{\Psi ^{(\text{i})} _{x _{2}} | [ \delta \hat a ^{\text{(i)}} _{x _{1}}  ]^{2} | \Psi ^{(\text{i})}_{x _{2}}  }
\nonumber \\
&\quad +  2 \braket{\Psi ^{(\text{i})} _{x _{2}} | \delta \hat a ^{\text{(i)}\dag } _{x _{1}}  \delta \hat a ^{\text{(i)}} _{x _{1}}   | \Psi ^{(\text{i})}_{x _{2}} } ^{2} + \braket{\Psi ^{(\text{i})} _{x _{2}} | \delta \hat a ^{\text{(i)}\dag } _{x _{1}}  \delta \hat a ^{\text{(i)}} _{x _{1}}   | \Psi ^{(\text{i})}_{x _{2}} } 
.
\end{align}
Substituting the above equation into Eq.~\eqref{neq:expt.n2.gen}, we obtain
\begin{align}
\braket{\Psi ^{(\text{i})} _{x _{2}} | [\hat a ^{\text{(i)}\dag } _{x _{1}}  \hat a ^{\text{(i)}} _{x _{1}}   ]^{2} | \Psi ^{(\text{i})}_{x _{2}} }
&= \left ( \braket{\Psi ^{(\text{i})} _{x _{2}}   | \hat a ^{\text{(i)}} _{x _{1}} | \Psi ^{(\text{i})}_{x _{2}}  } ^{2} + \braket{\Psi ^{(\text{i})} _{x _{2}} | [\delta \hat a ^{\text{(i)}} _{x _{1}}   ]^{2} | \Psi ^{(\text{i})}_{x _{2}}  } \right )  \left ( \braket{\Psi ^{(\text{i}) } _{x _{2}}   | \hat a ^{\text{(i)}\dag } _{x _{1}} | \Psi ^{(\text{i})}_{x _{2}}  } ^{2} + \braket{\Psi ^{(\text{i})} _{x _{2}} | [\delta \hat a ^{\text{(i)}\dag } _{x _{1}}   ]^{2} | \Psi ^{(\text{i})}_{x _{2}}  } \right )
\nonumber \\
&\quad +  \left ( 4 \left | \braket{\Psi ^{(\text{i})} _{x _{2}}   | \hat a ^{\text{(i)}} _{x _{1}} | \Psi ^{(\text{i})}_{x _{2}}  }\right | ^{2} + 2\braket{\Psi ^{(\text{i})} _{x _{2}} | \delta \hat a ^{\text{(i)}\dag } _{x _{1}}  \delta \hat a ^{\text{(i)}} _{x _{1}}   | \Psi ^{(\text{i})}_{x _{2}} }  \right ) 
\braket{\Psi ^{(\text{i})} _{x _{2}} | \delta \hat a ^{\text{(i)}\dag } _{x _{1}}  \delta \hat a ^{\text{(i)}} _{x _{1}}   | \Psi ^{(\text{i})}_{x _{2}} }
\nonumber \\
\label{neq:expt.n2.gaus}
&\quad+ \left | \braket{\Psi ^{(\text{i})} _{x _{2}}   | \hat a ^{\text{(i)}} _{x _{1}} | \Psi ^{(\text{i})}_{x _{2}}  }\right | ^{2} +\braket{\Psi ^{(\text{i})} _{x _{2}} | \delta \hat a ^{\text{(i)}\dag } _{x _{1}}  \delta \hat a ^{\text{(i)}} _{x _{1}}   | \Psi ^{(\text{i})}_{x _{2}} } 
.
\end{align}
By definition of the fluctuation operator $\delta \hat a ^{\text{(i)}} _{x _{1}} $ [Eq.~\eqref{app.eq:def.da}], the covariances satisfy
\begin{align}
\braket{\Psi ^{(\text{i})} _{x _{2}} | \hat a ^{\text{(i)}\dag } _{x _{1}}  \hat a ^{\text{(i)}} _{x _{1}}  | \Psi ^{(\text{i})}_{x _{2}} } & = \left | \braket{\Psi ^{(\text{i})} _{x _{2}} | \hat a ^{\text{(i)}} _{x _{1}} | \Psi ^{(\text{i})}_{x _{2}}  }\right | ^{2} + \braket{\Psi ^{(\text{i})} _{x _{2}} | \delta \hat a ^{\text{(i)}\dag } _{x _{1}}  \delta \hat a ^{\text{(i)}} _{x _{1}}  | \Psi ^{(\text{i})}_{x _{2}} } 
, \\
\braket{\Psi ^{(\text{i})} _{x _{2}} | \hat a ^{\text{(i)} } _{x _{1}}  \hat a ^{\text{(i)}} _{x _{1}}  | \Psi ^{(\text{i})}_{x _{2}} } & = \braket{\Psi ^{(\text{i})} _{x _{2}} | \hat a ^{\text{(i)}} _{x _{1}} | \Psi ^{(\text{i})}_{x _{2}}  } ^{2} + \braket{\Psi ^{(\text{i})} _{x _{2}} | [\delta \hat a ^{\text{(i)} } _{x _{1}}   ]^{2}  | \Psi ^{(\text{i})}_{x _{2}} } 
.
\end{align}
Substituting the above equations into Eq.~\eqref{neq:expt.n2.gaus} gives
\begin{align}
& \braket{\Psi ^{(\text{i})} _{x _{2}} | [\hat a ^{\text{(i)}\dag } _{x _{1}}  \hat a ^{\text{(i)}} _{x _{1}}   ]^{2} | \Psi ^{(\text{i})}_{x _{2}} }
\nonumber \\
= {} & \braket{\Psi ^{(\text{i})} _{x _{2}} | \hat a ^{\text{(i)} \dag } _{x _{1}}  \hat a ^{\text{(i)}\dag } _{x _{1}}  | \Psi ^{(\text{i})}_{x _{2}} }
\braket{\Psi ^{(\text{i})} _{x _{2}} | \hat a ^{\text{(i)} } _{x _{1}}  \hat a ^{\text{(i)}} _{x _{1}}  | \Psi ^{(\text{i})}_{x _{2}} }
+ \braket{\Psi ^{(\text{i})} _{x _{2}} | \hat a ^{\text{(i)}\dag } _{x _{1}}  \hat a ^{\text{(i)}} _{x _{1}}  | \Psi ^{(\text{i})}_{x _{2}} } 
\nonumber \\
& +  \left ( 2 \left | \braket{\Psi ^{(\text{i})} _{x _{2}}   | \hat a ^{\text{(i)}} _{x _{1}} | \Psi ^{(\text{i})}_{x _{2}}  }\right | ^{2} + 2\braket{\Psi ^{(\text{i})} _{x _{2}} |  \hat a ^{\text{(i)}\dag } _{x _{1}}   \hat a ^{\text{(i)}} _{x _{1}}   | \Psi ^{(\text{i})}_{x _{2}} }  \right ) 
\braket{\Psi ^{(\text{i})} _{x _{2}} | \delta \hat a ^{\text{(i)}\dag } _{x _{1}}  \delta \hat a ^{\text{(i)}} _{x _{1}}   | \Psi ^{(\text{i})}_{x _{2}} }
\\
= {} & \braket{\Psi ^{(\text{i})} _{x _{2}} | \hat a ^{\text{(i)} \dag } _{x _{1}}  \hat a ^{\text{(i)}\dag } _{x _{1}}  | \Psi ^{(\text{i})}_{x _{2}} }
\braket{\Psi ^{(\text{i})} _{x _{2}} | \hat a ^{\text{(i)} } _{x _{1}}  \hat a ^{\text{(i)}} _{x _{1}}  | \Psi ^{(\text{i})}_{x _{2}} }
+ \braket{\Psi ^{(\text{i})} _{x _{2}} | \hat a ^{\text{(i)}\dag } _{x _{1}}  \hat a ^{\text{(i)}} _{x _{1}}  | \Psi ^{(\text{i})}_{x _{2}} } 
\nonumber \\
& +  2 \left (  \braket{\Psi ^{(\text{i})} _{x _{2}} |  \hat a ^{\text{(i)}\dag } _{x _{1}}   \hat a ^{\text{(i)}} _{x _{1}}   | \Psi ^{(\text{i})}_{x _{2}} } + \left | \braket{\Psi ^{(\text{i})} _{x _{2}}   | \hat a ^{\text{(i)}} _{x _{1}} | \Psi ^{(\text{i})}_{x _{2}}  }\right | ^{2} \right ) 
\left ( \braket{\Psi ^{(\text{i})} _{x _{2}} |  \hat a ^{\text{(i)}\dag } _{x _{1}}   \hat a ^{\text{(i)}} _{x _{1}}   | \Psi ^{(\text{i})}_{x _{2}} }  - \left | \braket{\Psi ^{(\text{i})} _{x _{2}}   | \hat a ^{\text{(i)}} _{x _{1}} | \Psi ^{(\text{i})}_{x _{2}}  }\right | ^{2}  \right ) 
.
\end{align}
Using the Cauchy-Schwarz inequality and the fact that $ | \braket{\Psi ^{(\text{i})} _{x _{2}}   | \hat a ^{\text{(i)}} _{x _{1}} | \Psi ^{(\text{i})}_{x _{2}}  }  | \le \sqrt{\braket{\Psi ^{(\text{i})} _{x _{2}} | \hat a ^{\text{(i)}\dag } _{x _{1}}  \hat a ^{\text{(i)}} _{x _{1}}  | \Psi ^{(\text{i})}_{x _{2}} } }$, we derive the following upper bound:
\begin{align}
& \braket{\Psi ^{(\text{i})} _{x _{2}} (\tau) | [\hat a ^{\text{(i)}\dag } _{x _{1}}  (\tau ) \hat a ^{\text{(i)}} _{x _{1}}  (\tau )  ]^{2}  | \Psi ^{(\text{i})}_{x _{2}} (\tau) }
\nonumber \\
\le {}& \braket{\Psi ^{(\text{i})} _{x _{2}} (\tau) | \hat a ^{\text{(i)}} _{x _{1}}  (\tau ) \hat a ^{\text{(i)} \dag } _{x _{1}}  (\tau )  | \Psi ^{(\text{i})}_{x _{2}} (\tau) } \braket{\Psi ^{(\text{i})} _{x _{2}} (\tau) | \hat a ^{\text{(i)}\dag } _{x _{1}}  (\tau ) \hat a ^{\text{(i)}} _{x _{1}}  (\tau )  | \Psi ^{(\text{i})}_{x _{2}} (\tau) } 
\nonumber \\
& + \braket{\Psi ^{(\text{i})} _{x _{2}} (\tau) | \hat a ^{\text{(i)}\dag } _{x _{1}}  (\tau ) \hat a ^{\text{(i)}} _{x _{1}}  (\tau )  | \Psi ^{(\text{i})}_{x _{2}} (\tau) } + 2 \braket{\Psi ^{(\text{i})} _{x _{2}} (\tau) | \hat a ^{\text{(i)}\dag } _{x _{1}}  (\tau ) \hat a ^{\text{(i)}} _{x _{1}}  (\tau )  | \Psi ^{(\text{i})}_{x _{2}} (\tau) } ^{2} 
\nonumber \\
\le {}& 3 \braket{\Psi ^{(\text{i})} _{x _{2}} (\tau) | \hat a ^{\text{(i)}\dag } _{x _{1}}  (\tau ) \hat a ^{\text{(i)}} _{x _{1}}  (\tau )  | \Psi ^{(\text{i})}_{x _{2}} (\tau) } ^{2} + 2 \braket{\Psi ^{(\text{i})} _{x _{2}} (\tau) | \hat a ^{\text{(i)}\dag } _{x _{1}}  (\tau ) \hat a ^{\text{(i)}} _{x _{1}}  (\tau )  | \Psi ^{(\text{i})}_{x _{2}} (\tau) }
\nonumber \\
\label{appeq:bound.n2.gaus}
= {}& 3 \braket{\Psi ^{(\text{i})} _{x _{2}} (\tau) | \hat n ^{\text{(i)}  } _{x _{1}}  (\tau )    | \Psi ^{(\text{i})}_{x _{2}} (\tau) } ^{2} + 2 \braket{\Psi ^{(\text{i})} _{x _{2}} (\tau) | \hat n ^{\text{(i)}  } _{x _{1}}  (\tau )    | \Psi ^{(\text{i})}_{x _{2}} (\tau) }
.
\end{align}

Note that this upper bound only applies when the probe state stays Gaussian throughout the entire search protocol, which is the starting assumption in Sec.~\ref{nsec:Gaus.bound}.
Note also that the above upper bound holds for arbitrary choices of $x_{1}$ and $x_{2}$.

\section{Derivation of \texorpdfstring{Eq.~\eqref{neq:Ttot.lb.coh}: Lower}{lower} bound on query complexity using protocols involving only coherent-state probes}

\label{app.sec:bound.Ttot.lb.coh}

In the main text, we claim that the search time of any protocol using only coherent-state probes satisfies the following lower bound [Eq.~\eqref{neq:Ttot.lb.coh}]:
\begin{align}
T_{\text{tot}} > \frac{p ^{(0)}_{\text{th}} }{2B _{0} } \sqrt{\frac{N-1}{n _{\budgetlb}}}
.
\end{align}
In this Appendix, we provide a detailed proof of this result.
We begin by considering arbitrary search protocols such that the reduced, unconditional probe state is a coherent state throughout the protocol, with arbitrary oracle-independent controls and ancillas. By definition of coherent states, we have
\begin{align}
\hat a _{x} \ket{\Psi _{y} (\tau)} = \alpha _{x} (\tau; y) \ket{\Psi _{y} (\tau)} 
,
\end{align}
where $\alpha _{x} (\tau; y)$ denotes the coherent-state amplitude of mode $x$ at time $\tau$ evolving under the $y$th oracle and other oracle-independent controls.
Thus, in the interaction picture, we have
\begin{align}
\label{seq:def.coh.amps}
\hat a ^{\text{(i)}}_{x} (\tau)\ket{\Psi ^{(\text{i})}_{y} (\tau)} = \alpha _{x} (\tau; y) \ket{\Psi ^{(\text{i})}_{y} (\tau)}
.
\end{align}
Substituting the above equation into 
Eq.~\eqref{neq:int.overlap.time.deriv}, we obtain
\begin{align}
\partial _{\tau } \braket{\Psi ^{(\text{i})}_{x_{1}} (\tau ) | \Psi ^{(\text{i})}_{x_{2}}  (\tau ) } 
& = i \braket{\Psi ^{(\text{i})}_{x_{1}} (\tau )| [\hat H ^{\text{(i)}} _{x _{1}} (\tau ) - \hat H ^{\text{(i)}} _{x _{2}} (\tau ) ] | \Psi ^{(\text{i})}_{x_{2}} (\tau) }
\\
& = i B _{0} [\alpha _{x_{1}}^{*} (\tau; x_{1}) \alpha _{x_{1}} (\tau; x_{2}) - \alpha _{x_{2}}^{*} (\tau; x_{1}) \alpha _{x_{2}} (\tau; x_{2}) ] \braket{\Psi ^{(\text{i})}_{x_{1}} (\tau ) | \Psi ^{(\text{i})}_{x_{2}} (\tau) }
,
\end{align}
so that dividing both sides of the above equation by $\braket{\Psi ^{(\text{i})}_{x_{1}} (\tau ) | \Psi ^{(\text{i})}_{x_{2}} (\tau) }$, we have
\begin{align}
\partial _{\tau } \ln \braket{\Psi ^{(\text{i})}_{x_{1}} (\tau ) | \Psi ^{(\text{i})}_{x_{2}} (\tau) }
& = \frac{\partial _{\tau } \braket{\Psi ^{(\text{i})}_{x_{1}} (\tau ) | \Psi ^{(\text{i})}_{x_{2}}  (\tau ) } }{\braket{\Psi ^{(\text{i})}_{x_{1}} (\tau ) | \Psi ^{(\text{i})}_{x_{2}} (\tau) }}
\\
& = i B _{0} [\alpha _{x_{1}}^{*} (\tau; x_{1}) \alpha _{x_{1}} (\tau; x_{2}) - \alpha _{x_{2}}^{*} (\tau; x_{1}) \alpha _{x_{2}} (\tau; x_{2}) ] 
.
\end{align}
Taking the real part of the above equation, we 
find
\begin{align}
\partial _{\tau } \ln |\braket{\Psi ^{(\text{i})}_{x_{1}} (\tau ) | \Psi ^{(\text{i})}_{x_{2}} (\tau) }|
& = - B _{0} \text{Im} [\alpha _{x_{1}}^{*} (\tau; x_{1}) \alpha _{x_{1}} (\tau; x_{2}) - \alpha _{x_{2}}^{*} (\tau; x_{1}) \alpha _{x_{2}} (\tau; x_{2}) ] 
\\
& = B _{0} \text{Im} [\alpha _{x_{2}}^{*} (\tau; x_{1}) (\alpha _{x_{2}} (\tau; x_{2}) -\alpha _{x_{2}} (\tau; x_{1})) ] 
\nonumber \\
& \quad + B _{0} \text{Im} [ (\alpha _{x_{1}}^{*} (\tau; x_{2}) - \alpha _{x_{1}}^{*} (\tau; x_{1})) \alpha _{x_{1}} (\tau; x_{2})  ] 
,
\end{align}
so that from the triangle inequality, we obtain
\begin{align}
&\frac{1}{B _{0} }\left | \partial _{\tau } \ln |\braket{\Psi ^{(\text{i})}_{x_{1}} (\tau ) | \Psi ^{(\text{i})}_{x_{2}} (\tau) }| \right |
\nonumber \\
\label{neq:time.deriv.overlap.coh}
\le {}&  |\alpha _{x_{2}}^{*} (\tau; x_{1})| \cdot |\alpha _{x_{2}} (\tau; x_{2}) -\alpha _{x_{2}} (\tau; x_{1})|
+ |\alpha _{x_{1}} (\tau; x_{2}) | \cdot |\alpha _{x_{1}}^{*} (\tau; x_{2}) - \alpha _{x_{1}}^{*} (\tau; x_{1})|
.
\end{align}
Noting that the probes remain coherent states throughout the protocol, we can bound the overlap between the probe-ancilla states using the monotonicity of the fidelity under partial trace as 
\begin{align}
|\braket{\Psi ^{(\text{i})}_{x_{1}} (\tau ) | \Psi ^{(\text{i})}_{x_{2}} (\tau) }|
\le |\braket{ \boldsymbol{\alpha}(\tau; x_{1}) | \boldsymbol{\alpha}(\tau; x_{2}) }|
,
\end{align}
where $\ket{\boldsymbol{\alpha}(\tau; x_{1})}$ denotes the probe-mode coherent state corresponding to the $x_{1}$th oracle at time $\tau$. Since the overlap between two coherent states can be computed explicitly as 
\begin{align}
|\braket{ \boldsymbol{\alpha}(\tau; x_{1}) | \boldsymbol{\alpha}(\tau; x_{2}) }|
= e ^{- \frac{1}{2} \| \boldsymbol{\alpha}(\tau; x_{1})  - \boldsymbol{\alpha}(\tau; x_{2}) \|^{2} }
,
\end{align}
we can relate the vector norm of the difference between the coherent-state amplitudes $\| \boldsymbol{\alpha}(\tau; x_{1})  - \boldsymbol{\alpha}(\tau; x_{2}) \|$ to the state overlap function $\braket{\Psi ^{(\text{i})}_{x_{1}} (\tau ) | \Psi ^{(\text{i})}_{x_{2}} (\tau) }$ as
\begin{align}
\| \boldsymbol{\alpha}(\tau; x_{1})  - \boldsymbol{\alpha}(\tau; x_{2}) \|^{2} 
& = -2 \ln |\braket{ \boldsymbol{\alpha}(\tau; x_{1}) | \boldsymbol{\alpha}(\tau; x_{2}) }|
\\
& \le  -2 \ln |\braket{\Psi ^{(\text{i})}_{x_{1}} (\tau ) | \Psi ^{(\text{i})}_{x_{2}} (\tau) }|
.
\end{align}
Substituting the above relation into Eq.~\eqref{neq:time.deriv.overlap.coh}, we have
\begin{align}
& \left | \partial _{\tau } \ln |\braket{\Psi ^{(\text{i})}_{x_{1}} (\tau ) | \Psi ^{(\text{i})}_{x_{2}} (\tau) }| \right |
\le B _{0} (|\alpha _{x_{2}} (\tau; x_{1})| + |\alpha _{x_{1}} (\tau; x_{2})| ) \cdot 
\sqrt{-2 \ln |\braket{\Psi ^{(\text{i})}_{x_{1}} (\tau ) | \Psi ^{(\text{i})}_{x_{2}} (\tau) }|}
, 
\\
& \Rightarrow \left | 
\partial _{\tau } \sqrt{ - \ln |\braket{\Psi ^{(\text{i})}_{x_{1}} (\tau ) | \Psi ^{(\text{i})}_{x_{2}} (\tau) }| }
\right | 
\le \frac{B _{0} }{\sqrt{2}} (|\alpha _{x_{2}} (\tau; x_{1})| + |\alpha _{x_{1}} (\tau; x_{2})|)
.
\end{align}
Integrating the time derivative in the above equation from $\tau =0$ to $T_{\text{tot}}$ and applying the triangle inequality, we further have
\begin{align}
& \sqrt{- \ln |\braket{\Psi ^{(\text{i})}_{x_{1}} (T_{\text{tot}} ) | \Psi ^{(\text{i})}_{x_{2}} (T_{\text{tot}}) }| } -  \sqrt{- \ln |\braket{\Psi ^{(\text{i})}_{x_{1}} (0 ) | \Psi ^{(\text{i})}_{x_{2}} (0) }| }
\nonumber \\
\le { } &
\frac{B _{0} }{\sqrt{2}} \int ^{T_{\text{tot}}} _{0}d \tau (|\alpha _{x_{2}} (\tau; x_{1})| + |\alpha _{x_{1}} (\tau; x_{2})|)
.
\end{align}
Noting that $\braket{\Psi ^{(\text{i})}_{x_{1}} (0 ) | \Psi ^{(\text{i})}_{x_{2}} (0) } = 1$ and using the inequality $ 1-x \le-\ln x$ ($\forall x \in (0,1]$), we obtain the following bound on the probe states' overlap:
\begin{align}
\sqrt{- \ln |\braket{\Psi ^{(\text{i})}_{x_{1}} (T_{\text{tot}} ) | \Psi ^{(\text{i})}_{x_{2}} (T_{\text{tot}}) }| } 
& \le 
\frac{B _{0} }{\sqrt{2}} \int ^{T_{\text{tot}}} _{0}d \tau (|\alpha _{x_{2}} (\tau; x_{1})| + |\alpha _{x_{1}} (\tau; x_{2})|)
, \\
\Rightarrow \sqrt{1-|\braket{\Psi ^{(\text{i})}_{x_{1}} (T_{\text{tot}} ) | \Psi ^{(\text{i})}_{x_{2}} (T_{\text{tot}}) }|}
& \le 
\frac{B _{0} }{\sqrt{2}} \int ^{T_{\text{tot}}} _{0}d \tau (|\alpha _{x_{2}} (\tau; x_{1})| + |\alpha _{x_{1}} (\tau; x_{2})|)
.
\end{align}
Substituting the above inequality into Eq.~\eqref{neq:avg.bures.int.0}, we can relate the threshold probability to the integral over probe-state coherent amplitudes as 
\begin{align}
p ^{(0)}_{\text{th}} & < \sum _{x _{1} \ne x _{2}} \frac{\sqrt{ 2\left ( 1- |\! \braket{  \Psi ^{(\text{i})}_{x_{1}} (T_{\text{tot}} ) | \Psi ^{(\text{i})}_{x_{2}} (T_{\text{tot}} ) }\! | \right )  }}{N (N-1) } 
\\
\label{neq:coh.amp.int.0}
& \le \frac{B _{0} }{N (N-1) } \sum _{x _{1} \ne x _{2}} \int ^{T_{\text{tot}}} _{0}d \tau (|\alpha _{x_{2}} (\tau; x_{1})| + |\alpha _{x_{1}} (\tau; x_{2})|)
\\
& = \frac{2B _{0} }{N (N-1) } \sum _{x _{1} \ne x _{2}} \int ^{T_{\text{tot}}} _{0}d \tau |\alpha _{x_{2}} (\tau; x_{1})| 
.
\end{align}
In the last line, we have used the fact that the two terms in the integrand in Eq.~\eqref{neq:coh.amp.int.0} are symmetric under the exchange $x_{1} \leftrightarrow x_{2}$. From the Cauchy-Schwarz inequality, we can further bound the above expression as 
\begin{align}
p ^{(0)}_{\text{th}} & < \frac{2B _{0} }{N} \sum _{x _{1} }
\int ^{T_{\text{tot}}} _{0}d \tau \sqrt{ \frac{1}{N-1} \sum _{x_{2}\ne x_{1}} |\alpha _{x_{2}} (\tau; x_{1})| ^{2}}
\\
\label{neq:coh.amp.int.1}
&\le \frac{2B _{0} }{N} \sum _{x _{1} } \sqrt{
\frac{T_{\text{tot}}}{N-1} \int ^{T_{\text{tot}}} _{0}d \tau \sum _{x_{2}\ne x_{1}} |\alpha _{x_{2}} (\tau; x_{1})| ^{2}
}
.
\end{align}

Substituting the definition~\eqref{seq:def.coh.amps} of the coherent-state amplitudes of the probes into the boson-number constraint in Eq.~\eqref{neq:tavg.nph}, we derive the following constraint:
\begin{align}
\forall \, x_{1}: \quad 
& \frac{1}{T_{\text{tot}}} \sum _{x_{2}} \int ^{T_{\text{tot}}} _{0} 
\braket{\Psi ^{(\text{i})}_{x _{1}} | \hat H ^{\text{(i)}} _{x _{2}} (\tau ) | \Psi ^{(\text{i})}_{x _{1}} }  d \tau 
= \frac{B _{0} }{T_{\text{tot}}} \sum _{x_{2}} \int ^{T_{\text{tot}}} _{0} |\alpha _{x_{2}} (\tau; x_{1})|^{2} d \tau 
\le B _{0}  n _{\budgetlb} 
.
\end{align}
Substituting the above bound into Eq.~\eqref{neq:coh.amp.int.1}, we derive the following lower bound on the total search time $T_{\text{tot}} $ for any coherent-state protocol to identify the marked mode:
\begin{align}
p ^{(0)}_{\text{th}} < \frac{2B _{0} }{N} & \sum _{x _{1} } \sqrt{
\frac{T_{\text{tot}}}{N-1}  n _{\budgetlb} T_{\text{tot}}
}
= 2 B _{0} T_{\text{tot}} \sqrt{\frac{ n _{\budgetlb}}{N-1}}
,
\\
& \Rightarrow
T_{\text{tot}} > \frac{p ^{(0)}_{\text{th}} }{2B _{0} } \sqrt{\frac{N-1}{n _{\budgetlb}}}
.
\end{align}
In the $N\gg 1$ regime, we thus find the following lower bound for coherent-state strategies (where we change the subscript to emphasize the setting):
\begin{align}
\label{seq:Ttot.lb.coh}
T_{\text{coh}} = \Omega \left ( \frac{1}{B_0}\sqrt{\frac{N}{n_{\budgetlb}}} \right )
,
\end{align}
which proves Eq.~\eqref{neq:Ttot.lb.coh}.

\end{document}